\documentclass[notitlepage]{revtex4-1}
\usepackage[legalpaper, margin=1.15in]{geometry}
\usepackage{setspace}
\usepackage[utf8]{inputenc}
\usepackage[colorlinks=true,citecolor=blue,linkcolor=blue]{hyperref}
\usepackage[normalem]{ulem}
\usepackage{url}
\usepackage{graphicx,wrapfig,float,slashed,cancel}
\usepackage{amsmath,amssymb,epsfig,xcolor,stmaryrd}
\usepackage{bm}
\usepackage{enumitem}
\usepackage{hhline,multirow,tabularx}  
\usepackage{graphics}
\usepackage{latexsym}
\usepackage{rotating}
\usepackage{subfigure}
\usepackage{color}
\usepackage{blindtext}
\usepackage{lipsum}
\usepackage{physics}

\begin{document}


\title{Generalized parton distributions at zero skewness}

\author{Hadi Hashamipour$^{a}$}
\email{h\_hashamipour@ipm.ir}

\author{Muhammad Goharipour$^{b,a}$}
\email{muhammad.goharipour@ipm.ir}
\thanks{Corresponding author}

\author{K.~Azizi$^{b,c,a}$}
\email{kazem.azizi@ut.ac.ir}

\author{S.V. Goloskokov$^{d}$}
\email{goloskkv@theor.jinr.ru}

\affiliation{
$^{a}$School of Particles and Accelerators, Institute for Research in Fundamental Sciences (IPM), P.O. Box 19395-5746, Tehran, Iran\\
$^{b}$Department of Physics, University of Tehran, North Karegar Avenue, Tehran 14395-547, Iran\\
$^{c}$Department of Physics, Do\u gu\c s University, Ac{\i}badem-Kad{\i}k\"oy, 34722 Istanbul, Turkey\\
$^{d}$Bogoliubov Laboratory of Theoretical Physics, Joint Institute for Nuclear Research, Dubna 141980, Moscow region, Russia}

\date{\today}

\preprint{}

\begin{abstract}

We present a new determination of the generalized parton distributions (GPDs) with their uncertainties at zero skewness, $ \xi =0 $, through a simultaneous analysis of all available experimental data of the nucleon electromagnetic form factors (FFs), nucleon charge and magnetic radii, proton axial FFs (AFFs) and wide-angle Compton scattering (WACS) cross sections for the first time, and we investigate whether there is any tension between these data. This can be considered the most comprehensive analysis of GPDs at $ \xi=0 $ performed so far. We show that such an analysis provides the simultaneous determination of three kinds of GPDs, namely $ H^q $, $ \widetilde{H}^q $ and $ E^q $, considering also the sea-quark contributions. As a result, we find that the inclusion of the WACS and AFF data at larger values of the momentum transfer squared $ Q^2=-t $ can put new constraints on GPDs and change them drastically in some cases. We show that there is a considerable tension between the WACS and the proton magnetic form factor ($ G_M^p $) data, especially at larger values of $ -t $. However, we indicate that the results for the gravitational FF $ M_2 $ and the proton total angular momentum $ J^p $ calculated using the extracted GPDs are in relatively good agreement with the light-cone QCD sum rules (LCSRs) and lattice QCD predictions when the sea-quark contributions are considered and both WACS and $ G_M^p $ data are included in the analyses simultaneously.

\end{abstract}


\maketitle

\renewcommand{\thefootnote}{\#\arabic{footnote}}
\setcounter{footnote}{0}

\section{Introduction}\label{sec:one} 

The study of hadronic structure is a key problem of modern physics. Generalized parton 
distributions (GPDs), proposed in Refs.~\cite{Muller:1994ses,Radyushkin:1996nd,Ji:1996nm,Ji:1996ek,Burkardt:2000za}, are among the essential objects that give  broad information on the internal structure of hadrons. 
GPDs are nonperturbative objects which depend on three variables: $ x $, the longitudinal momentum 
fraction of the proton carried by partons, $ \xi $, the skewness that is the longitudinal 
momentum transfer; and $ t $, the momentum transfer squared. In the forward limit $ \xi=0 $ and $ t=0 $, GPDs are equal to the parton distribution functions (PDFs)~\cite{Alekhin:2012ig,ZEUS:2020ddd,Cocuzza:2021cbi,NNPDF:2021njg,Hou:2019efy,Bailey:2020ooq} that give information on the longitudinal hadron structure. On the other hand, moments of 
GPDs are associated with the hadron form factors (FFs)~\cite{Ji:1996ek} which are connected with 
the hadron distributions in the transverse plane.  Thus, GPDs give us information on 
the 3D structure of nucleons~\cite{Burkardt:2002hr,Belitsky:2003nz}. Recently, it has been indicated that GPDs can also be defined for spin-$ 3/2 $ particles~\cite{Fu:2022bpf}.

GPDs are an essential ingredient of the hard exclusive processes such as deeply 
virtual Compton scattering (DVCS)~\cite{Ji:1996nm,Radyushkin:1997ki,Collins:1998be,BessidskaiaBylund:2022qgg,Bhattacharya:2022xxw,Braun:2022qly,Semenov-Tian-Shansky:2023bsy}, deeply virtual meson production (DVMP)~\cite{Goeke:2001tz,Vanderhaeghen:1999xj,Goloskokov:2005sd,Goloskokov:2007nt,CLAS:2022iqy}, wide-angle Compton scattering (WACS)~\cite{Radyushkin:1998rt,Diehl:1998kh}, exclusive photoproduction of a $ \gamma\rho $ pair~\cite{Duplancic:2023kwe},  exclusive heavy-vector-meson production (HVMP)~\cite{Dutrieux:2023qnz}, and also single diffractive hard exclusive processes (SDHEPs)~\cite{Qiu:2022pla}. A factorization theorem~\cite{Ji:1996nm,Ji:1996ek,Collins:1998be,Radyushkin:1996ru,Collins:1996fb} allows one to express the amplitudes of these reactions as the convolution of a hard subprocess amplitude and a soft part that is expressed in terms of GPDs. At zero skewness, GPDs are connected with the electromagnetic FFs~\cite{Guidal:2004nd}, which makes them also an essential ingredient of the elastic electron-nucleon scattering. 
One can find more information on GPDs, e.g., in the review papers~\cite{Goeke:2001tz,Diehl:2003ny,Belitsky:2005qn}.

Access to GPDs from exclusive reactions is indirect because their contribution to 
the process amplitude appears in the integrated forms. As a result, one needs to 
use an appropriate model of GPDs with further fitting of their parameters from 
experiment. We would like to mention some model approaches that were used to extract information on GPDs from DVCS and FF data. There are some dynamical models of hadron structure for example, the Reggeized spectator model, which gives a parametrization of GPDs $ H $, $ E $, and corresponding polarized GPDs,~\cite{Gonzalez-Hernandez:2012xap,Kriesten:2019jep,Kriesten:2021sqc} as well as conformal-moment-based models that were used to analyze GPD properties~\cite{Kumericki:2007sa,Kumericki:2009uq,Guo:2022upw,Guo:2023ahv}. Results of these models are discussed in a review paper (Ref.~\cite{Kumericki:2016ehc}). The light-front approaches
can also be utilized to explore GPDs and hadron structure~\cite{Mondal:2019jdg,Ahmady:2021qed,Xu:2021wwj,Nair:2023lir}.

The lattice analyses of PDFs and GPDs were summarized in Refs.~\cite{Lin:2017snn,Constantinou:2020hdm} and give access mainly to the distribution moments. The new large-momentum effective theory 
gives us the possibility to extend lattice methods to estimate the $ x $ dependence of the PDFs and GPDs~\cite{Ji:2014gla,Ji:2020ect,Lin:2021brq}.
The GPD models based on the double-distribution (DD) representation~\cite{Musatov:1999xp} were 
successfully used for phenomenological analyses of DVCS~\cite{Guidal:2013rya} and DVMP~\cite{Goeke:2001tz,Vanderhaeghen:1999xj,Goloskokov:2005sd,Goloskokov:2007nt}. 
The DD generates the $ \xi $ dependence of GPDs by integration of the DD function. It 
connects generalized distributions with GPDs at $ \xi=0 $, which is a research topic of 
the present paper.

There are several observables which are related to GPDs at $ \xi=0 $~\cite{Diehl:2013xca}, including the nucleon Sachs FFs $ G_E $ and $ G_M $, the axial FF (AFF) $ G_A $, the charge and magnetic radii $ \left<r_{E}^2\right> $ and $ \left<r_{M}^2\right> $, and the WACS cross section. Their measurements can be used to extract the nucleon GPDs by means of a $ \chi^2 $ analysis considering a suitable phenomenological framework~\cite{Diehl:2004cx,Diehl:2013xca,Hashamipour:2019pgy,Hashamipour:2020kip,Hashamipour:2021kes}. For example, in Ref.~\cite{Hashamipour:2019pgy}, the authors have independently determined the polarized GPDs $ \widetilde{H}^q $, where $ q $ denotes up ($ u $) and down ($ d $) quarks, by analyzing the world experimental measurements of $ G_A $. Note that wherever we use the superscript $ q $ without a  subscript $ v $ pointing out the valence quarks, we mean both the valence and sea-quark ($ \bar{q} $) contributions.
A simultaneous analysis of the $ G_A $ and WACS data has been performed in Ref.~\cite{Hashamipour:2020kip} to extract $ \widetilde{H}^q $ as well as the unpolarized GPDs $ {H}^q $, taking GPDs $ E^q $ from an older analysis of the world electron scattering data~\cite{Diehl:2004cx}. A new determination of the GPDs $ H_v^q $ and $ E_v^q $ through the $ \chi^2 $ analysis of the nucleon Sachs FFs data in addition to the measurements of the nucleon charge and magnetic radii has also been performed recently~\cite{Hashamipour:2021kes}. 

In the present study, we are going to perform the most comprehensive analysis of GPDs at $ \xi =0$, by including all available experimental data of the nucleon FFs $ G_E $ and $ G_M $ (or their ratios), nucleon radii $ \left<r_{E}^2\right> $ and $ \left<r_{M}^2\right> $, proton AFFs $ G_A $, and the WACS cross section for the first time. This provides us a simultaneous extraction of three kinds of GPDs: namely $ H^q $, $ \widetilde{H}^q $, and $ E^q $. It is expected that such an analysis will lead to more universal and precise GPDs, although we show that there is a considerable tension between the WACS and the proton $ G_M^p $ data (not other electromagnetic data), especially at larger values of $ -t $, so that a set of universal GPDs fails to provide a desirable description of these data simultaneously.

The content of the present paper is as follows: In Sec.~\ref{sec:two}, we 
introduce the phenomenological framework that we use to extract GPDs from the experimental measurements. 
The datasets which are included in the present study and related remarks are presented in Sec.~\ref{sec:three}. Section~\ref{sec:four} is devoted to presenting the results obtained and investigating the 
goodness of fits. By performing several analyses and comparing them with each other, the impacts of various datasets on the extracted GPDs are studied. Moreover, we calculate other quantities such as the gravitational FF $ M_2 $ and the total angular momentum of the proton $ J^p $ (as well as its individual quark contributions) using our final sets of GPDs and compare them with the corresponding ones obtained from other studies. We summarize our results and conclusions in Sec.~\ref{sec:five}.

%
\section{ Phenomenological framework}\label{sec:two}

The phenomenological method that we use in the present study is similar to our previous study~\cite{Hashamipour:2021kes} in which we extracted the GPDs $ H_v^q $ and $ E_v^q $ by analyzing the nucleon Sachs FFs and radii data simultaneously. However, there are some differences, since we are going to include the proton AFF and WACS data in our new analysis too. First, we must also parametrize the polarized GPDs, $ \widetilde{H}^q $, because of their presence in the theoretical calculations of $ G_A $~\cite{Hashamipour:2019pgy} and the WACS cross section~\cite{Hashamipour:2020kip,Huang:2001ej}. 
Second, we must consider the contributions of the sea-quark distributions in addition to the valence sectors for the same reason.

In the present study, we use the same ansatz proposed in Refs.~\cite{Diehl:2004cx,Diehl:2013xca} and reused in Refs.~\cite{Hashamipour:2019pgy,Hashamipour:2020kip,Hashamipour:2021kes} to express GPDs at $ \xi =0$, for both valence and sea-quark distributions:
\begin{align}
H_v^q(x,t,\mu^2)= q_v(x,\mu^2)\exp [tf_v^q(x)],  \nonumber \\ 
E_v^q(x,t,\mu^2)= e_v^q(x,\mu^2)\exp [tg_v^q(x)], \nonumber \\ 
\widetilde{H}_v^q(x,t,\mu^2)= \Delta q_v(x,\mu^2)\exp [t\widetilde{f}_v^q(x)],  \nonumber \\ 
H^{\bar{q}}(x,t,\mu^2)= \bar{q}(x,\mu^2)\exp [tf^{\bar{q}}(x)], \nonumber \\ 
E^{\bar{q}}(x,t,\mu^2)= e^{\bar{q}}(x,\mu^2)\exp [tg^{\bar{q}}(x)], \nonumber \\ 
\widetilde{H}^{\bar{q}}(x,t,\mu^2)= \Delta \bar{q}(x,\mu^2)\exp [t\widetilde{f}^{\bar{q}}(x)],
\label{Eq1}
\end{align}
where $ f $, $ g $, and $ \widetilde{f} $ are profile functions. Note that the strange-quark contribution is ignored as suggested by Diehl and Kroll (DK13)~\cite{Diehl:2013xca}. The forward limits of GPDs $ H^q $ and $ \widetilde{H}^q $ namely, the unpolarized and polarized PDFs $ q(x,\mu^2) $ and $ \Delta q(x,\mu^2) $ are taken from the {NNPDF} analyses~\cite{NNPDF:2021njg,Nocera:2014gqa} at the next-to-leading order (NLO) and scale $ \mu=2 $ GeV,  utilizing the \texttt{LHAPDF} package~\cite{Buckley:2014ana}. Note that such a choice has the advantage that both $ q(x) $ and $ \Delta q(x) $ have been determined using the same methodology. Moreover, in Refs.~\cite{Hashamipour:2019pgy,Hashamipour:2020kip} the authors have shown that the dependence of the ansatz introduced in Eq.~(\ref{Eq1}) on the choice of PDFs and polarized PDFs is negligible. 

For the case of GPDs $ E^q $, two points should be mentioned. First, their forward limits are not available from the analysis of the high energy experimental measurements, so we must determine them from the present analysis. Second, the sea-quark contributions of GPDs $ E^q $ namely $ E^{\bar q}(x,t,\mu^2) $, do not play a role in the theoretical calculations of the nucleon Sachs FFs and radii~\cite{Hashamipour:2021kes}, and on the other hand, they are not significant in the theoretical calculations of the WACS cross section~\cite{Hashamipour:2020kip,Huang:2001ej}. So, there are not enough constraints from data to extract them. Considering these facts, we do not consider $ E^{\bar q}(x,t,\mu^2) $ contributions in the present study.

For the forward limits of GPDs $ E_v^q $ in Eq.~(\ref{Eq1}) i.e., $ e_v^q(x,\mu^2) $ we take the same parametrization proposed in the  DK13~\cite{Diehl:2013xca} study and used in our previous analysis~\cite{Hashamipour:2021kes}:
\begin{equation}
\label{Eq2}
e_v^q(x)=\kappa_q N_q x^{-\alpha_q} (1-x)^{\beta_q} (1+\gamma_q\sqrt{x}),
\end{equation}
which is defined at $ \mu=2 $ GeV. Here, $ \kappa_u=1.67 $ and  $ \kappa_d=-2.03 $ are calculated
from the measured magnetic moments of the proton and neutron in the units of nuclear magnetons~\cite{ParticleDataGroup:2018ovx}, and for the normalization factor $ N_q $ we have
\begin{equation}
\label{Eq3}
\int_0^1 dx e_v^q(x)=\kappa_q.
\end{equation}

For the profile functions in Eq.~(\ref{Eq1}), we use the general form~\cite{Diehl:2004cx,Diehl:2013xca}
\begin{equation}
\label{Eq4}
{\cal F}(x)=\alpha^{\prime}(1-x)^3\log\frac{1}{x}+B(1-x)^3 + Ax(1-x)^2,
\end{equation}
which is flexible enough and leads to a better fit of the data~\cite{Hashamipour:2019pgy}. It should be noted that the forward limits of GPDs and the profile functions must imply a positivity condition as  
follows~\cite{Diehl:2013xca}:
\begin{equation}
\label{Eq5}
\frac{[e_v^q(x)]^2}{8m^2} \leq \exp(1) \Big[\frac{g_v^q(x)}{f_v^q(x)}\Big]^3[f_v^q(x)-g_v^q(x)]\times \big\{[q_v(x)]^2-[\Delta q_v(x)]^2\big\},
\end{equation}
where $ m $ is the nucleon mass. This relation obviously leads to the condition $ g_v^q(x) < f_v^q(x) $.

The minimization procedure, as well as the method for calculating uncertainties that we use to extract GPDs and their uncertainties from data, are as in our recent study~\cite{Hashamipour:2021kes}. To be more precise, we follow the same approach (Scenario 2) used in Ref.~\cite{Hashamipour:2021kes}, where the positivity condition Eq.~(\ref{Eq5}) is preserved automatically in a wide range of the $ x $ values [by implementing condition $ g_v^q(x) < f_v^q(x) $ in the main body of the fit program], and a parametrization scan procedure is also used to find the optimum values of the unknown parameters using the CERN program library \texttt{MINUIT}~\cite{James:1975dr}. To calculate the uncertainties of the extracted GPDs and also related observables we use the standard Hessian approach~\cite{Pumplin:2001ct}.

%
%
\section{Data selection}\label{sec:three}

 The main constraints on GPDs $ H_v^q $ and $ E_v^q $ at $ \xi =0$ come from the measurements of the elastic electron-nucleon scattering where the electric and magnetic FFs of the nucleons (proton $ p $ and neutron $ n $) or their ratios can be extracted~\cite{Arrington:2007ux,Ye:2017gyb,Punjabi:2015bba}. In addition, the charges and magnetic radii of the nucleons can provide crucial information about their small-$ t $ behavior. In the present study, following our previous analysis~\cite{Hashamipour:2021kes}, we use 
the data of the Ye, Arrington, Hill and Lee ({YAHL18}) analysis~\cite{Ye:2017gyb} for the electromagnetic FFs where the two-photon exchange (TPE) corrections have also been incorporated. To be more precise, we use 69, 38, and 33 data points of the world  $ R^p=\mu_p G_E^p/G_M^p $ polarization, the $ G_E^n $, and the $ G_M^n/\mu_n G_D $ measurements, respectively, where $ G_D=(1+Q^2/\Lambda^2)^{-2} $, with $ \Lambda^2=0.71 $ GeV$ ^2 $, and $ Q^2=-t $. The data of the charge and magnetic radii of the nucleons (four data points) are taken from the Review of Particle Physics~\cite{ParticleDataGroup:2018ovx} as quoted in Eq.~(18) of Ref.~\cite{Hashamipour:2021kes}. For a thorough review on $ r_E^p $, see also Ref.~\cite{Xiong:2023zih}. However, in the present study, in order to put further constraints on GPDs $ H_v^q $ and $ E_v^q $ at both smaller and larger values of $ -t $, we use also $ G_M^p $ measurements of an older world data analysis by Arrington, Melnitchouk, and Tjon (AMT07)~\cite{Arrington:2007ux} as well as the {Mainz} data~\cite{A1:2013fsc}, which contain 56 and 77 data points, respectively. Note that the {Mainz} data have also been included in the {YAHL18} analysis to extract $ R^p $, but the authors have not extracted $ G_M^p $ values independently. 

on the contrary, the measurements of the nucleon AFF $ G_A $ provide us with the main constraints on the polarized GPDs  $ \widetilde{H}^q $ at $ \xi =0$. In this case, we follow the same approach used in Ref.~\cite{Hashamipour:2019pgy} and consider a reduced set of the world measurements of the proton $ G_A $ (see Refs.~\cite{Bernard:2001rs,Schindler:2006jq} for a review of AFF experimental data until 2007). This set includes more recent measurements~\cite{Butkevich:2013vva,Esaulov:1978ed,DelGuerra:1975uiy,Bloom:1973fn,Joos:1976ng,Choi:1993vt} and also contains the most accurate data points between cases with the same value of $ -t $. This decreases the number of $ G_A $ data points included in the analysis to 34. We investigate the impact of CLAS Collaboration measurements at higher values of $ -t $~\cite{CLAS:2012ich} (five data points) on GPDs, especially $ \widetilde{H}^q $, as a separate analysis in Sec.~\ref{sec:four-two}.

As described in Refs.~\cite{Hashamipour:2020kip,Huang:2001ej}, the WACS cross section is related theoretically to three kinds of GPDs: namely $ H^q $, $ \widetilde{H}^q $, and $ E^q $ at $ \xi =0$. Then, by including the measurements of the WACS cross section in the analysis besides the other data introduced above, it is possible to determine the GPDs $ H^q $, $ \widetilde{H}^q $, and $ E^q $ simultaneously. Although the WACS measurements constrain mainly the  GPDs $ H^q $, since they have more contributions to the WACS cross section, important information about GPDs $ \widetilde{H}^q $ and $ E^q $ at larger values of $ -t $ can also be found from these measurements. In the present study, just like Ref.~\cite{Hashamipour:2020kip}, we use the measurements by the Jefferson Lab (JLab) Hall A Collaboration~\cite{Danagoulian:2007gs} containing 25 data points. The data are belonging to four different values of the Mandelstam variables $ s $ namely, $ s=4.82, 6.79, 8.90, $ and 10.92 GeV$ ^2 $ and they cover an energy range of $ 1.65 \leq -t \leq 6.46 $ GeV$^2 $.

%
%
\section{Results}\label{sec:four}
 This section is devoted to presenting our results obtained for the $ \chi^2 $ analysis of the experimental data introduced in Sec.~\ref{sec:three}. To this aim, we first perform some analyses of the nucleon electromagnetic FFs, AFFs, and charge and magnetic radii data, simultaneously, to construct our base fit. We investigate the impact of the CLAS data of $ G_A $ at high $ -t $ on the extracted GPDs as a separate analysis. As the next step, we include also the data of the WACS cross section in the analysis to investigate their impact on GPDs {as well as the possible tensions between them and other data}. Finally, we compare our results obtained for other quantities related to GPDs such as the gravitational FFs and the total angular momentum carried by the quarks inside the nucleon with the corresponding ones obtained from other studies.

\subsection{The base fit}\label{sec:four-one}

 In our previous study~\cite{Hashamipour:2021kes}, we determined the GPDs $ H_v^q $ and $ E_v^q $ (just valence sectors) by analyzing the nucleon electromagnetic FF data presented in the {YAHL18} paper~\cite{Ye:2017gyb} and the nucleon radii data taken from the Review of Particle Physics~\cite{ParticleDataGroup:2018ovx}. The nucleon electromagnetic FF data were contained the $ R^p=\mu_p G_E^p/G_M^p $ polarization,  $ G_E^n $, and $ G_M^n/\mu_n G_D $ measurements (note that the extraction of the $ G_M^p $ data from the measurements of the elastic electron-nucleon scattering has not been performed in {YAHL18} analysis). As a result, we found that the inclusion of the radii data in the analysis can put further constraints on GPDs, especially in the case of $ E_v^q $. Moreover, we indicated that it is necessary to include more experimental data in the analysis in order to get more universal GPDs. 

In this subsection, as a first step, we are going to improve our previous analysis by 
\begin{itemize}
\item [(i)] including the world $ G_M^p $ measurements from the {AMT07} analysis~\cite{Arrington:2007ux} as well as the {Mainz} data~\cite{A1:2013fsc}. This increases constraints on GPDs at both smaller and larger values of $ -t $, since the $ G_M^p $ measurements cover a $ -t $ range from 0.007 to 32.2 GeV$ ^{2} $.
\item [(ii)] including a reduced set of the world $ G_A $ measurements (see Sec.~\ref{sec:three} and Refs.~\cite{Hashamipour:2019pgy,Hashamipour:2020kip} for more information) to extract also polarized GPDs $ \widetilde{H}^q $ (both valence and sea-quark sectors). Actually, this makes possible a simultaneous determination of three kinds of GPDs if there are some relations between them. Otherwise, it does not make sense to perform a simultaneous analysis of the axial and electromagnetic FF data, since $ G_A $ measurements will put constraints on GPDs $ \widetilde{H}^q $ separately. In the following, we see that by utilizing a standard parametrization scan procedure~\cite{Hashamipour:2021kes} to find unknown parameters of GPDs of Eq.~(\ref{Eq1}), some relations between them can be achieved automatically. 
\end{itemize}

It should be noted that it is not possible to determine GPDs $ H^{\bar q} $ at this stage, since the data included in the analysis are just related to the valence sectors of the unpolarized GPDs ($ H $ and $ E $). However, both valence and sea-quark sectors of the polarized GPDs $ \widetilde{H}^q $ can be extracted from data because they are both contributing to the theoretical calculations of $ G_A $~\cite{Hashamipour:2019pgy}. The determination of $ H^{\bar q} $ is postponed to Sec.~\ref{sec:four-three} where we include also the WACS data in the analysis.

Our procedure for obtaining the optimum values of the fit parameters is as follows. We first consider the profile functions $ g_v^q(x) $ and $ \widetilde{f}_v^q(x) $ to be equal to $ f_v^q(x) $ and perform the parametrization scan to find the optimum values of the parameters $ \alpha^{\prime} $, $ A $, and $ B $ in Eq.~(\ref{Eq4}) for the profile functions $ f_v^u(x) $ and $ f_v^d(x) $, and also parameters $ \alpha $ and $ \beta $ in Eq.~(\ref{Eq2}) for the forward limits $ e_v^u(x) $ and $ e_v^d(x) $. Note that the parameters $ \gamma_u $ and $ \gamma_d $ are set to zero at this stage. Then, we continue the parametrization scan to find the optimum values of the remaining parameters, one by one, considering the fact that the profile functions $ g_v^q(x) $ and $ \widetilde{f}_v^q(x) $ can be different from $ f_v^q(x) $, and also assuming $ \widetilde{f}^{\bar q}(x)=\widetilde{f}_v^q(x) $. The procedure is continued until the minimum value of $ \chi^2 $ is reached. Note that at the end of the parametrization scan, some parameters may be equal or zero, since we are looking for the lowest value of $\chi^2$ per number of degrees of freedom ($\chi^2$/d.o.f.). To be more precise, sometimes it happens that releasing a parameter does not affect the value of the total $\chi^2$ or its decrease is so small that it does not reduce the $\chi^2$/d.o.f. value.

It should be noted that the condition $ g_v^q(x) < f_v^q(x) $ is implemented in the main body of the fit program and considered during the parametrization scan, in order to preserve the positivity property of GPDs introduced in Eq.~(\ref{Eq5}), automatically. This holds Eq.~(\ref{Eq5}) to a great extent so that there are just some violations at very large values of $ x $.

\begin{figure}[!htb]
    \centering
\includegraphics[scale=0.9]{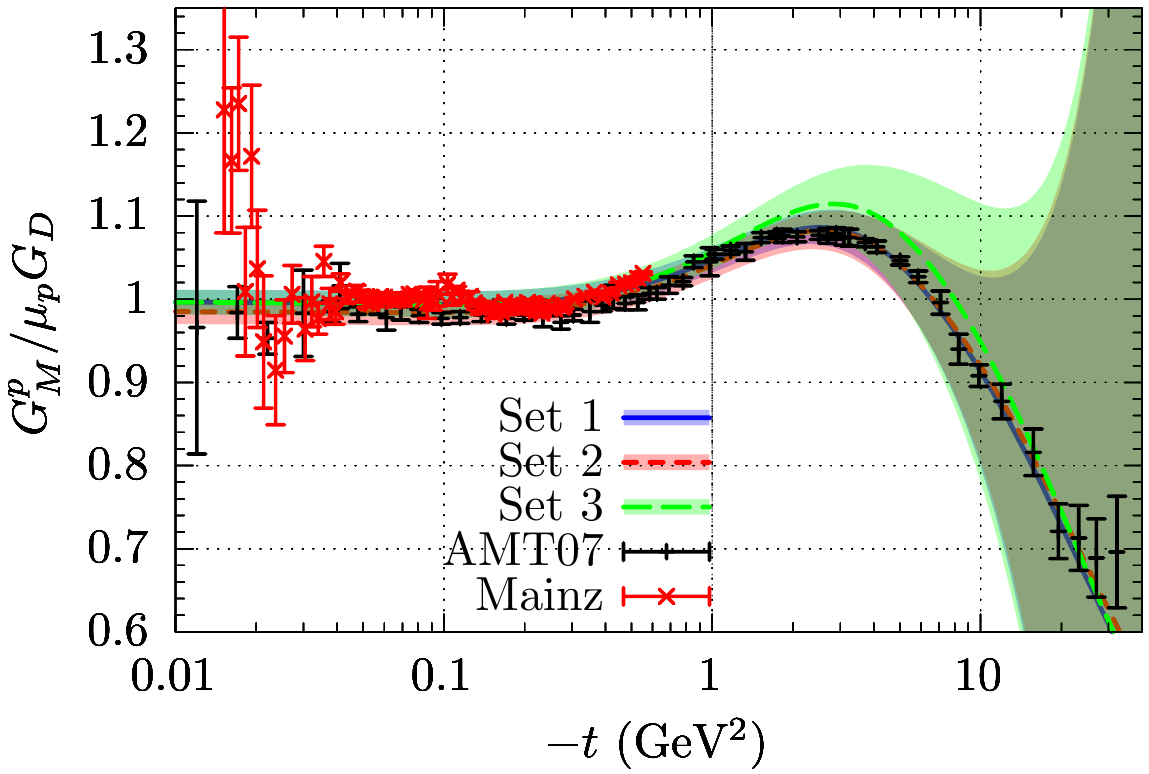}    
    \caption{A comparison between our results for $ G_M^p/\mu_p G_D $ obtained from three analyses described in Sec.~\ref{sec:four-one} and the related experimental data of the {AMT07}\cite{Arrington:2007ux} and {Mainz}~\cite{A1:2013fsc}. }
\label{fig:GMp}
\end{figure}

Following the procedure described above, we find a set of GPDs with $ \alpha^{\prime}_{\widetilde{f}_v^u}=\alpha^{\prime}_{f_v^u} $, $ \alpha^{\prime}_{\widetilde{f}_v^d}=\alpha^{\prime}_{f_v^d} $, $ A_{g_v^u}=A_{\widetilde{f}_v^u}=A_{f_v^u} $, $ A_{\widetilde{f}_v^d}=A_{f_v^d} $, $  B_{g_v^d}=B_{g_v^u} $, and $ \gamma_u= \gamma_d=0 $, which is called Set 1. The value of $\chi^2$/d.o.f. is 3.05 for 311 data points, which seems relatively large at first glance. By examining this issue,
one finds that the reason is the large value of $\chi^2$ obtained for two sets of data namely, $ G_M^p $ and $ G_A $ measurements. We expected a large $\chi^2$ for $ G_A $ data, since different measurements~\cite{Butkevich:2013vva,Esaulov:1978ed,DelGuerra:1975uiy,Bloom:1973fn,Joos:1976ng,Choi:1993vt} do not have good agreement with each other and actually form a spectrum as discussed in Ref.~\cite{Hashamipour:2019pgy}. For the case of $ G_M^p $ data, the large value of $\chi^2$ comes mainly from the {Mainz} data (it is 463 for 77 data points), which are concentrated in small $ -t $ values. Actually, there is a tension between these data and the world data taken from the {AMT07} analysis in this region (see Fig.~\ref{fig:GMp}). However, it should be note that the {AMT07} data have an acceptable $\chi^2$ value (it is 114 for 56 data points), which means that they are well fitted. Accordingly, it is interesting to perform also two other analyses, one by including just the {AMT07} data and excluding the {Mainz} data, and the other vice versa. We call the sets of GPDs obtained from these two analyses Set~2 and Set~3, respectively.

The values of the optimum parameters obtained from three analyses described above are listed in Table~\ref{tab:par}. Note again that some parameters obtained are equal to another parameter or equal to zero through the parametrization scan, automatically. Comparing Sets~1 and 2, one finds that by excluding the {Mainz} data from the analysis, significant changes are happening in the values of parameters of the profile function $ g_v^u $ and forward limit $ e_v^u $, while the other distributions are less affected than before. So, it can be concluded that the {Mainz} data have the most impact on GPDs $ E_v^u $.  In contrast, comparing Sets~1 and 3 shows that if we exclude the {AMT07} data and maintain the {Mainz} data, the main changes are happening in profile functions $ f_v^u $ and $ g_v^u $ (less than $ f $),  which indicates the importance of the \texttt{AMT07} data in constraining the whole $ -t $ behavior of the up-quark unpolarized GPDs, especially $ H_v^u $. 
\begin{table}[th!]
\scriptsize
\setlength{\tabcolsep}{8pt} 
\renewcommand{\arraystretch}{1.4} 
\caption{The values of the optimum parameters obtained from three analyses described in Sec.~\ref{sec:four-one}. Note that $ \widetilde{f}^{\bar q}(x)=\widetilde{f}_v^q(x) $.}\label{tab:par}
\begin{tabular}{lcccc}
\hline
\hline
 Distribution &  Parameter           &  Set 1            &  Set 2            &  Set 3  \\
\hline 
\hline
$ f_v^u(x) $  & $ \alpha^{\prime} $  & $ 0.687\pm0.007 $ & $ 0.678\pm0.008 $ & $ 0.710\pm0.026 $  \\
			  &	$ A $                & $ 0.884\pm0.023 $ & $ 0.831\pm0.035 $ & $ 0.926\pm0.258 $  \\
			  &	$ B $                & $ 0.968\pm0.024 $ & $ 1.009\pm0.026 $ & $ 0.871\pm0.122 $  \\
\hline
$ f_v^d(x) $  & $ \alpha^{\prime} $  & $ 0.453\pm0.022 $ & $ 0.463\pm0.071 $ & $ 0.483\pm0.090 $  \\
			  &	$ A $                & $ 3.075\pm0.430 $ & $ 3.093\pm1.055 $ & $ 3.102\pm1.577 $  \\
			  &	$ B $                & $ 1.167\pm0.127 $ & $ 1.127\pm0.389 $ & $ 1.023\pm0.531 $  \\
\hline
$ g_v^u(x) $  & $ \alpha^{\prime} $  & $ 0.989\pm0.097 $ & $ 0.632\pm0.045 $ & $ 1.025\pm0.096 $  \\
			  &	$ A $                & $ A_{f_v^u} $     & $ A_{f_v^u} $     & $ A_{f_v^u} $  \\
			  &	$ B $                & $-0.497\pm0.135 $ & $-0.572\pm0.142 $ & $-0.586\pm0.124 $  \\
\hline
$ g_v^d(x) $  & $ \alpha^{\prime} $  & $ 0.814\pm0.058 $ & $ 0.831\pm0.087 $ & $ 0.832\pm0.059 $  \\
			  &	$ A $                & $ 3.235\pm0.328 $ & $ 3.249\pm0.624 $ & $ 3.258\pm0.302 $  \\
			  &	$ B $                & $ B_{g_v^u} $     & $ B_{g_v^u} $     & $ B_{g_v^u} $  \\
\hline
$\widetilde{f}_v^u(x)$  & $\alpha^{\prime}$  & $ \alpha^{\prime}_{f_v^u} $ & $ \alpha^{\prime}_{f_v^u} $ & $ \alpha^{\prime}_{f_v^u} $ \\
			  &	$ A $                & $ A_{f_v^u} $     & $ A_{f_v^u} $     & $ A_{f_v^u} $  \\
			  &	$ B $                & $-0.500\pm0.270 $ & $-0.635\pm0.187 $ & $-0.589\pm0.339 $  \\
\hline
$\widetilde{f}_v^d(x)$  & $\alpha^{\prime}$  & $ \alpha^{\prime}_{f_v^d} $ & $ \alpha^{\prime}_{f_v^d} $ & $ \alpha^{\prime}_{f_v^d} $ \\
			  &	$ A $                & $ A_{f_v^d} $     & $ A_{f_v^d} $     & $ A_{f_v^d} $  \\
			  &	$ B $                & $ 0.366\pm0.558 $ & $ 0.743\pm0.573 $ & $ 0.370\pm0.743 $  \\
\hline
$ e_v^u(x) $  & $ \alpha $           & $ 0.489\pm0.070 $ & $ 0.741\pm0.022 $ & $ 0.480\pm0.070 $  \\
			  &	$ \beta $            & $ 6.868\pm0.657 $ & $ 8.761\pm1.561 $ & $ 6.853\pm0.794 $  \\
			  &	$ \gamma $           & $ 0.000 $         & $ 0.000 $         & $ 0.000 $  \\
\hline
$ e_v^d(x) $  & $ \alpha $           & $ 0.607\pm0.028 $ & $ 0.614\pm0.046 $ & $ 0.603\pm0.030 $  \\
			  &	$ \beta $            & $ 4.609\pm0.938 $ & $ 4.638\pm1.223 $ & $ 4.987\pm0.759 $  \\
			  &	$ \gamma $           & $ 0.000 $         & $ 0.000 $         & $ 0.000 $  \\
\hline 		 	
\hline 	
\end{tabular}
\end{table}

The results of three analyses described above are presented in Table~\ref{tab:chi2}. The first column contains datasets used in the analysis with their references. The data are separated according to their related observables. The range of $ -t $ which is covered by data is reported in the second column. For each dataset, the value of $ \chi^2 $ divided by the number of data points, $\chi^2$/$ N_{\textrm{pts.}} $, is presented, which can be considered as a criterion for the goodness of fit. The last row of Table~\ref{tab:chi2} presents the values of $\chi^2$/d.o.f. for three analyses which show the goodness of the fit in total.
\begin{table}[th!]
\scriptsize
\setlength{\tabcolsep}{8pt} 
\renewcommand{\arraystretch}{1.4} 
\caption{The results of three analyses described in Sec.~\ref{sec:four-one}.}\label{tab:chi2}
\begin{tabular}{lcccc}
\hline
\hline
  Observable         &  -$t$ (GeV$^2$)   &  \multicolumn{3}{c}{ $\chi^2$/$ N_{\textrm{pts.}} $  }  \\
                     &                   &     Set 1    &      Set 2     &     Set 3               \\
\hline 		 	
\hline 
$G_{M}^p$~\cite{A1:2013fsc}                        & $0.0152-0.5524$& $463.4 / 77$  & $ \cdots $         &  $424.7 / 77 $ \\
$G_M^p/\mu_p G_D$~\cite{Arrington:2007ux}                & $0.007-32.2$   & $113.8 / 56$  & $ 51.5 / 56$  &  $ \cdots $        \\
$R^p = \mu_p G_{E}^p / G_{M}^p$~\cite{Ye:2017gyb}  & $0.162-8.49$   & $107.8 / 69$  & $119.8 / 69$  &  $113.0 / 69$ \\
$G_{E}^n$~\cite{Ye:2017gyb}                        & $0.00973-3.41$ & $27.9 / 38$   & $ 26.0 / 38$  &  $ 27.0 / 38$ \\
$G_M^n/\mu_n G_D$~\cite{Ye:2017gyb}                & $0.071-10.0$   & $45.2 / 33$   & $45.9 / 33 $  &  $47.9 / 33 $ \\
$G_{A}$~\cite{Butkevich:2013vva,Esaulov:1978ed,DelGuerra:1975uiy,Bloom:1973fn,Joos:1976ng,Choi:1993vt}                          & $0.025-1.84$   & $130.0 / 34$  & $129.3  / 34$ &  $ 129.5 / 34$ \\
$\sqrt{\left<r_{pE}^2\right>}$~\cite{ParticleDataGroup:2018ovx}   & $ 0 $          & $0.0 / 1$     & $0.2 / 1 $    &  $0.0 / 1 $     \\
$\sqrt {\left<r_{pM}^2\right>}$~\cite{ParticleDataGroup:2018ovx}  & $ 0 $          & $0.0 / 1$     & $2.2 / 1 $    &  $1.2 / 1 $   \\
$\left<r_{nE}^2\right>$~\cite{ParticleDataGroup:2018ovx}          & $ 0 $          & $0.2 / 1$     & $0.1 / 1 $    &  $0.4 / 1 $   \\
$\sqrt {\left<r_{nM}^2\right>}$~\cite{ParticleDataGroup:2018ovx}  & $ 0 $          & $12.5 / 1$    & $14.9 / 1 $   &  $9.9 / 1 $   \\	
\hline
Total $\chi^2 /\mathrm{d.o.f.} $ &                & $900.8 / 295$ & $389.9 / 218$   &  $753.6 / 239$   \\
\hline
\hline
\end{tabular}
\end{table}

Comparing the results obtained for three analysis, one concludes that the main difference between them comes from the proton electromagnetic FF data, since the neutron and axial FFs as well as the radii data have almost the same $\chi^2$ from one analysis to another.   
Overall, the values of $\chi^2$/d.o.f. for two analyses including the {Mainz} dataset, namely Sets 1  and 3 are not very desirable (they are 3.05 and 3.15, respectively). By excluding the {Mainz} data from the analysis (Set 2), the value of $\chi^2$/d.o.f. is significantly decreased (from 3.05 to 1.79) due to the significant decrease in the $\chi^2$ of the {AMT07} data as well. On the other hand, although the $\chi^2$ of the {Mainz} data is slightly decreased by excluding the {AMT07} data (from 463 to 425), it still remains large. Considering these facts together, at first glance, one may conclude that there is a tension between the {Mainz} data not only with {AMT07} data, but also with other data presented in the analysis. However, this issue needs to be further explored, since it can be due to the model incompatibility with the {Mainz} data, especially at smaller values of $ -t $, where they have a different trend compared to the {AMT07} data (see Fig.~\ref{fig:GMp}). An interesting thing that should be mentioned is that the $\chi^2$ of $ R^p $ data is the lowest in the analysis in which both the {AMT07} and {Mainz} data are included, while it is increased in the analyses which contain only one or the other. The neutron data are well fitted in all three analyses, especially $ G_E^n $ data, and their goodness of fit is not affected by changing the content of the $ G_M^p $ data.
Note that the large value of $\chi^2$ obtained for $ G_A $ data (it is about 129 for 34 data points) is due to the fact that there is not a good agreement between different experimental measurements, as mentioned before. For the case of nucleon radii data, a very good description is obtained except for $\sqrt {\left<r_{nM}^2\right>}$. This can be due to the lack of the experimental information on the neutron magnetic radius~\cite{Meissner:2022rsm}, especially considering the fact that the other radii data are well fitted. Anyway, our prediction for $\sqrt {\left<r_{nM}^2\right>}$ obtained from all three analyses underestimate the value presented in Ref.~\cite{ParticleDataGroup:2018ovx} which is the average of just two measurements.

Figure~\ref{fig:GMp} shows a comparison between our results for $ G_M^p/\mu_p G_D $ obtained from three analyses described above and the related experimental data of {AMT07} and {Mainz}. Note that the original {Mainz} data~\cite{A1:2013fsc} have been presented as $ G_M^p $ not as the ratio to $ \mu_p G_D $. So, in order to make them comparable with the {AMT07} data, we show them as $ G_M^p/\mu_p G_D $ in Fig.~\ref{fig:GMp}, though their original values have been included in our analyses. As can be seen, the {Mainz} data are concentrated in the small $ -t $ region and follow a different trend at very small values of $ -t $, in the contrast to the {AMT07} data, which cover a wide range of $ -t $ up to 32 GeV$ ^2 $. They are also located at the top of the {AMT07} data at almost at all values of $ -t $. These differences leads to a tension between the {AMT07} and {Mainz} data and make their simultaneous description difficult, as is obvious from Table~\ref{tab:chi2} (Set 1). However, regardless of the {Mainz} data at $ -t<0.01 $, these descriptions of data are acceptable considering the uncertainty bands of the predictions. Overall, the predictions obtained from the analyses including the {Mainz} data (Sets 1 and 3) are larger in magnitude at small and medium $ -t $ and fall off with a steeper slope  at larger values of $ -t $. An interesting thing is that Set~3, which has been obtained from the analysis contained just the {Mainz} data, has a similar prediction to Sets 1 and 2 at larger values of $ -t $, while the {Mainz} data only cover $ -t\lesssim 0.6 $. This can be attributed to the presence of $ R^p $ data in the analysis.

\begin{figure}[!htb]
    \centering
\includegraphics[scale=0.9]{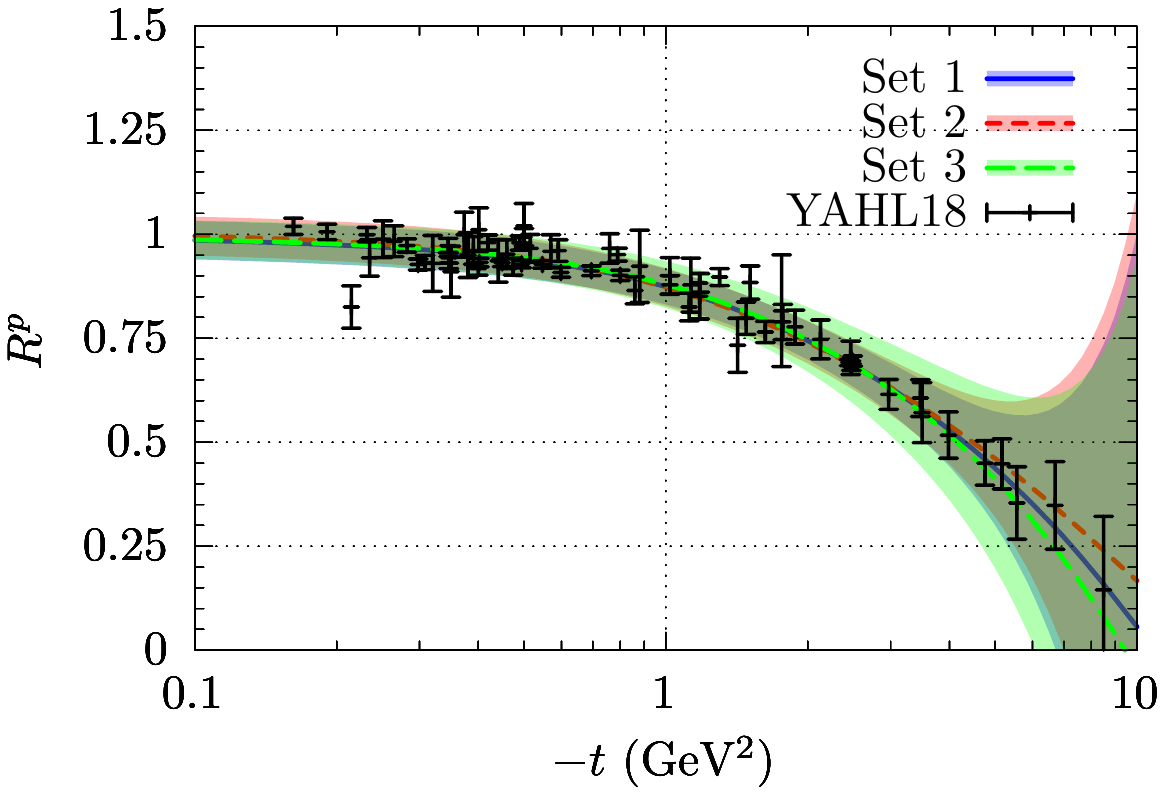}    
    \caption{A comparison between our results for $ R^p $ obtained from three analyses described in Sec.~\ref{sec:four-one} and the corresponding experimental data of the {YAHL18} analysis~\cite{Ye:2017gyb}. }
\label{fig:Rp}
\end{figure}

A comparison between our results for $ R^p $ and the corresponding data included in the analysis from {YAHL18}~\cite{Ye:2017gyb} is shown in Fig.~\ref{fig:Rp}. As can be seen, the data are well fitted in all three analyses, especially Set 1, which has better description of data at larger values of $ -t $. It is worth noting in this context, as is also clear from Table~\ref{tab:chi2}, that including both the {AMT07} and {Mainz} data of $ G_M^p $ in the analysis leads to better description of $ R^p $ data. Note also that the $ R^p $  data have a lesser $\chi^2$ when the analysis includes just the {Mainz} data (Set 3) compared with the case in which it includes just $ G_M^p $ from the {AMT07} (Set 2). Considering these facts, one can realize that the inclusion of both {AMT07} and {Mainz} data in the analysis is preferred, though it leads to a large $\chi^2$ for them. The analyses including the {Mainz} data have lesser $\chi^2$ values for the neutron magnetic radius, too. This indicates again the importance of the inclusion of the {Mainz} data in the analysis.

\begin{figure}[!htb]
    \centering
\includegraphics[scale=0.9]{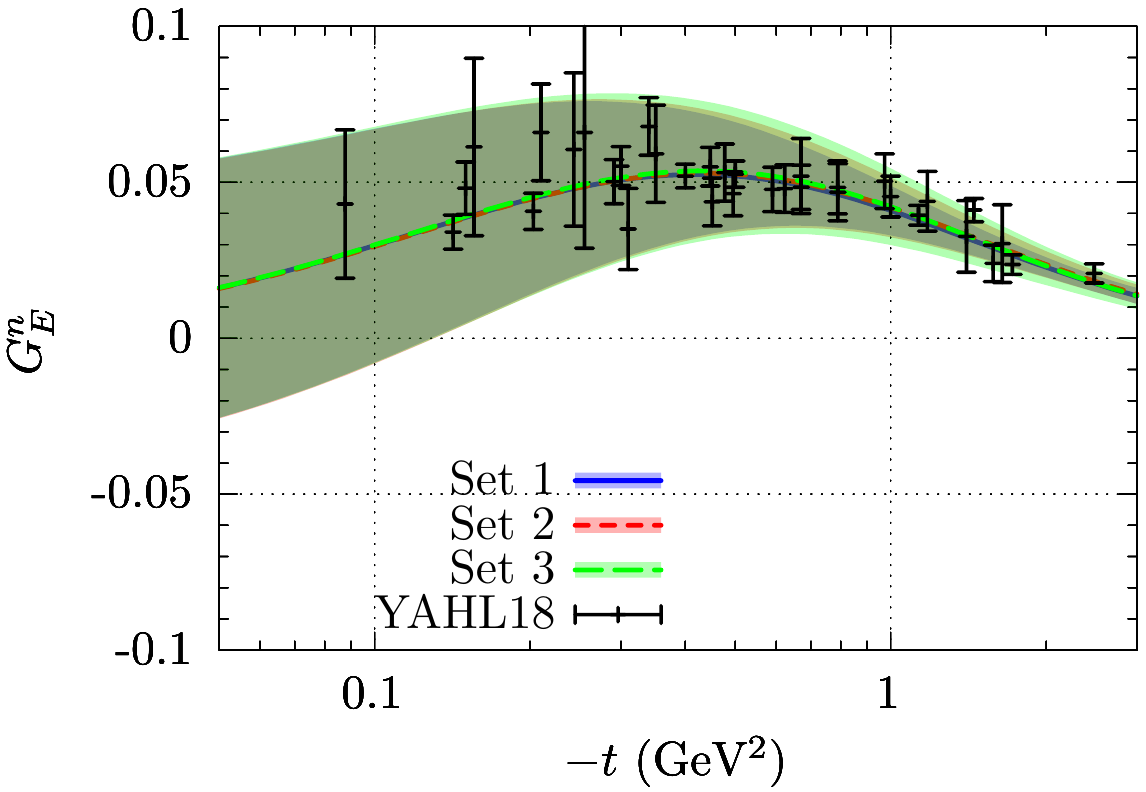}    
    \caption{A comparison between our results for $ G_E^n $ obtained from three analyses described in Sec.~\ref{sec:four-one} and the corresponding experimental data of the {YAHL18} analysis~\cite{Ye:2017gyb}. }
\label{fig:GEn}
\end{figure}

\begin{figure}[!htb]
    \centering
\includegraphics[scale=0.9]{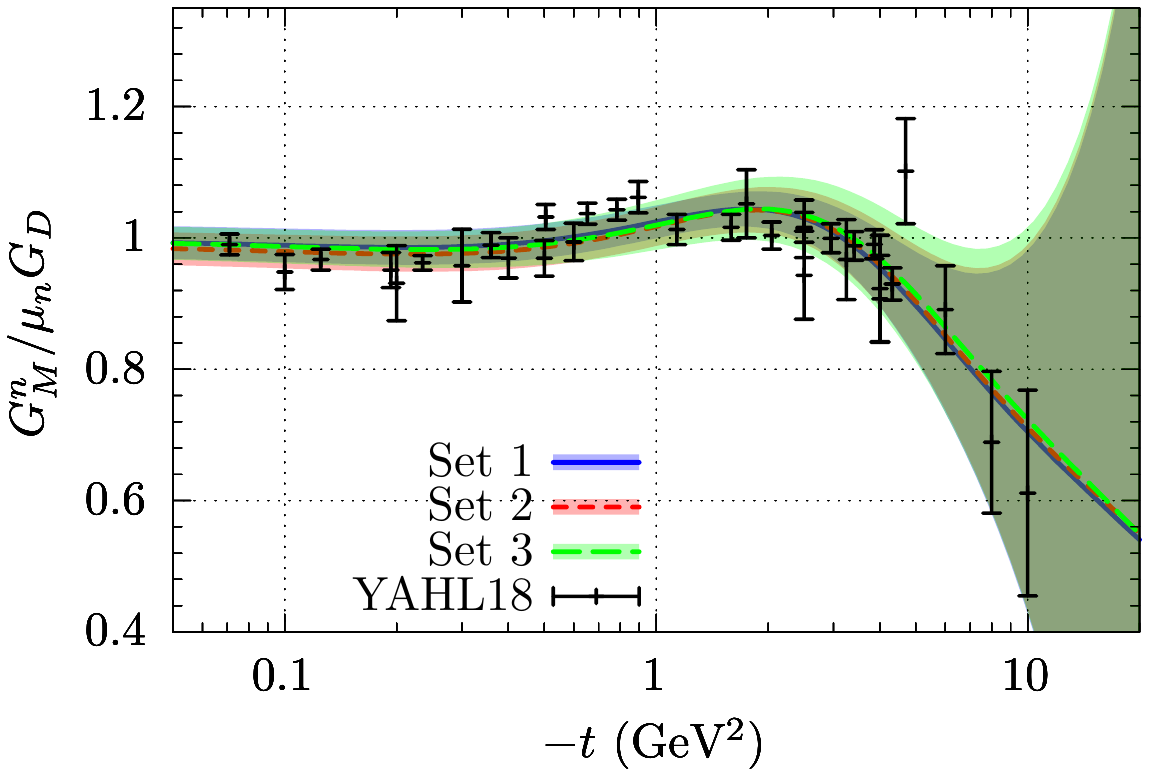}    
    \caption{A comparison between our results for $ G_M^n/\mu_n G_D $ obtained from three analyses described in Sec.~\ref{sec:four-one} and the corresponding experimental data of the {YAHL18} analysis~\cite{Ye:2017gyb}. }
\label{fig:GMn}
\end{figure}

Figures~\ref{fig:GEn} and~\ref{fig:GMn} show the same comparisons as Fig.~\ref{fig:Rp} but for $ G_E^n $ and $ G_M^n/\mu_n G_D $. In both cases, the results are in good agreement with data considering uncertainties. These figures clearly show that the different approaches to include the $ G_M^p $ data in the analysis do not affect significantly the description of the neutron data, since all sets lead to almost the same prediction. There are just some little differences for the case of neutron magnetic FF in Fig.~\ref{fig:GMn}. Overall, Sets 1 and 3 are in better consistency with each other. 

\begin{figure}[!htb]
    \centering
\includegraphics[scale=0.9]{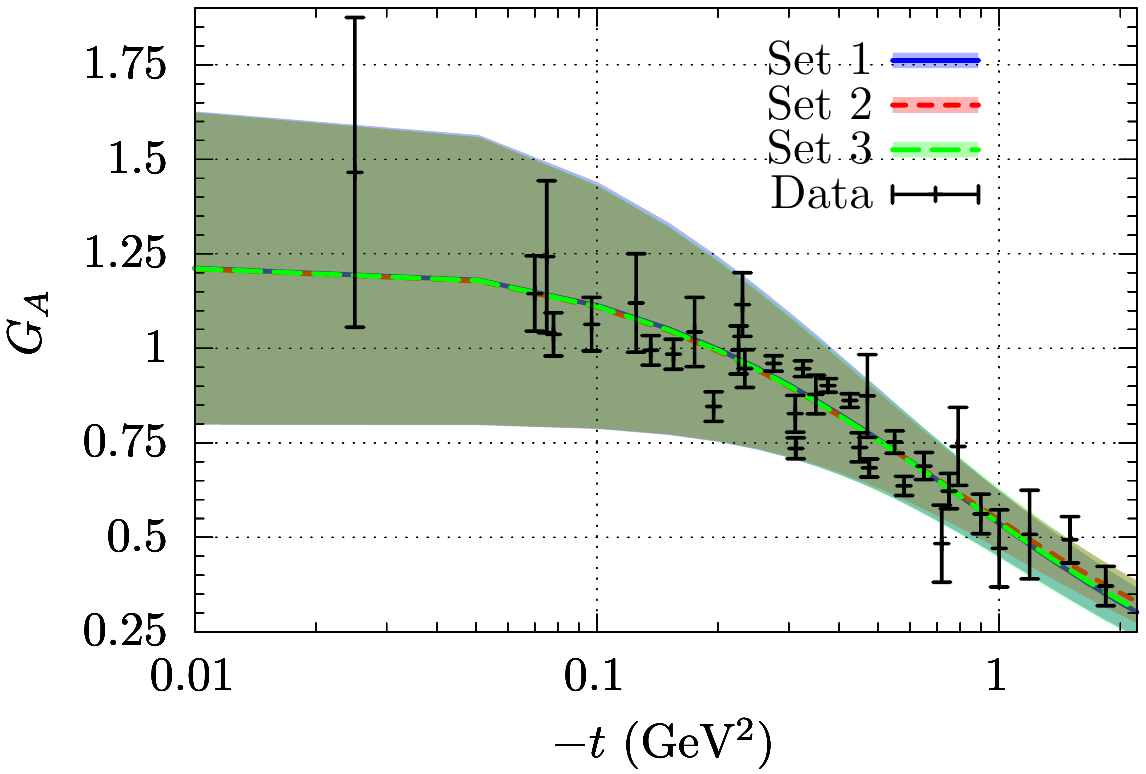}    
    \caption{A comparison between our results for $ G_A $ obtained from three analyses described in Sec.~\ref{sec:four-one} and the corresponding experimental data of Refs.~\cite{Butkevich:2013vva,Esaulov:1978ed,DelGuerra:1975uiy,Bloom:1973fn,Joos:1976ng,Choi:1993vt}.}
\label{fig:GA}
\end{figure}

We have compared our results obtained for $ G_A $ and fitted data~\cite{Butkevich:2013vva,Esaulov:1978ed,DelGuerra:1975uiy,Bloom:1973fn,Joos:1976ng,Choi:1993vt} in Fig.~\ref{fig:GA}. Although data are belonging to different measurements and there is not a good consistency between them over almost the whole range of $ -t $, a reasonable description has been obtained considering the uncertainties. The results indicate that there is not any correlation between the polarized GPD $ \widetilde{H}^q $ and $ G_M^p $ data, since they remain unaffected by removing part of the $ G_M^p $ data from the analysis.  

\begin{figure}[!htb]
    \centering
\includegraphics[scale=0.9]{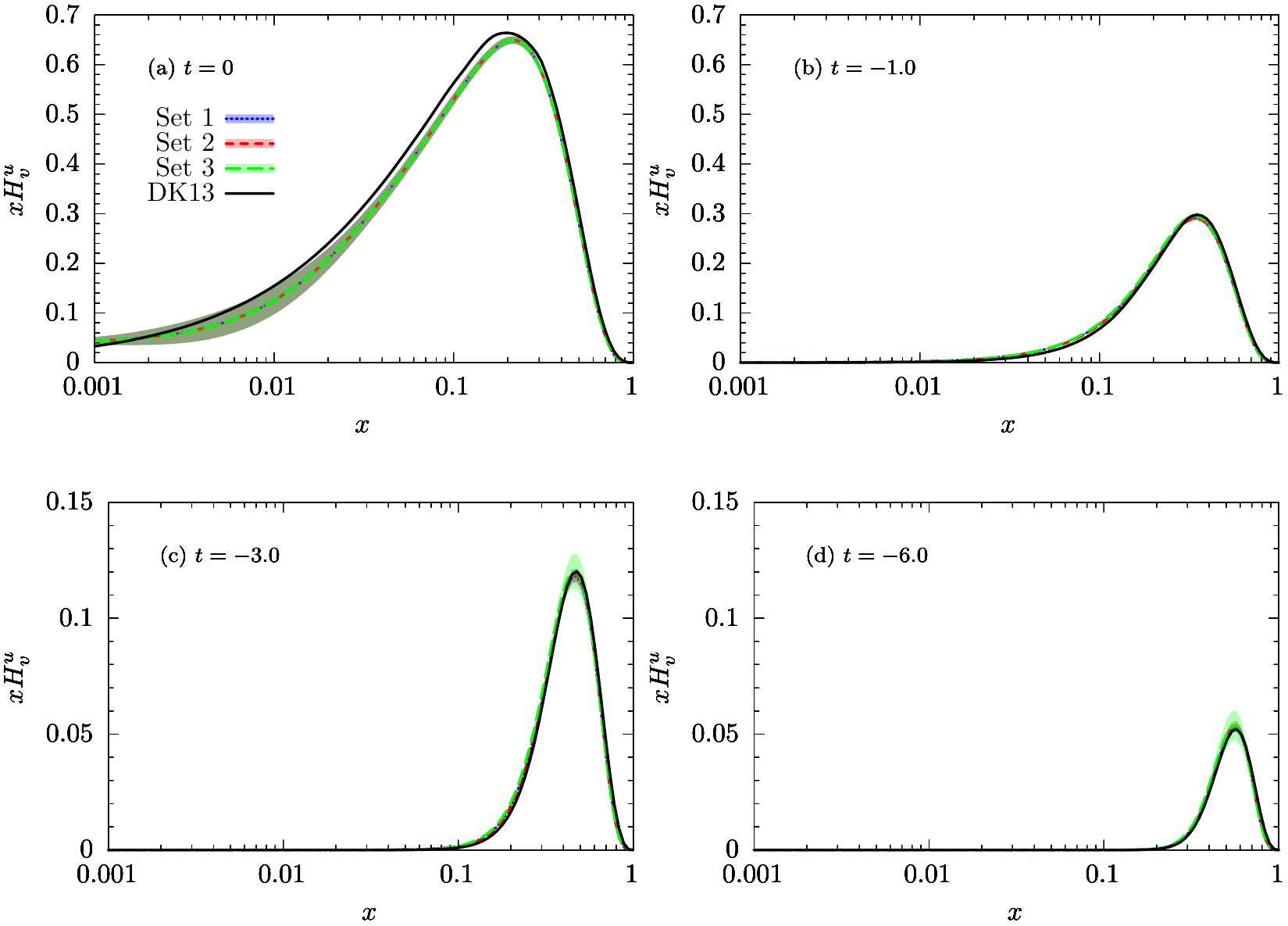}   
    \caption{A comparison between our results for GPDs $ xH_v^u(x) $ and the corresponding results from DK13 analysis~\cite{Diehl:2013xca} at four $ t $ values shown in panels (a) $ t=0$, (b)
    	$t=-1$, (c) $t=-3$, and (d) $t=-6 $ GeV$ ^2 $. See Sec.~\ref{sec:four-one}  for more details.}
\label{fig:Hu}
\end{figure}

A comparison between our results for GPD $ xH_v^u(x) $ with their uncertainties and the result of the DK13 analysis~\cite{Diehl:2013xca} at four $ t $ values $ t=0,-1,-3,-6 $ GeV$ ^2 $ has been shown in Fig.~\ref{fig:Hu}. The little difference observed at $ t=0 $ between our results and DK13 is due to the fact that, in Eq.~(\ref{Eq1}), we have taken PDFs from the {NNPDF} group~\cite{NNPDF:2021njg}, while they have been taken from the Alekhin, Blumlein, and Moch ({ABM11})~\cite{Alekhin:2012ig} in the DK13 analysis. Overall, the results are very similar at all values of $ -t $ 
that show the crucial constraints of the electromagnetic FF data on the unpolarized up valence GPDs $ H_v^u $. 

\begin{figure}[!htb]
    \centering
\includegraphics[scale=0.9]{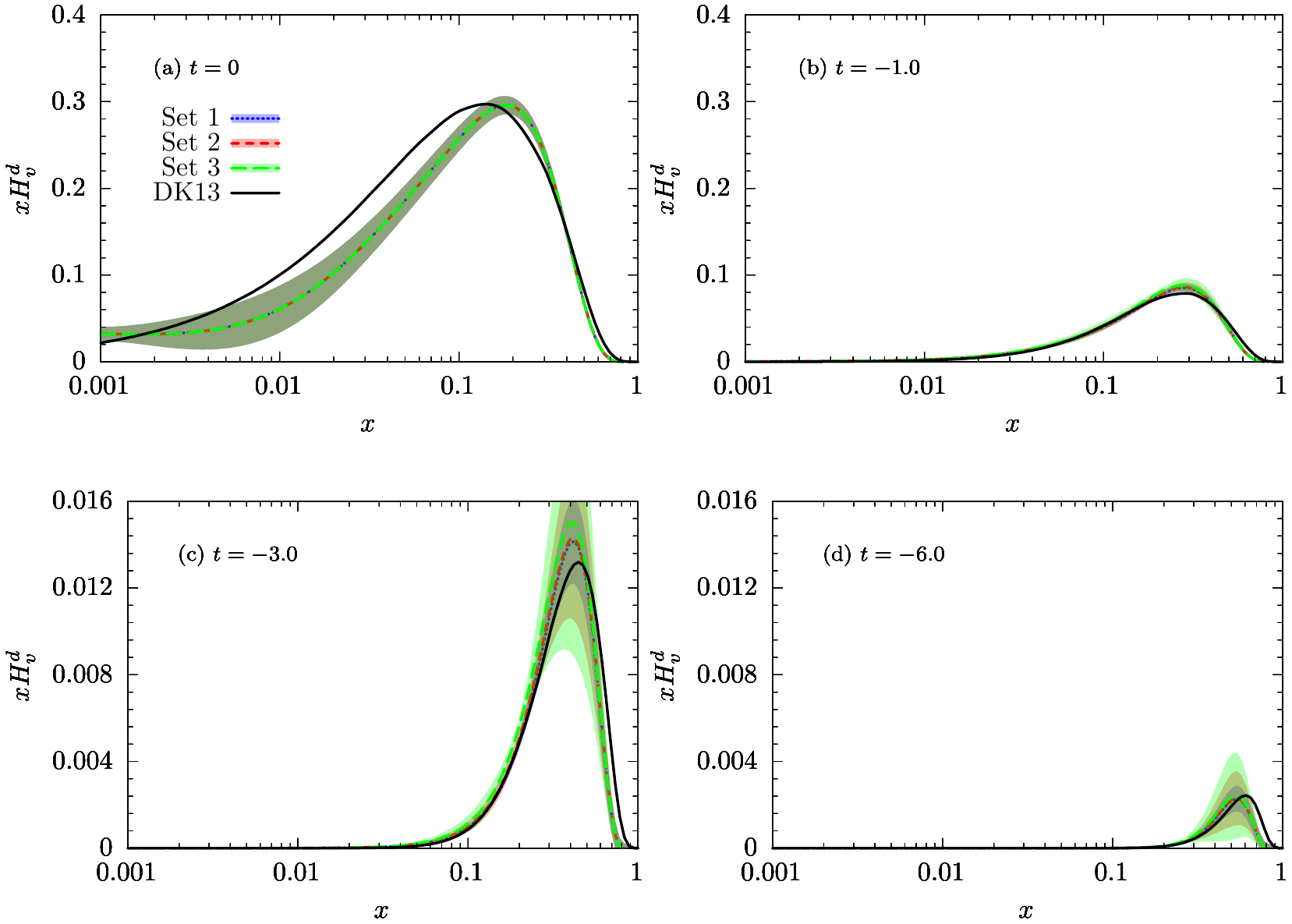}    
    \caption{Same as Fig.~\ref{fig:Hu} but for GPDs $ xH_v^d(x) $. }
\label{fig:Hd}
\end{figure}

Figure~\ref{fig:Hd} shows the same results as Fig.~\ref{fig:Hu}, but for GPDs $ xH_v^d(x) $. Here, the difference between our results and those of  DK13 is significant when comparing with the case of $ xH_v^u(x) $ 
in Fig.~\ref{fig:Hu}. Although the difference between the forward limits ({NNPDF} and {ABM11} PDFs) plays an important role in this case, as can be seen from Fig.~\ref{fig:Hd}(a), it is obvious from the figure that the $ t $ dependence of our GPDs is also somewhat different with DK13. By increasing the absolute value of $ t $, our results  for $ xH_v^d(x) $ are first increased in magnitude and somewhat shifted to smaller $ x $ values (e.g., at $ t=-1 $ and $ -3 $ GeV$ ^2 $). If the increase in $ -t $ continues (e.g., at $ t=-6 $ GeV$ ^2 $), the peak of our results goes down again, but it shifts more to the smaller $ x $. Note that the uncertainties from PDFs have also been considered in the error calculations of GPDs $ H_v^q $ in Figs.~\ref{fig:Hu} and~\ref{fig:Hd}, as well as the other plots which are presented in the following. Overall, Set 3 has larger uncertainties, especially with $ -t $ growing, since it does not contain the {AMT07} data of $ G_M^p $, which cover a wide range of $ -t $. 

\begin{figure}[!htb]
    \centering
\includegraphics[scale=0.9]{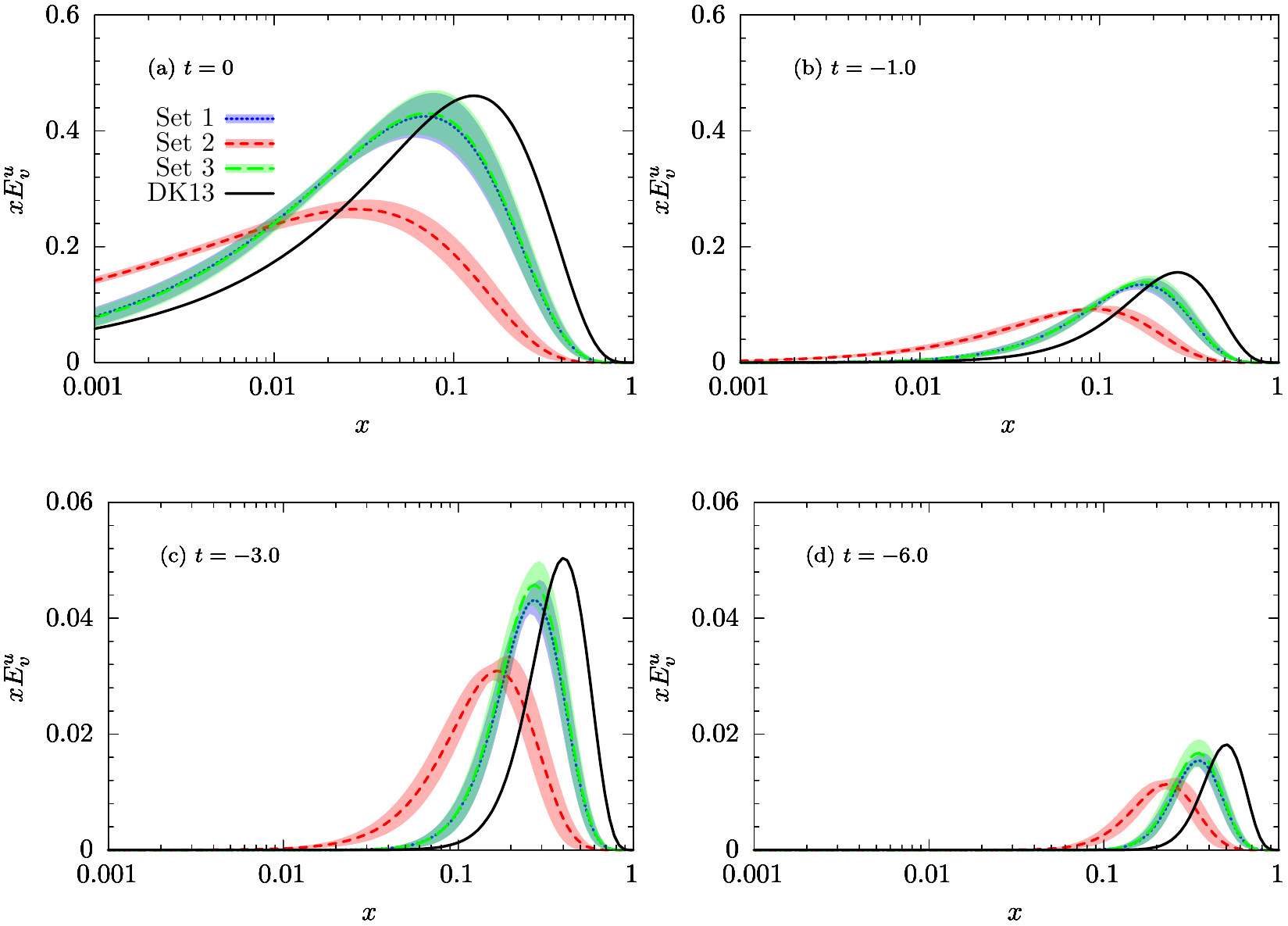}    
    \caption{Same as Fig.~\ref{fig:Hu} but for GPDs $ xE_v^u(x) $. }
\label{fig:Eu}
\end{figure}

Our results obtained for GPDs $ xE_v^u(x) $ have been shown in Fig.~\ref{fig:Eu} and compared again with DK13 at $ t=0,-1,-3,-6 $ GeV$ ^2 $. As one can see, Sets 1 and 3, which have been obtained from the analyses including the {Mainz} data are very similar at all values of $ -t $. However, they differ significantly from DK13 and also Set 2, which has been obtained from the analysis excluding the {Mainz} data and including the world $ G_M^p $ data from {AMT07}. Overall, our results tend toward the smaller values of $ x $ rather than DK13. Among different sets, the results of Set 2 have the lowest peak and are most inclined to small $ x $. The egregious difference between Set 2 and DK13 may be strange at first glance, since they contain the same data of $ G_M^p $. Actually, this can be due to the inclusion of different data, for the case of other observables, in these two analyses (see Sec.~IV A of our previous study~\cite{Hashamipour:2021kes} to get full information on the differences and similarities of our data selection with DK13 for the cases of $ R^p $, $ G_E^n $, $ G_M^p $, and nucleon radii data). However, we have explored this issue further and found that another important factor is the different producers used to preserve the positivity condition of Eq.~(\ref{Eq5}). 
According to the results obtained, it can be concluded that the difference between the forward limits of $ E_v^u $ (i.e., $ e_v^u $) that have been shown in  Fig.~\ref{fig:Eu}(a) plays a more important role in the difference between the final GPDs at larger values of $ -t $, rather than the difference between profile functions.

\begin{figure}[!htb]
    \centering
\includegraphics[scale=0.9]{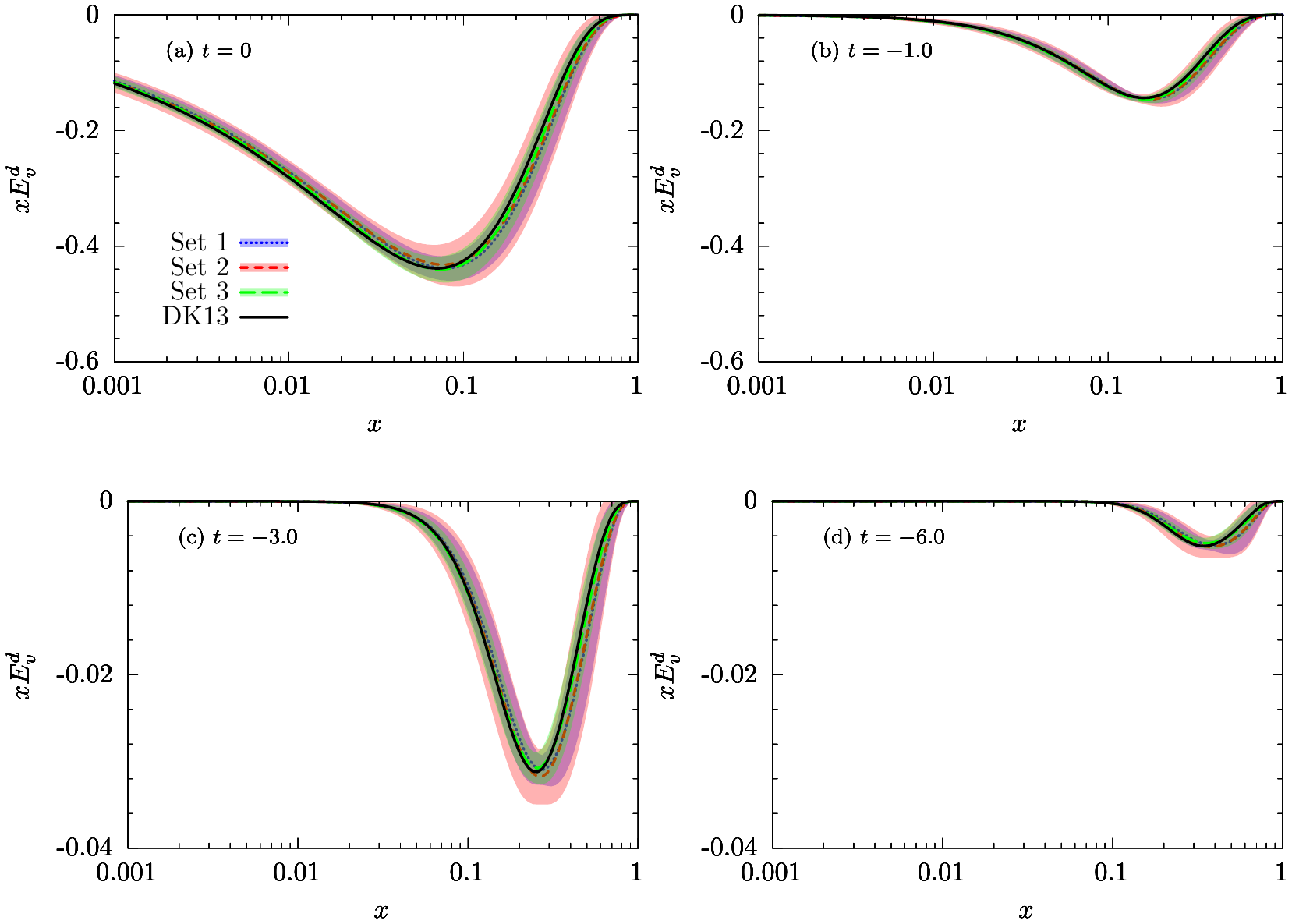}
    \caption{Same as Fig.~\ref{fig:Hu} but for GPDs $ xE_v^d(x) $. }
\label{fig:Ed}
\end{figure}

Figure~\ref{fig:Ed} shows the same results as Fig.~\ref{fig:Eu}, but for GPDs $ xE_v^d(x) $. In this case, 
all results obtained are in very good agreement with each other and also with DK13. This is a reflection of the fact that the constraints on GPDs $ E_v^d $ come mainly from the neutron data, while we are discussing here about including or excluding the {Mainz} or {AMT07} data for the proton magnetic FF, $ G_M^p $ in the analysis. Note that although GPDs $ E_v^q $ are contributed in both $ G_E $ and $ G_M $ FFs, they play a more important role in $ G_M $ (as well as $ G_E $, but at larger values of $ -t $). So, it is expected that $ G_M^p $ data will be more impressive on GPDs $ E_v^u $, as can  be clearly seen from Fig.~\ref{fig:Eu}. Finally, since the polarized GPDs $ \widetilde{H}^q $ are not affected by the $ G_M^p $ data and thus there are not any significant differences between various sets obtained from the analyses performed in this subsection, we do not compare them here, and instead postpone drawing GPDs $ x\widetilde{H}^q(x) $ to the next subsection, where we investigate the impact of the CLAS data of $ G_A $~\cite{CLAS:2012ich} on the final results of GPDs.

\subsection{Impact of CLAS data}\label{sec:four-two}

As mentioned in Sec.~\ref{sec:three}, the main constraints on the polarized GPDs $ \widetilde{H}^q $ at $ \xi=0 $ come from the measurements of the nucleon AFF $ G_A $. At present, there are just two kinds of processes that have been used to extract $ G_A $ data; the (anti)neutrino scattering off nucleons and charged pion electroproduction. Although the latter provides most of the available data, the related measurements are old and in most cases incompatible. In the analyses performed in the previous subsection, we used the data from Refs.~\cite{Esaulov:1978ed,DelGuerra:1975uiy,Bloom:1973fn,Joos:1976ng,Choi:1993vt} which are more recent between all measurements and also most accurate between data points with the same value of $ -t $. For the case of (anti)neutrino scattering, we used data from the MiniBooNE experiment~\cite{Butkevich:2013vva} which are much newer compared to the world data from the charged pion electroproduction. 

In this subsection, we are going to investigate the impact of CLAS Collaboration measurements of $ G_A $~\cite{CLAS:2012ich} on the extracted GPDs, especially $ \widetilde{H}^q $. Note that these data have been extracted from the charged pion electroproduction. However, compared with data used before, they are more recent and accurate, and they also cover higher values of $ -t $. Thus, including them in the analysis may significantly affect the final results of GPDs. 
Following the procedure described in the previous subsection, we find a set of GPDs with $ \alpha^{\prime}_{\widetilde{f}_v^u}=\alpha^{\prime}_{f_v^u} $, $ \alpha^{\prime}_{\widetilde{f}_v^d}=\alpha^{\prime}_{f_v^d} $, $ A_{g_v^u}=A_{f_v^u} $, $ A_{g_v^d}=A_{f_v^d} $, $  B_{g_v^d}=B_{g_v^u} $,  and $ \gamma_u= \gamma_d=0 $ which is called Set 4.
Note again that $ \widetilde{f}^{\bar q}(x)=\widetilde{f}_v^q(x) $.
The value of $\chi^2 /\mathrm{d.o.f.} $ for this analysis is 3.24 which is considerably larger than the corresponding value of Set 1 (3.05). Actually, the results show a crucial tension between the CLAS data and the other world data of $ G_A $. The CLAS data are severely suppressed at larger values of $ -t $ and thus are not compatible with the trend of other data. This tension can show that the CLAS data have some problems with normalization. In fact, it is not possible to obtain a simultaneously good description of all $ G_A $ data without normalization. Consequently, we also obtain another set of GPDs by introducing a normalization factor $ {\cal N}_\mathrm{CL} $ for the CLAS data. Actually, we treat $ {\cal N}_\mathrm{CL} $ as a free parameter of the fit procedure. Following the parameterization scan, one finds a set of GPDs with $ \alpha^{\prime}_{\widetilde{f}_v^u}=\alpha^{\prime}_{f_v^u} $, $ \alpha^{\prime}_{\widetilde{f}_v^d}=\alpha^{\prime}_{f_v^d} $, $ A_{g_v^u}=A_{\widetilde{f}_v^u}=A_{f_v^u} $, $ A_{g_v^d}=A_{\widetilde{f}_v^d}=A_{f_v^d} $, $  B_{g_v^d}=B_{g_v^u} $, $ \gamma_u= \gamma_d=0 $, and $ {\cal N}_\mathrm{CL}=2.09 $. We call this Set 5. Note again that releasing parameters $ A_{\widetilde{f}_v^u} $ and $ A_{\widetilde{f}_v^d} $ does not lead to any improvement in the fit quality. Moreover, we have fixed $ {\cal N}_\mathrm{CL} $ on its best value in the last run.

The values of the optimum parameters of Sets 4 and 5 are listed in Table~\ref{tab:par2} and compared with the corresponding values of Set 1 from the previous subsection.  Comparing Set 1  and Set 4, one finds that the original CLAS data have the greatest impact on the polarized profile functions $ \widetilde{f}_v^q $, as expected. In addition, they affect the down-quark distributions more than up-quark distributions. Actually, the large value obtained for $ A $ parameters in $ \widetilde{f}_v^q $ is a reflection of the fact that the CLAS data need a significant suppression of GPDs $ \widetilde{H}^q $ at larger values of $ -t $ to be well fitted. However, by normalizing the CLAS data (Set 5) the results will be more compatible with Set 1 where the CLAS data have not been included.
\begin{table}[th!]
\scriptsize
\setlength{\tabcolsep}{8pt} 
\renewcommand{\arraystretch}{1.4} 
\caption{A comparison between the values of the optimum parameters obtained from the analysis without including CLAS data~\cite{CLAS:2012ich} (Set 1), the analysis including the original CLAS data (Set 4), and the analysis including the normalized CLAS data (Set 5). See Sec.~\ref{sec:four-two} for more details.}\label{tab:par2}
\begin{tabular}{lcccc}
\hline
\hline
 Distribution &  Parameter           &  Set 1            &  Set 4            &  Set 5  \\
\hline 
\hline
$ f_v^u(x) $  & $ \alpha^{\prime} $  & $ 0.687\pm0.007 $ & $ 0.690\pm0.006 $ & $ 0.692\pm0.007 $  \\
			  &	$ A $                & $ 0.884\pm0.023 $ & $ 0.901\pm0.026 $ & $ 0.905\pm0.032 $  \\
			  &	$ B $                & $ 0.968\pm0.024 $ & $ 0.963\pm0.023 $ & $ 0.956\pm0.025 $  \\
\hline
$ f_v^d(x) $  & $ \alpha^{\prime} $  & $ 0.453\pm0.022 $ & $ 0.523\pm0.014 $ & $ 0.517\pm0.028 $  \\
			  &	$ A $                & $ 3.075\pm0.430 $ & $ 4.100\pm0.285 $ & $ 3.911\pm0.471 $  \\
			  &	$ B $                & $ 1.167\pm0.127 $ & $ 0.788\pm0.074 $ & $ 0.833\pm0.154 $  \\
\hline
$ g_v^u(x) $  & $ \alpha^{\prime} $  & $ 0.989\pm0.097 $ & $ 0.996\pm0.054 $ & $ 0.998\pm0.072 $  \\
			  &	$ A $                & $ A_{f_v^u} $     & $ A_{f_v^u} $     & $ A_{f_v^u} $  \\
			  &	$ B $                & $-0.497\pm0.135 $ & $-0.449\pm0.122 $ & $-0.458\pm0.142 $  \\
\hline
$ g_v^d(x) $  & $ \alpha^{\prime} $  & $ 0.814\pm0.058 $ & $ 0.792\pm0.036 $ & $ 0.796\pm0.043 $  \\
			  &	$ A $                & $ 3.235\pm0.328 $ & $ A_{f_v^d} $     & $ A_{f_v^d} $  \\
			  &	$ B $                & $ B_{g_v^u} $     & $ B_{g_v^u} $     & $ B_{g_v^u} $  \\
\hline
$\widetilde{f}_v^u(x)$  & $\alpha^{\prime}$  & $ \alpha^{\prime}_{f_v^u} $ & $ \alpha^{\prime}_{f_v^u} $ & $ \alpha^{\prime}_{f_v^u} $ \\
			  &	$ A $                & $ A_{f_v^u} $     & $ 9.939\pm1.028 $ & $ A_{f_v^u} $  \\
			  &	$ B $                & $-0.500\pm0.270 $ & $-1.411\pm0.180 $ & $0.241\pm0.138 $  \\
\hline
$\widetilde{f}_v^d(x)$  & $\alpha^{\prime}$  & $ \alpha^{\prime}_{f_v^d} $ & $ \alpha^{\prime}_{f_v^d} $ & $ \alpha^{\prime}_{f_v^d} $ \\
			  &	$ A $                & $ A_{f_v^d} $     & $13.347\pm1.556 $ & $ A_{f_v^d} $  \\
			  &	$ B $                & $ 0.366\pm0.558 $ & $-1.969\pm0.091 $ & $-1.056\pm0.167 $  \\
\hline
$ e_v^u(x) $  & $ \alpha $           & $ 0.489\pm0.070 $ & $ 0.483\pm0.039 $ & $ 0.483\pm0.053 $  \\
			  &	$ \beta $            & $ 6.868\pm0.657 $ & $ 6.544\pm0.550 $ & $ 6.570\pm0.656 $  \\
			  &	$ \gamma $           & $ 0.000 $         & $ 0.000 $         & $ 0.000 $  \\
\hline
$ e_v^d(x) $  & $ \alpha $           & $ 0.607\pm0.028 $ & $ 0.648\pm0.020 $ & $ 0.639\pm0.023 $  \\
			  &	$ \beta $            & $ 4.609\pm0.938 $ & $ 3.256\pm0.610 $ & $ 3.532\pm1.016 $  \\
			  &	$ \gamma $           & $ 0.000 $         & $ 0.000 $         & $ 0.000 $  \\
\hline 		 	
\hline 	
\end{tabular}
\end{table}

Table~\ref{tab:chi22} presents a comparison between the results obtained from the analyses of Sets 1, 4, and 5. As can be seen, including the original CLAS data in the analysis leads to a significant increase in the value of $ \chi^2 $ from the world $ G_A $ data. Although the CLAS data have relatively large $ \chi^2 $ values (about 22 for five data points), it is acceptable considering the small uncertainties of the data. Note that the presence of the CLAS data in the analysis also somewhat increases the $ \chi^2 $ of the $ R^p $ and $ G_E^n $ data. However, it does not significantly affect the $ \chi^2 $ of the $ G_M $ (whether for the proton or neutron) or the nucleon radii data. Another point that should be noted is that by introducing a normalization factor for the CLAS data, the value of $ \chi^2 $ of the world $ G_A $ data becomes the same as the one corresponding to the analysis of Set 1 where the CLAS data have not been included. Moreover, the value of the $ \chi^2 $ of the CLAS data decreases from 22 to 7. These facts clearly indicate that by normalizing the CLAS data the tension between them and other $ G_A $ data can be resolved.
\begin{table}[th!]
\scriptsize
\setlength{\tabcolsep}{8pt} 
\renewcommand{\arraystretch}{1.4} 
\caption{The results of the analysis without including CLAS data~\cite{CLAS:2012ich} (Set 1), the analysis including the original CLAS data (Set 4), and the analysis including the normalized CLAS data (Set 5). See Sec.~\ref{sec:four-two} for more details.}\label{tab:chi22}
\begin{tabular}{lcccc}
\hline
\hline
  Observable         &  -$t$ (GeV$^2$)   &  \multicolumn{3}{c}{ $\chi^2$/$ N_{\textrm{pts.}} $  }  \\
                     &                   &     Set 1    &      Set 4     &     Set 5               \\
\hline 		 	
\hline 
$G_{M}^p$~\cite{A1:2013fsc}                        & $0.0152-0.5524$& $463.4 / 77$  & $ 463.5 / 77 $         &  $462.7 / 77 $ \\
$G_M^p/\mu_p G_D$~\cite{Arrington:2007ux}                & $0.007-32.2$   & $113.8 / 56$  & $ 115.3 / 56$  &  $ 116.6 /56$        \\
$R^p = \mu_p G_{E}^p / G_{M}^p$~\cite{Ye:2017gyb}  & $0.162-8.49$   & $107.8 / 69$  & $114.5 / 69$  &  $112.7 / 69$ \\
$G_{E}^n$~\cite{Ye:2017gyb}                        & $0.00973-3.41$ & $27.9 / 38$   & $ 33.0 / 38$  &  $ 31.5 / 38$ \\
$G_M^n/\mu_n G_D$~\cite{Ye:2017gyb}                & $0.071-10.0$   & $45.2 / 33$   & $44.5 / 33 $  &  $44.8 / 33 $ \\
$G_{A}$~\cite{Butkevich:2013vva,Esaulov:1978ed,DelGuerra:1975uiy,Bloom:1973fn,Joos:1976ng,Choi:1993vt}                          & $0.025-1.84$   & $130.0 / 34$  & $161.5  / 34$ &  $ 131.1 / 34$ \\
$G_{A}$~\cite{CLAS:2012ich}                        & $2.12-4.16$    & $ \cdots $         & $22.2  / 5$   &  $7.4 / 5 $ \\
$\sqrt{\left<r_{pE}^2\right>}$~\cite{ParticleDataGroup:2018ovx}   & $ 0 $          & $0.0 / 1$     & $0.0 / 1 $    &  $0.0 / 1 $     \\
$\sqrt {\left<r_{pM}^2\right>}$~\cite{ParticleDataGroup:2018ovx}  & $ 0 $          & $0.0 / 1$     & $1.3 / 1 $    &  $1.3 / 1 $   \\
$\left<r_{nE}^2\right>$~\cite{ParticleDataGroup:2018ovx}          & $ 0 $          & $0.2 / 1$     & $0.8 / 1 $    &  $0.8 / 1 $   \\
$\sqrt {\left<r_{nM}^2\right>}$~\cite{ParticleDataGroup:2018ovx}  & $ 0 $          & $12.5 / 1$    & $12.8 / 1 $   &  $13.0 / 1 $   \\	
\hline
Total $\chi^2 /\mathrm{d.o.f.} $ &                & $900.8 / 295$ & $969.4 / 299$   &  $921.9 / 301$   \\
\hline
\hline
\end{tabular}
\end{table}

\begin{figure}[!htb]
    \centering
\includegraphics[scale=0.9]{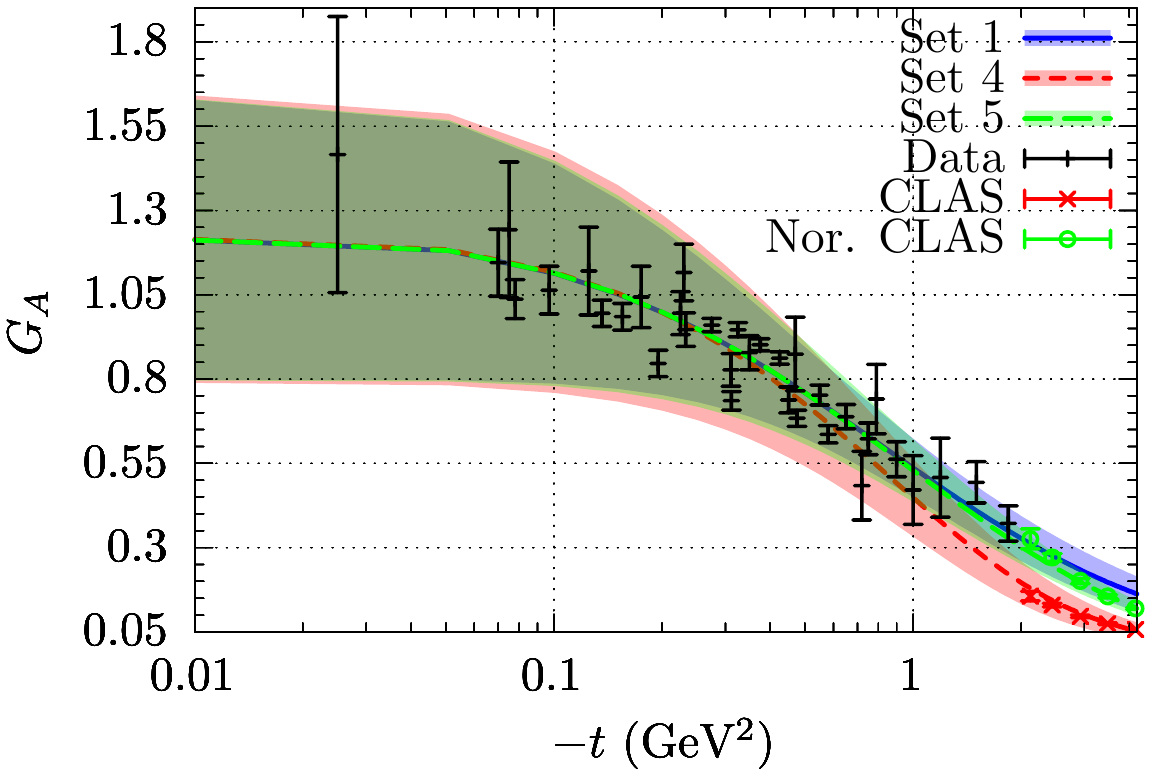}    
    \caption{A comparison between our results for $ G_A $ obtained using Sets 1, 4, Set 5 and fitted data from Refs.~\cite{Butkevich:2013vva,Esaulov:1978ed,DelGuerra:1975uiy,Bloom:1973fn,Joos:1976ng,Choi:1993vt} (black dash symbol) as well as the original (red cross symbol) and normalized (green circle symbol) CLAS data~\cite{CLAS:2012ich}.}
\label{fig:GA2}
\end{figure}

Figure~\ref{fig:GA2} shows a comparison between the theoretical predictions of $ G_A $ obtained using GPDs of Sets 1, 4, and 5 and the fitted data taken from Refs.~\cite{Butkevich:2013vva,Esaulov:1978ed,DelGuerra:1975uiy,Bloom:1973fn,Joos:1976ng,Choi:1993vt} (black dash symbol) as well as the original (red cross symbol) and normalized (green circle symbol) CLAS data~\cite{CLAS:2012ich}. As can be seen, including the original CLAS data in the analysis leads to a considerable change in the results at larger values of $ -t $, so that the theoretical prediction falls off faster than before with $ -t $ growing. Another point that should be mentioned is that the CLAS data also decrease the theoretical uncertainties at larger values of $ -t $. This was predictable because they are very precise compared with other $ G_A $ data. The increase in uncertainties at $ -t\lesssim 2 $ GeV$ ^2 $ can be attributed to the fact that Set 4 has one more free parameter than Set 1. Another result that can be concluded from Fig.~\ref{fig:GA2} is that by normalizing the CLAS data the results obtained become entirely compatible with each other, considering uncertainties. Note that here we have not shown the results obtained for other observables, since the inclusion of the CLAS data (whether original or normalized) in the analysis has not significantly affected them. 

\begin{figure}[!htb]
    \centering
\includegraphics[scale=0.9]{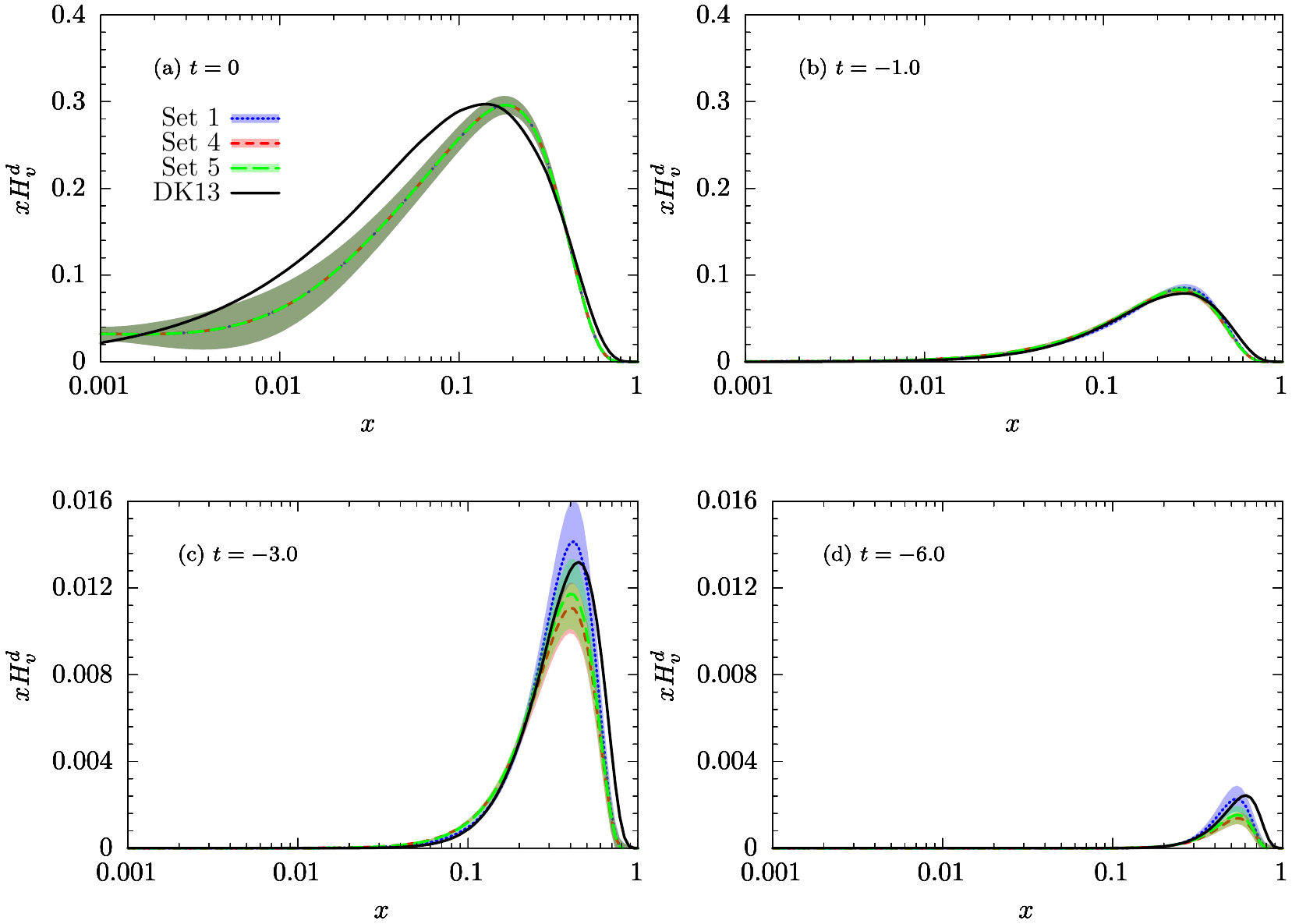}   
    \caption{A comparison between our results for GPDs $ xH_v^d(x) $ obtained from the analysis excluding the CLAS data~\cite{CLAS:2012ich} (Set 1), the analysis including the original CLAS data (Set 4), and the analysis including the normalized CLAS data (Set 5) as well as the DK13 results at four $ t $ values shown in panels (a) $ t=0$, (b) $t=-1$, (c) $t=-3$, and (d) $t=-6 $ GeV$ ^2 $. See Sec.~\ref{sec:four-two} for more details.}
\label{fig:HdCLAS}
\end{figure}

\begin{figure}[!htb]
    \centering
\includegraphics[scale=0.9]{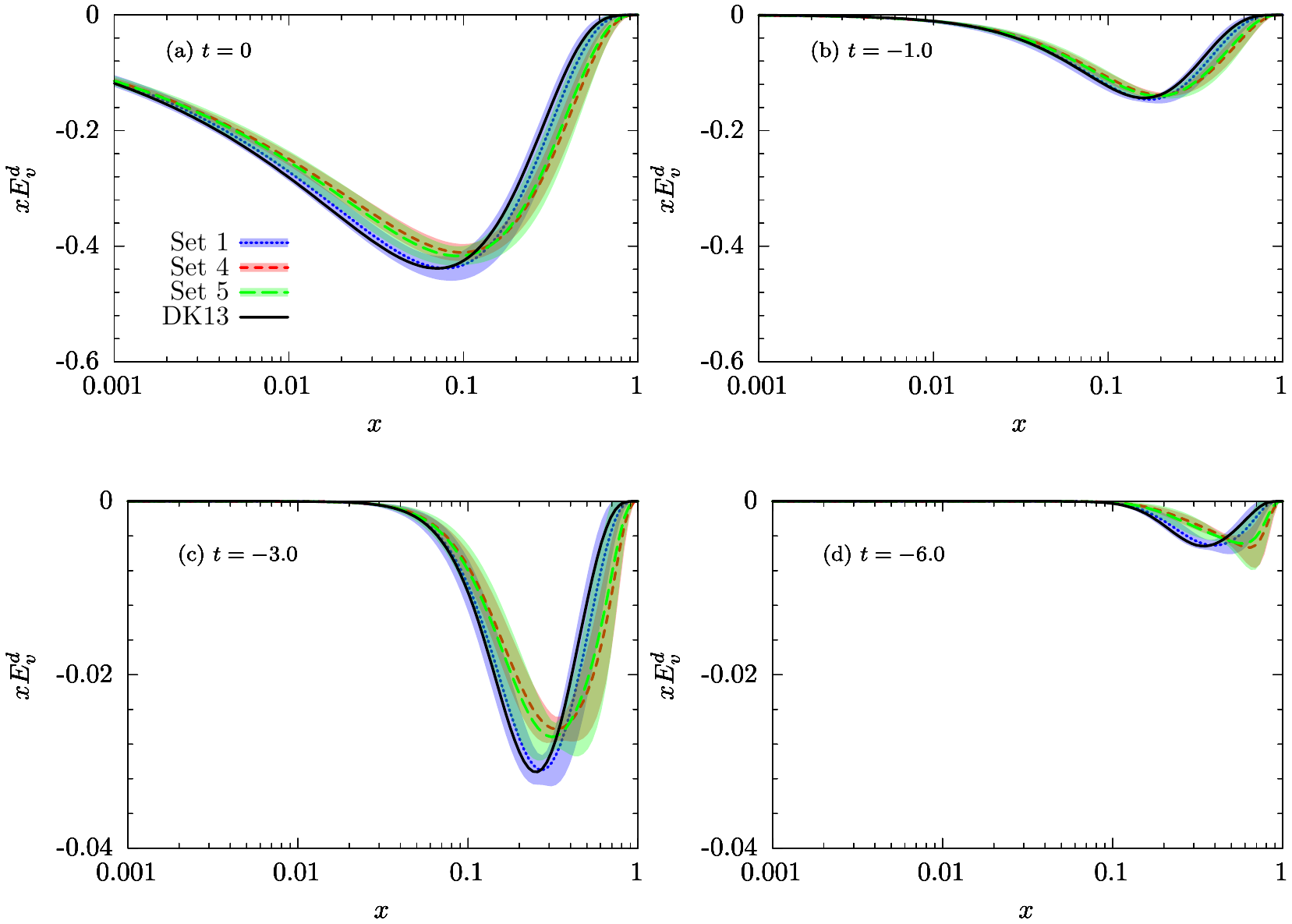}
    \caption{Same as Fig.~\ref{fig:HdCLAS} but for GPDs $ xE_v^d(x) $. }
\label{fig:EdCLAS}
\end{figure}

For the case of unpolarized GPDs $ H_v^q $ and $ E_v^q $, the significant changes are happening just for the down-quark distributions; the CLAS data (whether original or normalized) do not considerably affect the up-quark unpolarized GPDs (here we mean the shape of the distributions; not their uncertainties, since including the CLAS data also leads to a decrease in the uncertainties of $ H_v^u $ and $ E_v^u $). In Figs.~\ref{fig:HdCLAS} and~\ref{fig:EdCLAS}, we have compared, respectively, the $ xH_v^d(x) $ and $ xE_v^d(x) $ GPDs of Sets 1, 4, and 5 as well as the corresponding ones from the analysis of DK13~\cite{Diehl:2013xca}. As can be seen, by including the original CLAS data in the analysis (Set 4), $ xH_v^d(x) $ is more suppressed with $ -t $ growing than before. However, by considering a normalization factor for the CLAS data, the results (Set 5) become somewhat more compatible with Set 1 and DK13. Overall, Set 1 is in more consistency with DK13 because both of them have been obtained without considering the CLAS data. Such a situation is observed also for $ xE_v^d(x) $. However, in this case,  Sets 4 and 5 are first suppressed at small and medium $ -t $ compared to Set 1 and DK13, and they are increased again at large $ -t $ and shifted to the larger values of $ x $. Note that the large uncertainties of Sets 4 and 5 in the last panel of Fig~\ref{fig:EdCLAS}, where the distributions have been compared at $t=-6 $ GeV$ ^2 $, are due to the fact that the CLAS data contain data points with $ -t $ values only up to about 4  GeV$ ^2 $.

\begin{figure}[!htb]
    \centering
\includegraphics[scale=0.9]{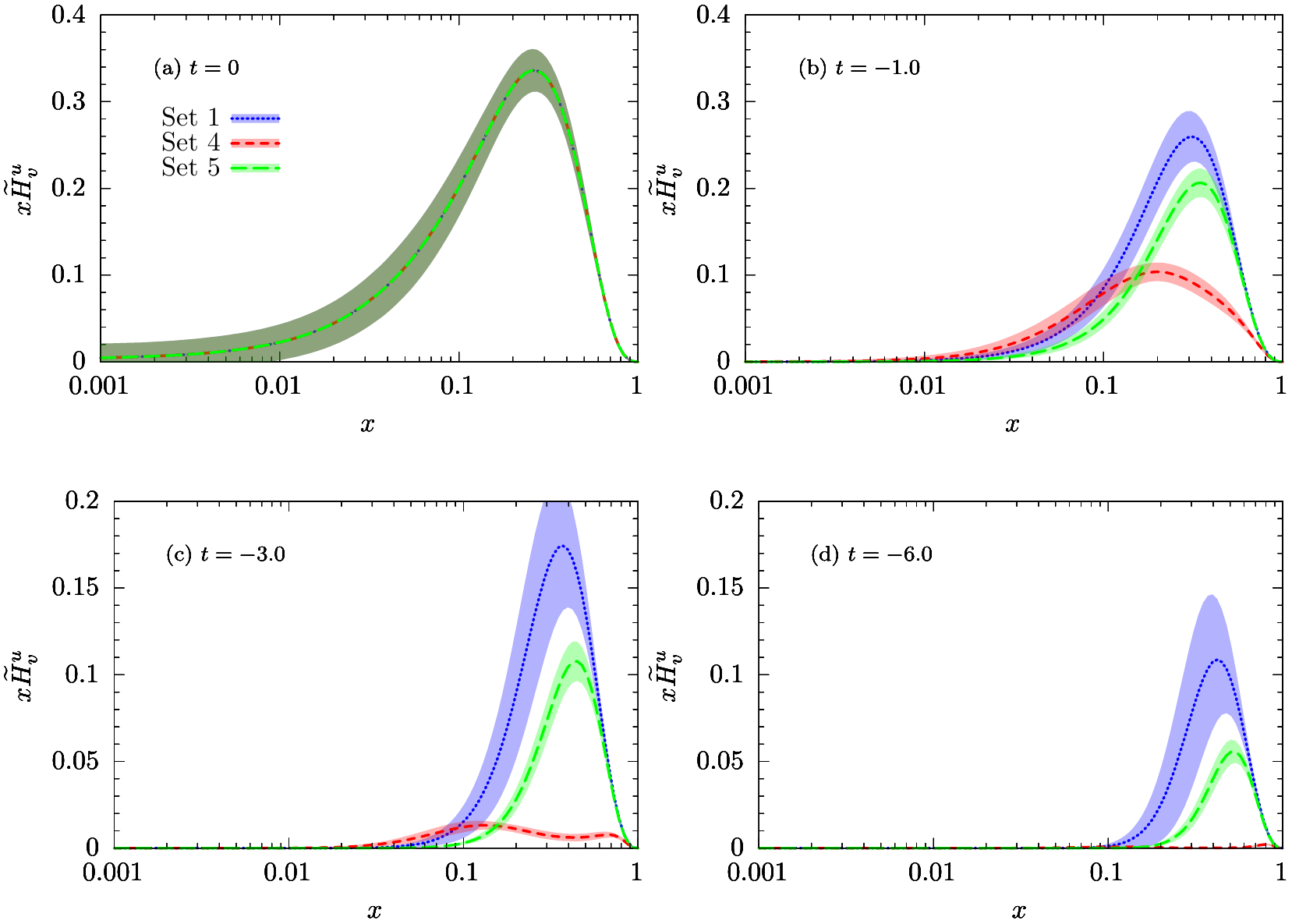}   
    \caption{A comparison between our results for polarized GPDs $ x\widetilde{H}_v^u(x) $ obtained from three analyses: excluding the CLAS data~\cite{CLAS:2012ich} (Set 1), including the original CLAS data (Set 4), and including the normalized CLAS data (Set 5) at four $ t $ values shown in panels (a) $ t=0$, (b) $t=-1$, (c) $t=-3$, and (d) $t=-6 $ GeV$ ^2 $. See Sec.~\ref{sec:four-two} for more details.}
\label{fig:HTu}
\end{figure}

\begin{figure}[!htb]
    \centering
\includegraphics[scale=0.9]{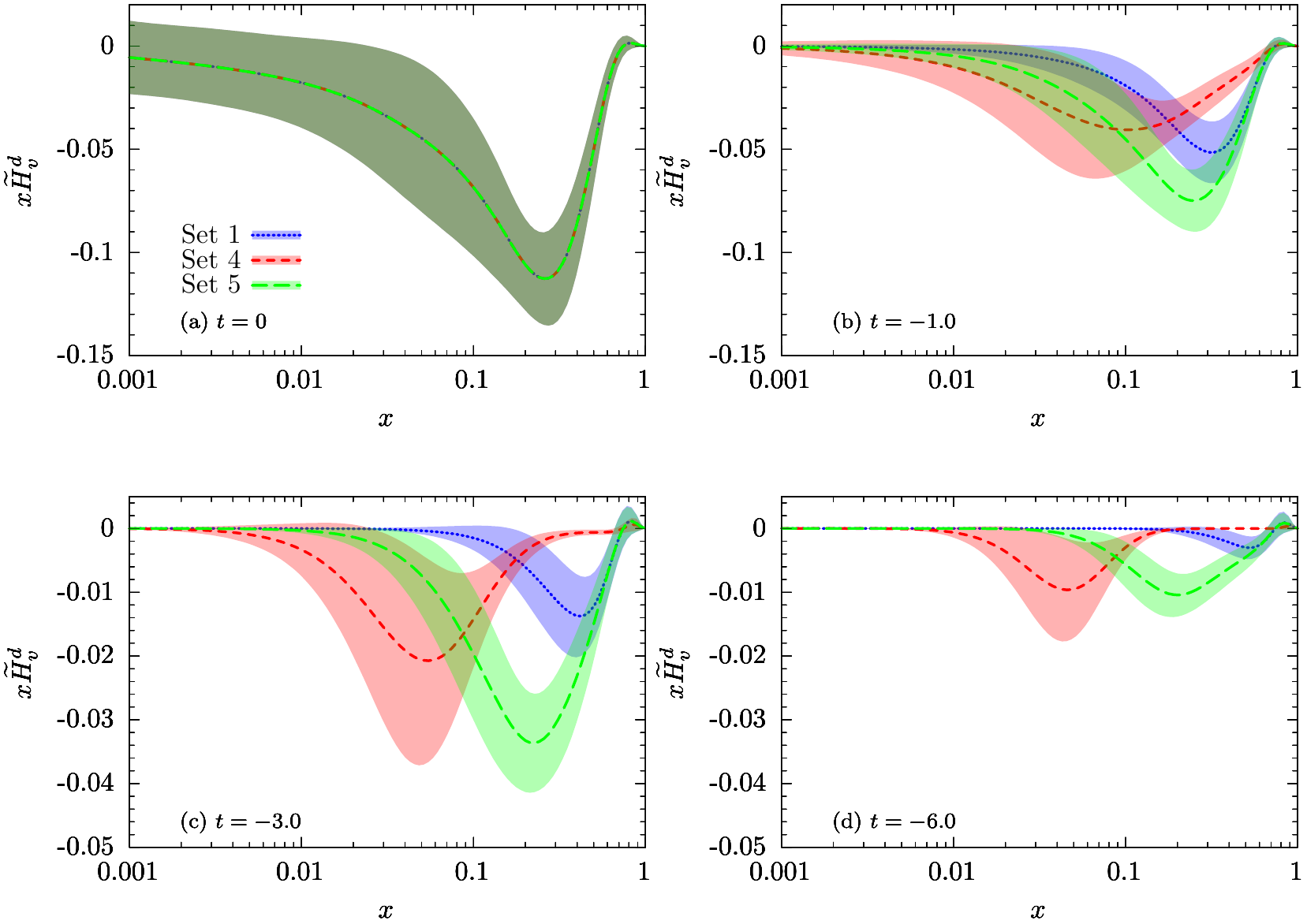}   
    \caption{Same as Fig.~\ref{fig:HTu}, but for polarized GPDs $ x\widetilde{H}_v^d(x) $.}
\label{fig:HTd}
\end{figure}

The results obtained for polarized GPDs $ x\widetilde{H}_v^u(x) $ and $ x\widetilde{H}_v^d(x) $ have been compared, respectively, in Figs.~\ref{fig:HTu} and~\ref{fig:HTd} at four $ t $ values, $ t=0,-1,-3$ and $-6 $ GeV$ ^2 $. Note that in these cases, a comparison with the DK13 analysis is not possible, since the extraction of polarized GPDs has not been performed in that analysis. This figures clearly indicate the significant impact of the CLAS data on polarized GPDs as expected. For the case of the up valence quark, when the CLAS data are included in the analysis, the magnitude of the distribution is considerably decreased with $ -t $ growing. This reduction is much greater for the analysis containing the original CLAS data (Set 4) and introducing a normalization factor makes the results more compatible with Set 1 as before.
The extracted distributions are also inclined to  larger values of $ x $ with $ -t $ growing when compared with Set 1. According to Fig.~\ref{fig:GA2}, the reduction of $ x\widetilde{H}_v^u(x) $ after including the CLAS data in the analysis can be attributed to the significant suppression of these data at larger values of $ -t $. The CLAS data also have crucial impact on the down valence quark GPDs, as it is clear from Fig.~\ref{fig:HTd}. In this case, both Sets 4 and 5 are shifted to smaller values of $ x $ with $ -t $ growing. In terms of magnitude, Set 4 first decreases compared with Set 1, but then increases at larger values of $ -t $. Although Set 5 has the largest magnitude, it is more compatible again with Set 1 than with Set 4.

\begin{figure}[!htb]
    \centering
\includegraphics[scale=0.9]{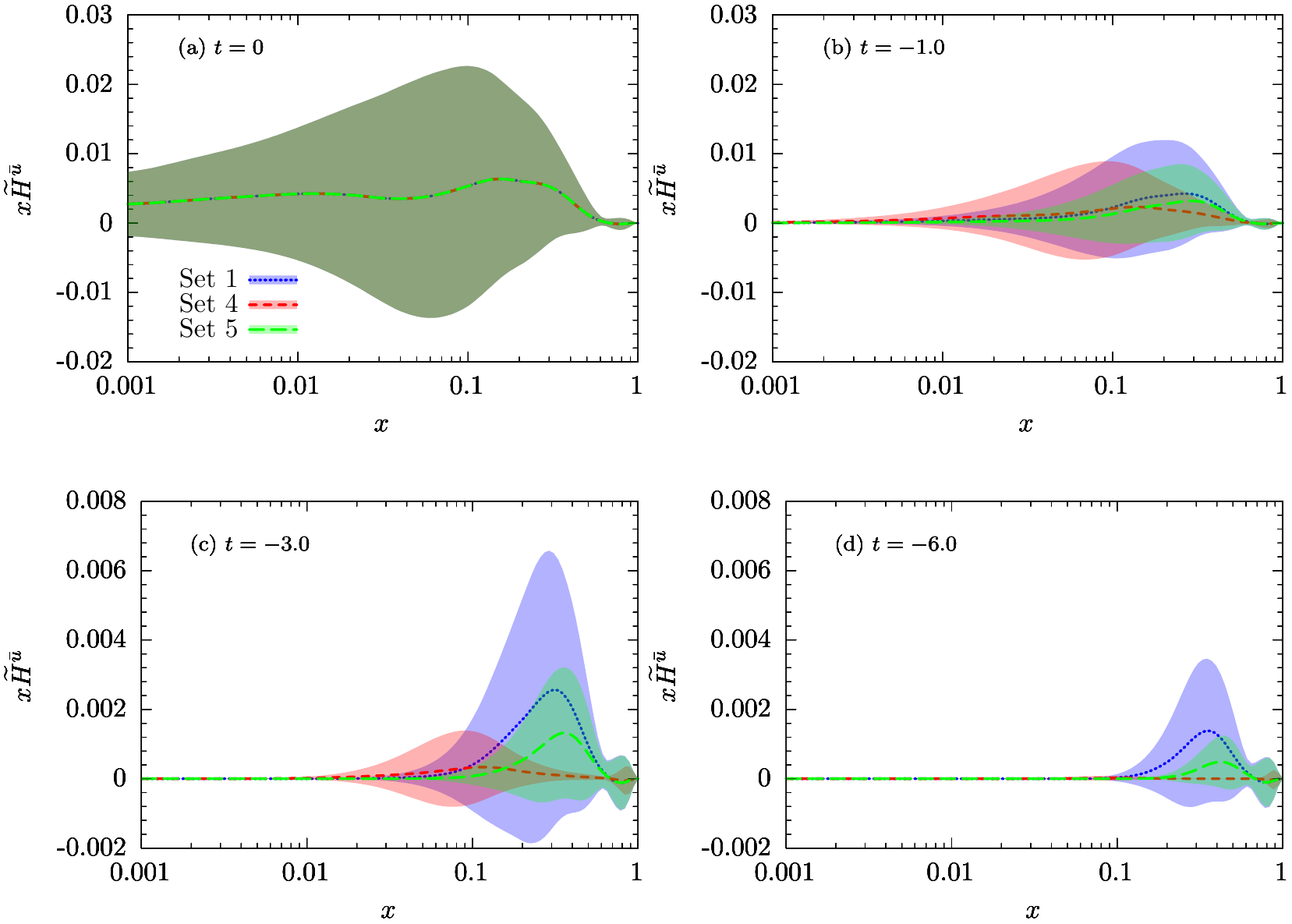}   
    \caption{Same as Fig.~\ref{fig:HTu}, but for polarized GPDs $ x\widetilde{H}^{\bar u}(x) $.}
\label{fig:HTubar}
\end{figure}

\begin{figure}[!htb]
    \centering
\includegraphics[scale=0.9]{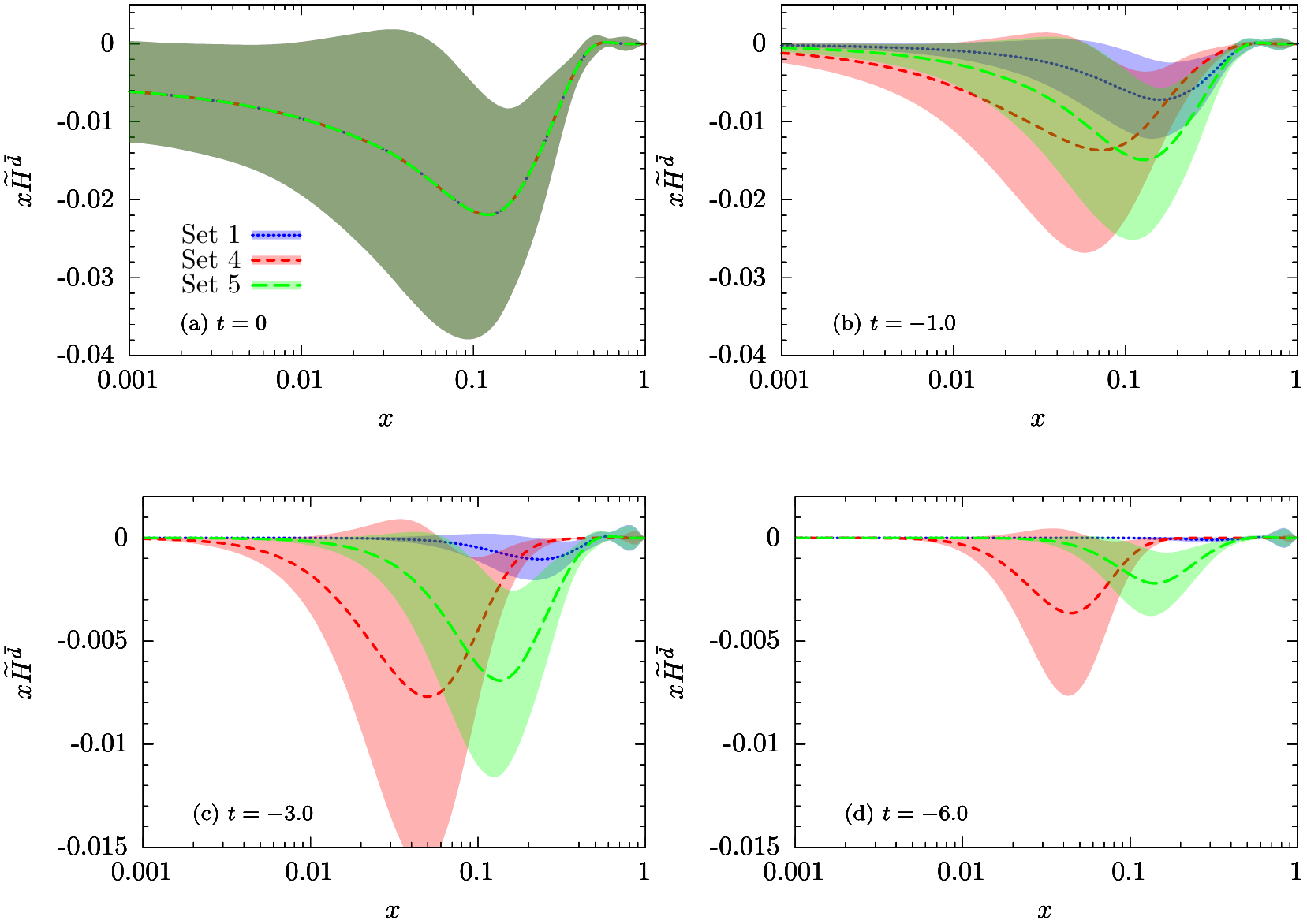}   
    \caption{Same as Fig.~\ref{fig:HTu}, but for polarized GPDs $ x\widetilde{H}^{\bar d}(x) $.}
\label{fig:HTdbar}
\end{figure}

As mentioned before, since the theoretical calculation of the AFF $ G_A $ also includes the sea-quark contributions of the polarized GPDs,  $ x\widetilde{H}^{\bar q}(x) $, it is possible to extract them from the analyses performed in this subsection and the previous one. However, since they have smaller contributions in $ G_A $ than the valence sectors, and on the other hand the data do not provide enough constraints for them, we have decided to consider the assumption  $ \widetilde{f}^{\bar q}(x)=\widetilde{f}_v^q(x) $. Figures~\ref{fig:HTubar} and~\ref{fig:HTdbar} show the same comparisons as Figs.~\ref{fig:HTu} and~\ref{fig:HTd}, respectively, but for $ x\widetilde{H}^{\bar u}(x) $ and $ x\widetilde{H}^{\bar d}(x) $.
As  is clear from these figures, the sea-quark polarized GPDs behave almost the same as valance polarized GPDs by including the CLAS data in the analysis. However, they have larger uncertainties as expected due to the large uncertainties of the {NNPDF} sea-quark polarized PDFs~\cite{Nocera:2014gqa}.

\subsection{Impact of WACS data}\label{sec:four-three}

As pointed out before, the WACS cross section is related to three kinds of GPDs: namely, $ H^q $, $ E^q $, and $ \widetilde{H}^q $~\cite{Huang:2001ej}. So, including the WACS data in the analysis can provide new and important information about GPDs, especially at large values of $ -t $. An advantage of these data is that they can also provide some information about the sea-quark contributions of GPDs. In this subsection, we are going to investigate the impact of the JLab Hall A Collaboration data~\cite{Danagoulian:2007gs} on GPDs obtained in the previous subsections. Note that a simultaneous analysis of the world $ G_A $ (excluding the CLAS data) and JLab WACS data was performed in Ref.~\cite{Hashamipour:2020kip}, but it did not include the nucleon Sachs FFs and radii data. Moreover, the GPDs $ E_v^q $ were fixed from an older analysis of the world electron scattering data~\cite{Diehl:2004cx}. 

For the theoretical calculation of the WACS cross section at NLO, we use the formula presented in Ref.~\cite{Huang:2001ej} (for a compact review, see Ref.~\cite{Hashamipour:2020kip}). An important point that should be noted is that there are three scenarios to relate the Mandelstam variables at the partonic level and those of the whole process~\cite{Diehl:2002ee}. Actually, if one neglects the mass of the proton, they will be equal (Scenario 1). However, considering the proton mass we have two more scenarios:
\begin{align}  
{\label{Eq:6}}
&\mathrm{Scenario\ 2:}\quad \hat{s}=s-m^2,\quad \hat{t}=t,\quad \hat{u}=u-m^2.\nonumber\\
&\mathrm{Scenario\ 3:}\quad \hat{s}=s-m^2,\quad \hat{t}=-\frac{\hat{s}}{2}\left(1-\cos \theta_\mathrm{cm}\right), \quad \hat{u}=-\hat{s}-\hat{t}. &
\end{align}  

In Ref.~\cite{Hashamipour:2020kip}, the authors showed that Scenario 3 leads to a better description of the WACS data. However, because of the differences between their analysis and the present study mentioned above, we considered both Scenarios 2 and 3 to investigate which of them leads to better results when a wide range of the experimental data is included in the analysis. As a result, we found that Scenario 3 works better and makes $ \chi^2 $ smaller in all analyses which include different datasets. Therefore, in the following, we present just the results obtained using Scenario 3.
On the other hand, according to our initial investigations, there are some tensions between the CLAS data not only with other $ G_A $ data (as has been shown in the previous subsection), but also with WACS data. In this way, here we are going to perform three analyses: (i) by excluding the CLAS data, (ii) by including the original CLAS data, and (iii) by including the normalized CLAS data. We call these three analyses Set 9, Set 10, and Set 11 respectively. We also repeat the third analysis by removing the $ G_M^p $ data, which leads to Set 12. Its motivations and results will be introduced later.

The best parametrization forms, as well as the optimum values of unknown parameters are obtained utilizing the parametrization scan as before. To this aim, we consider again $ \widetilde{f}^{\bar q}(x)=\widetilde{f}_v^q(x) $ and use the same assumption for the sea-quark contributions of GPDs $ H $, but by fixing the $ \alpha^{\prime} $ parameters as $ \alpha^{\prime}_{f^{\bar u}}=\alpha^{\prime}_{f^{\bar d}}=2 $. The reason to choose such a large value for these parameters is the problem of overestimating the WACS data at smaller values of $ -t $. To be more precise, if one considers $ \alpha^{\prime}_{f^{\bar q}}=\alpha^{\prime}_{f_v^q} $ just as for polarized GPDs $ \widetilde{H} $, it leads to a significant increase in the $ \chi^2 $ value of the WACS data due to the poor description of data points in the small $ -t $ region. In fact, releasing these parameters leads to an even larger value that is unphysical. So, we decided to consider a value that is not far from unity and, on the other hand, leads to an acceptable description of data (note that considering, for example, $ \alpha^{\prime}_{f^{\bar u}}=\alpha^{\prime}_{f^{\bar d}}=10 $ just decreases the total $ \chi^2 $ up to 15 units compared to $ \alpha^{\prime}_{f^{\bar u}}=\alpha^{\prime}_{f^{\bar d}}=2 $). Finally, we found GPDs with $ \alpha^{\prime}_{\widetilde{f}_v^u}=\alpha^{\prime}_{f_v^u} $, $ \alpha^{\prime}_{\widetilde{f}_v^d}=\alpha^{\prime}_{f_v^d} $, and $ \gamma_u= \gamma_d=0 $. Other parameters of the valence GPDs have been obtained from the fit. 
The normalization factor $ {\cal N}_\mathrm{CL} $ of the CLAS data obtained from the analysis of Set 11 is equal to 1.67 (it is 2.16 for Set 12).

Table~\ref{tab:chi23} presents a comparison between the results obtained from the aforementioned analyses.
Regardless of Set 12, one can see that the best $ \chi^2 $ belongs to the analysis in which the CLAS data have been excluded (Set 9). However, utilizing a normalization factor for these data makes the situation better (it decreases the total $ \chi^2 $ from 1417 to 1302). Overall, the results do not show a desirable description of the WACS data even after excluding the CLAS data (the value of $ \chi^2 $ is 285 for 25 data points), in comparison with Ref.~\cite{Hashamipour:2020kip}, where just the AFF and WACS data were analyzed. For investigating the pure impact of the WACS data on GPDs, one can compare the results of Set 9 with the corresponding results of Set 1 from Table~\ref{tab:chi2}. As a result, one can see that the inclusion of the WACS data in the analysis makes the description of the neutron data worse. Moreover, the $ \chi^2 $ value of the {Mainz} data is increased up to about 23 units. 
\begin{table}[th!]
\scriptsize
\setlength{\tabcolsep}{5.7pt} 
\renewcommand{\arraystretch}{1.4} 
\caption{The results of the analyses including the WACS data~\cite{Danagoulian:2007gs} considering Scenario 3 of Eq.~(\ref{Eq:6}). See Sec.~\ref{sec:four-three} for more details.}\label{tab:chi23}
\begin{tabular}{lccccc}
\hline
\hline
  Observable         &  -$t$ (GeV$^2$)   &  \multicolumn{4}{c}{ $\chi^2$/$ N_{\textrm{pts.}} $  }  \\
                     &                   &     Set 9    &      Set 10     &     Set 11  &     Set 12  \\
\hline 		 	
\hline 
$G_{M}^p$~\cite{A1:2013fsc}                        & $0.0152-0.5524$ &  $486.6 / 77$  & $ 485.3 / 77 $  
												   & $ 488.6 / 77 $  & $ - $ \\
$G_M^p/\mu_p G_D$~\cite{Arrington:2007ux}          & $0.007-32.2$    & $110.7 / 56$   & $ 107.8 / 56$ 
												   &  $ 109.0 /56$   &  $ - $  \\
$R^p = \mu_p G_{E}^p / G_{M}^p$~\cite{Ye:2017gyb}  & $0.162-8.49$    &  $107.5 / 69$  & $106.8 / 69$ 
												   &  $104.9 / 69$   &  $101.6 / 69$  \\
$G_{E}^n$~\cite{Ye:2017gyb}                        & $0.00973-3.41$  &  $52.1 / 38$   & $ 51.8 / 38$  
												   &  $ 45.1 / 38$   &  $ 32.8 / 38$  \\
$G_M^n/\mu_n G_D$~\cite{Ye:2017gyb}                & $0.071-10.0$    &  $54.5 / 33$   & $53.3 / 33 $  
												   &  $54.0 / 33 $   &  $58.3 / 33 $  \\
$G_{A}$~\cite{Butkevich:2013vva,Esaulov:1978ed,DelGuerra:1975uiy,Bloom:1973fn,Joos:1976ng,Choi:1993vt}                          												   & $0.025-1.84$    &  $129.8 / 34$  & $162.6  / 34$ 
												   &  $ 138.2 / 34$  &  $ 131.8 / 34$ \\
$G_{A}$~\cite{CLAS:2012ich}                        & $2.12-4.16$     &  $ - $         & $49.0  / 5$   
												   &  $3.0 / 5 $     &  $13.7 / 5 $    \\
$\sqrt{\left<r_{pE}^2\right>}$~\cite{ParticleDataGroup:2018ovx}   
												   & $ 0 $           &  $0.0 / 1 $    &  $0.0 / 1 $   
												   &  $0.0 / 1 $     &  $0.0 / 1 $    \\
$\sqrt {\left<r_{pM}^2\right>}$~\cite{ParticleDataGroup:2018ovx}  
												   & $ 0 $           &  $1.6 / 1 $    &  $1.5 / 1 $   
												   &  $1.7 / 1 $     &  $1.8 / 1 $    \\
$\left<r_{nE}^2\right>$~\cite{ParticleDataGroup:2018ovx}          
												   & $ 0 $           &  $2.6 / 1$     & $4.1 / 1 $   
												   &  $4.3 / 1 $     &  $0.0 / 1 $    \\
$\sqrt {\left<r_{nM}^2\right>}$~\cite{ParticleDataGroup:2018ovx}   
												   & $ 0 $           &  $17.6 / 1$    & $23.1 / 1 $  
												   &  $22.4 / 1 $    &  $14.0 / 1 $   \\	
$d\sigma/dt$ (WACS)~\cite{Danagoulian:2007gs}
												   & $1.65-6.46$     &  $284.7 / 25$  & $371.5 / 25 $  
												   &  $330.4 / 25 $  &  $119.9 / 25 $ \\
\hline
Total $\chi^2 /\mathrm{d.o.f.} $ &                 &  $1247.7 / 316$ & $1416.8 / 321$   &  $1301.6 / 321$ 
                                                   &  $473.9 / 188$   \\
\hline
\hline
\end{tabular}
\end{table}

The values of the optimum parameters are listed in Table~\ref{tab:par22} and compared with the corresponding values of Set 1 presented in Sec.~\ref{sec:four-one}.  
Comparing Sets 1 and 9, one finds that the WACS data have the greatest impact on GPDs $ E_v^d(x)$ and also the polarized profile functions $ \widetilde{f}_v^q $. As before, including the original CLAS data in the analysis dramatically changes the polarized profile functions $ \widetilde{f}_v^q $ (Set 10), while their impacts become less drastic after considering a normalization factor (Set 11). We discuss Set 12 and the impact of removing $ G_M^p $ data on the extracted GPDs later.
\begin{table}[th!]
\scriptsize
\setlength{\tabcolsep}{8pt} 
\renewcommand{\arraystretch}{1.4} 
\caption{A comparison between the values of the optimum parameters obtained from the analyses 
performed in this subsection namely Sets 9, 10, 11, and 12, as well as the corresponding values of Set 1 presented in Sec.~\ref{sec:four-one}. See Sec.~\ref{sec:four-three} for more details.}\label{tab:par22}
\begin{tabular}{lcccccc}
\hline
\hline
 Distribution &  Parameter           &  Set 1            &  Set 9            &  Set 10    
                                     &  Set 11           &  Set 12    \\
\hline 
\hline
$ f_v^u(x) $  & $ \alpha^{\prime} $  & $ 0.687\pm0.007 $ & $ 0.657\pm0.008 $ & $ 0.655\pm0.007 $  
									 & $ 0.659\pm0.007 $ & $ 0.627\pm0.005 $ \\
			  &	$ A $                & $ 0.884\pm0.023 $ & $ 0.797\pm0.035 $ & $ 0.786\pm0.031 $ 
									 & $ 0.807\pm0.036 $ & $ 1.544\pm0.045 $ \\
			  &	$ B $                & $ 0.968\pm0.024 $ & $ 1.085\pm0.030 $ & $ 1.092\pm0.027 $  
									 & $ 1.070\pm0.029 $ & $ 1.129\pm0.019 $ \\
\hline
$ f_v^d(x) $  & $ \alpha^{\prime} $  & $ 0.453\pm0.022 $ & $ 0.418\pm0.030 $ & $ 0.382\pm0.028 $  
									 & $ 0.409\pm0.024 $ & $ 0.454\pm0.012 $ \\
			  &	$ A $                & $ 3.075\pm0.430 $ & $ 3.796\pm0.507 $ & $ 3.259\pm0.442 $  
									 & $ 3.525\pm0.449 $ & $ 5.368\pm0.299 $ \\
			  &	$ B $                & $ 1.167\pm0.127 $ & $ 1.240\pm0.167 $ & $ 1.441\pm0.151 $  
									 & $ 1.293\pm0.139 $ & $ 0.973\pm0.083 $ \\
\hline
$ g_v^u(x) $  & $ \alpha^{\prime} $  & $ 0.989\pm0.097 $ & $ 0.944\pm0.135 $ & $ 1.019\pm0.158 $  
									 & $ 1.025\pm0.122 $ & $ 0.860\pm0.048 $ \\
			  &	$ A $                & $ A_{f_v^u} $     & $ 0.925\pm0.192 $ & $ 0.949\pm0.150 $  
									 & $ 0.967\pm0.171 $ & $ 1.691\pm0.218 $ \\
			  &	$ B $                & $-0.497\pm0.135 $ & $-0.237\pm0.177 $ & $-0.590\pm0.145 $  
									 & $-0.621\pm0.115 $ & $-0.381\pm0.066 $\\
\hline
$ g_v^d(x) $  & $ \alpha^{\prime} $  & $ 0.814\pm0.058 $ & $ 0.788\pm0.049 $ & $ 0.743\pm0.041 $  
									 & $ 0.743\pm0.035 $ & $ 0.760\pm0.016 $ \\
			  &	$ A $                & $ 3.235\pm0.328 $ & $ 4.048\pm0.771 $ & $ 3.450\pm0.748 $  
									 & $ 3.780\pm0.670 $ & $ 1.837\pm0.288 $ \\
			  &	$ B $                & $ B_{g_v^u} $     & $-1.353\pm0.194 $ & $-1.198\pm0.256 $  
									 & $-1.325\pm0.118 $ & $-0.402\pm0.066 $ \\
\hline
$\widetilde{f}_v^u(x)$  & $\alpha^{\prime}$  & $ \alpha^{\prime}_{f_v^u} $ & $ \alpha^{\prime}_{f_v^u} $ & $ \alpha^{\prime}_{f_v^u} $   & $ \alpha^{\prime}_{f_v^u} $ & $ \alpha^{\prime}_{f_v^u} $ \\
			  &	$ A $                & $ A_{f_v^u} $     & $ 1.325\pm0.514 $ & $14.713\pm0.756 $  
									 & $ 5.442\pm0.987 $ & $ 1.597\pm0.315 $ \\
			  &	$ B $                & $-0.500\pm0.270 $ & $-0.390\pm0.136 $ & $-1.967\pm0.063 $  
									 & $-0.208\pm0.216 $ & $-0.528\pm0.059 $ \\
\hline
$\widetilde{f}_v^d(x)$  & $\alpha^{\prime}$  & $ \alpha^{\prime}_{f_v^d} $ & $ \alpha^{\prime}_{f_v^d} $ & $ \alpha^{\prime}_{f_v^d} $   & $ \alpha^{\prime}_{f_v^d} $  & $ \alpha^{\prime}_{f_v^d} $ \\
			  &	$ A $                & $ A_{f_v^d} $     & $-1.231\pm2.509 $ & $ 3.323\pm0.915 $  
									 & $ 4.892\pm0.369 $ & $ 3.578\pm0.517 $ \\
			  &	$ B $                & $ 0.366\pm0.558 $ & $ 0.855\pm0.467 $ & $-0.547\pm0.210 $  
									 & $-1.167\pm0.087 $ & $ 0.302\pm0.119 $ \\
\hline
$ e_v^u(x) $  & $ \alpha $           & $ 0.489\pm0.070 $ & $ 0.509\pm0.100 $ & $ 0.488\pm0.113 $  
									 & $ 0.482\pm0.096 $ & $ 0.458\pm0.050 $ \\
			  &	$ \beta $            & $ 6.868\pm0.657 $ & $ 5.726\pm0.772 $ & $ 6.950\pm1.165 $  
									 & $ 7.126\pm0.559 $ & $10.651\pm0.445 $ \\
			  &	$ \gamma $           & $ 0.000 $         & $ 0.000 $         & $ 0.000 $  & $ 0.000 $ 
			                         & $ 0.000 $\\			  
\hline
$ e_v^d(x) $  & $ \alpha $           & $ 0.607\pm0.028 $ & $ 0.613\pm0.027 $ & $ 0.612\pm0.026 $  
									 & $ 0.620\pm0.025 $ & $ 0.647\pm0.011 $ \\
			  &	$ \beta $            & $ 4.609\pm0.938 $ & $14.918\pm2.699 $ & $15.573\pm3.986 $  
									 & $ 17.889\pm2.995 $ & $ 4.663\pm0.686 $ \\
			  &	$ \gamma $           & $ 0.000 $         & $ 0.000 $         & $ 0.000 $  & $ 0.000 $ 
			                         & $ 0.000 $\\
\hline 		 	
\hline 	
\end{tabular}
\end{table}

\begin{figure}[!htb]
    \centering
\includegraphics[scale=1.3]{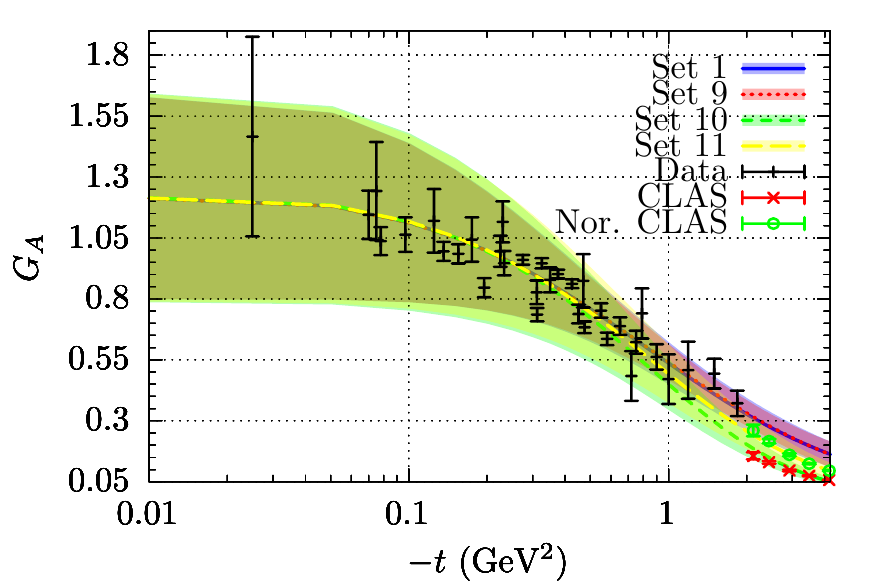}    
    \caption{A comparison between our results for $ G_A $ obtained using Sets 1, 9, 10, and 11 and fitted data from Refs.~\cite{Butkevich:2013vva,Esaulov:1978ed,DelGuerra:1975uiy,Bloom:1973fn,Joos:1976ng,Choi:1993vt} (black dash symbol), as well as the original (red cross symbol) and normalized (green cross symbol) CLAS data~\cite{CLAS:2012ich}.}
\label{fig:GA3}
\end{figure}

Figure~\ref{fig:GA3} shows the same comparison as Fig.~\ref{fig:GA2} but for Sets 1, 9, 10, and 11. Note that here the normalization factor $ {\cal N}_\mathrm{CL} $ for the normalized CLAS data is equal to 1.67. As can be seen, the inclusion of the WACS data in the analysis does not considerably affect the description of the $ G_A $ data. This could also be inferred from Table.~\ref{tab:chi23}, where Sets 1 and 9 have the same $ \chi^2 $ for the AFF data. Comparing Figs.~\ref{fig:GA2} and~\ref{fig:GA3}, one can conclude that the presence of the WACS data makes the normalization of the CLAS data milder; i.e., the result of Set 11 is almost between Sets 9 and 10 at larger values of $ -t $, while without the WACS data, there is no significant difference between the results of Sets 1 and 5 in Fig.~\ref{fig:GA2}. 

\begin{figure}[!htb]
    \centering
\includegraphics[scale=0.9]{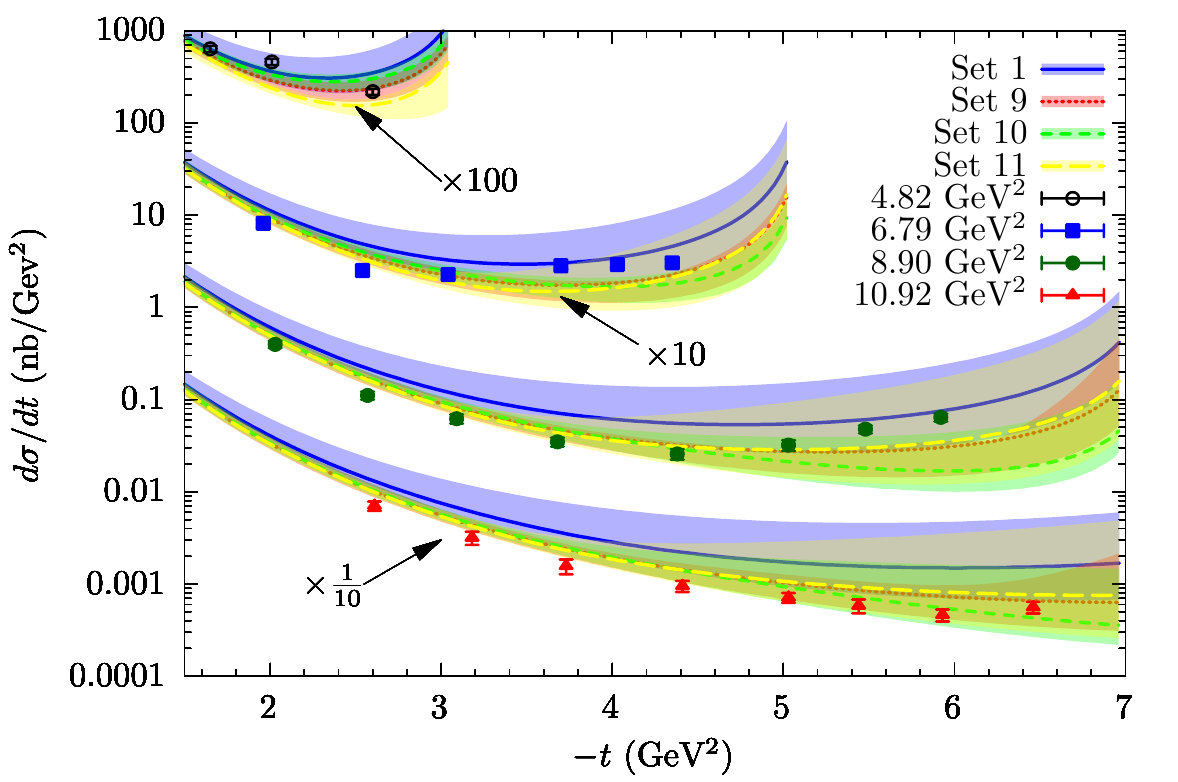}    
    \caption{A comparison between the experimental data of the WACS cross section ($ d\sigma/dt $)~\cite{Danagoulian:2007gs} with the related theoretical predictions obtained using GPDs of Sets 1, 9, 10, and 11. The data points belong to four different values of $ s $, namely $ s=4.82, 6.79, 8.90,$ and $ 10.92 $ GeV$ ^2 $, which are shown by the circle, square, filled circle, and triangle symbols, respectively. The indicated factors are used to distinguish the graphs.}
\label{fig:WACS}
\end{figure}

In Fig.~\ref{fig:WACS}, we have compared the theoretical predictions of the WACS cross section calculated using Sets 9, 10, and 11 and the related experimental data from JLab~\cite{Danagoulian:2007gs} at four different values of $ s $ namely, $ s=4.82, 6.79, 8.90,$ and $ 10.92$ GeV$ ^2 $, that have been multiplied by appropriate factors to make the results more distinguishable. In order to investigate the pure impact of the WACS data on GPDs we have also presented the result of Set 1 in this figure. Comparing Sets 1 and 9, one can realize that the WACS data need GPDs that are smaller in magnitude to be described at smaller values of $ -t $, since all sets overestimate data in this region for all values  of $ s $ except $ s=4.82 $ GeV$ ^2 $. However, in the case of medium values of $ s $, the results obtained in this subsection underestimate data at larger values of $ -t $. Actually, in this case, the data clearly prefer Set 1, which indicates that they are more compatible with other data at this kinematic (medium $ s $ and large $ -t $). Among the three sets of GPDs obtained in this subsection, Set 10, which includes the CLAS data without considering a normalization factor has the worst description of the WACS data. This implies that tension exists between the original CLAS data not only with the world $ G_A $ data but also with the WACS data. Overall, the results of Sets 9 and 11 are acceptable, considering uncertainties. 

Here we do not present comparisons to other observables. Actually, in the case of Sachs FFs, there are no considerable differences between the theoretical predictions of Sets 9, 10, and 11. We just note that for the neutron Sachs FFs, they are a little different from the corresponding predictions of Set 1 at medium values of $ -t $. This leads to a worse description of these data than before as it can be inferred by comparing Tables~\ref{tab:chi23} and~\ref{tab:chi2}. 

Before discussing the impact of WACS data on the extracted GPDs, it is necessary to address the question of why the fit of WACS data is not as satisfying. Actually, the obtained $ \chi^2 $ for these dataset is 284.7 for 25 data points in the best situation (except for Set 12), showing strong deviation in the data. Generally, there are three possibilities to explain this unfavorable description of the WACS data: (i) there is a problem with the Jefferson Lab measurements, (ii) there is a problem with the theoretical description of the WACS process using the handbag mechanism, and (iii) there is a problem with our phenomenological framework. The first of these is almost impossible, and besides, it cannot be checked until the new measurements of the WACS data are performed. The second possibility should be investigated by revisiting the related theoretical calculations, which is out of the scope of the present study. For example, maybe the NNLO contributions should be considered, or it is necessary to use a new scenario to relate the Mandelstam variables at the partonic level and those of the whole process in Eq.~(\ref{Eq:6}). It is also possible that the handbag approach is not a good case for describing the WACS process, although we do not have a better approach at present, since the perturbative QCD predictions are at least 10 times smaller than the experimental data as reported in the JLab paper~\cite{Danagoulian:2007gs}. However, we can check the third possibility from different points of view. Performing a comprehensive investigation on the poor description of the WACS data in the present global analysis may also shed light on the first and second ones. In the following, we have tried to deeply investigate this issue.

From a phenomenological point of view, we have performed the following steps:
\begin{itemize}
\item We checked the flexibility of all parametrizations by releasing more parameters as much as possible, in particular for the sea-quark distributions. However, there was no improvement in the value of $\chi^2$/d.o.f.. This indicates that the poor descriptions of the WACS data in various kinematics (different values of -$ t $ and $ s $) are not due to neglecting or making a bad estimation of the GPD contributions. Perhaps one could say that the ansatz introduced in Eq.~(\ref{Eq1}) or the parametrization form of the profile function introduced in Eq.~(\ref{Eq4}) are not good choices. But they have been used in different studies concerning GPDs~\cite{Diehl:2004cx,Diehl:2013xca,Hashamipour:2019pgy,Hashamipour:2020kip,Hashamipour:2021kes} and are compatible with the bulk of the experimental data.
\item We changed the input PDFs to check if the results are PDF dependent. Just as in previous studies~\cite{Hashamipour:2019pgy,Hashamipour:2020kip}, we found that changing input PDFs does not lead to a considerable change in the predictions.
\item We checked if changing the value of the scale $ \mu $ in which PDFs are chosen can lead to a better description of the WACS data. By fixing $ \mu $ at different values below and above the default $ \mu=2 $ GeV, we found no significant improvement in the $\chi^2$ of the WACS data, or in the total $\chi^2$ of the analysis.
\item Since the application of the handbag approach is questionable at smaller values of $ -t $~\cite{Huang:2001ej}, we put a cut on the WACS data at smaller values of $ -t $ and removed data points with $ -t < 2.6 $ GeV$ ^2 $ (six data points). By repeating the analysis, we found that although the $\chi^2$ of the WACS data reduces from 330 (Set 11, for example) to 266, the value of $\chi^2$ per data point increases from 13.2 to 14. This indicates that putting a cut on small values of $ -t $ cannot solve the problem raised.
\end{itemize}

\begin{figure}[!htb]
    \centering
\includegraphics[scale=0.9]{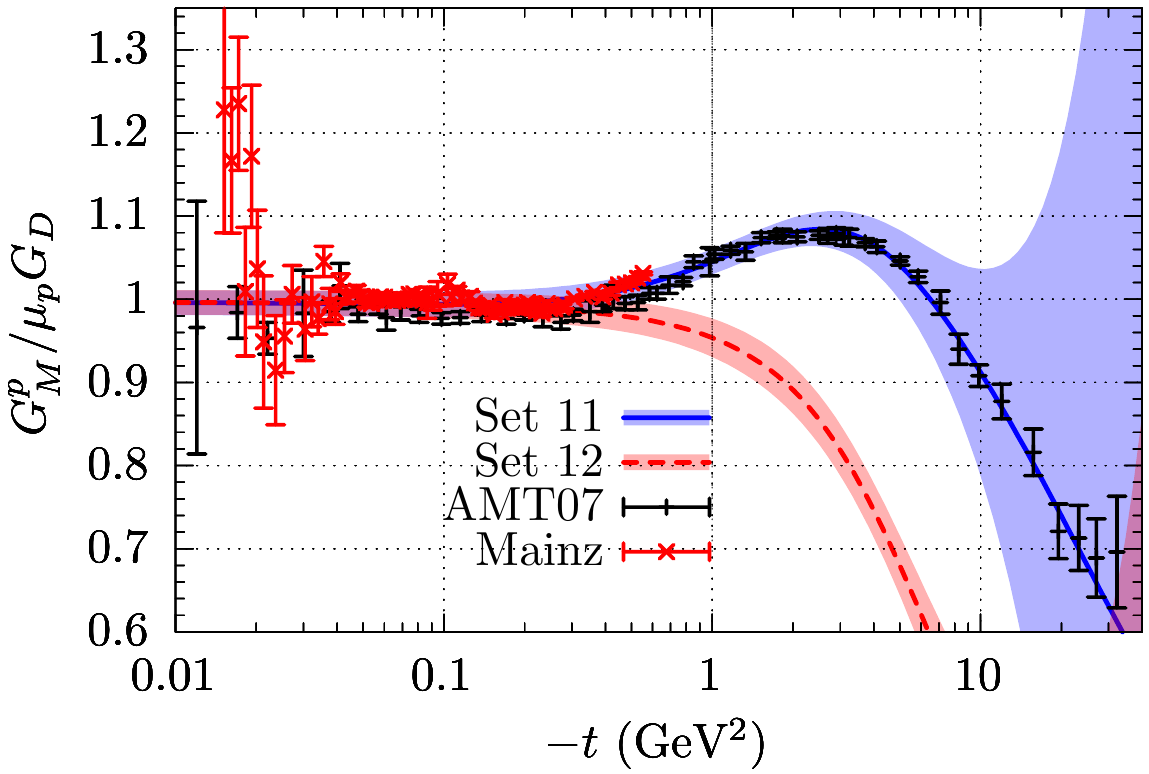}    
    \caption{A comparison between our results for $ G_M^p/\mu_p G_D $ obtained using GPDs of Sets 11 and 12 and the related experimental data of the {AMT07}\cite{Arrington:2007ux} and {Mainz}~\cite{A1:2013fsc}.}
\label{fig:GMp2}
\end{figure}

\begin{figure}[!htb]
    \centering
\includegraphics[scale=0.9]{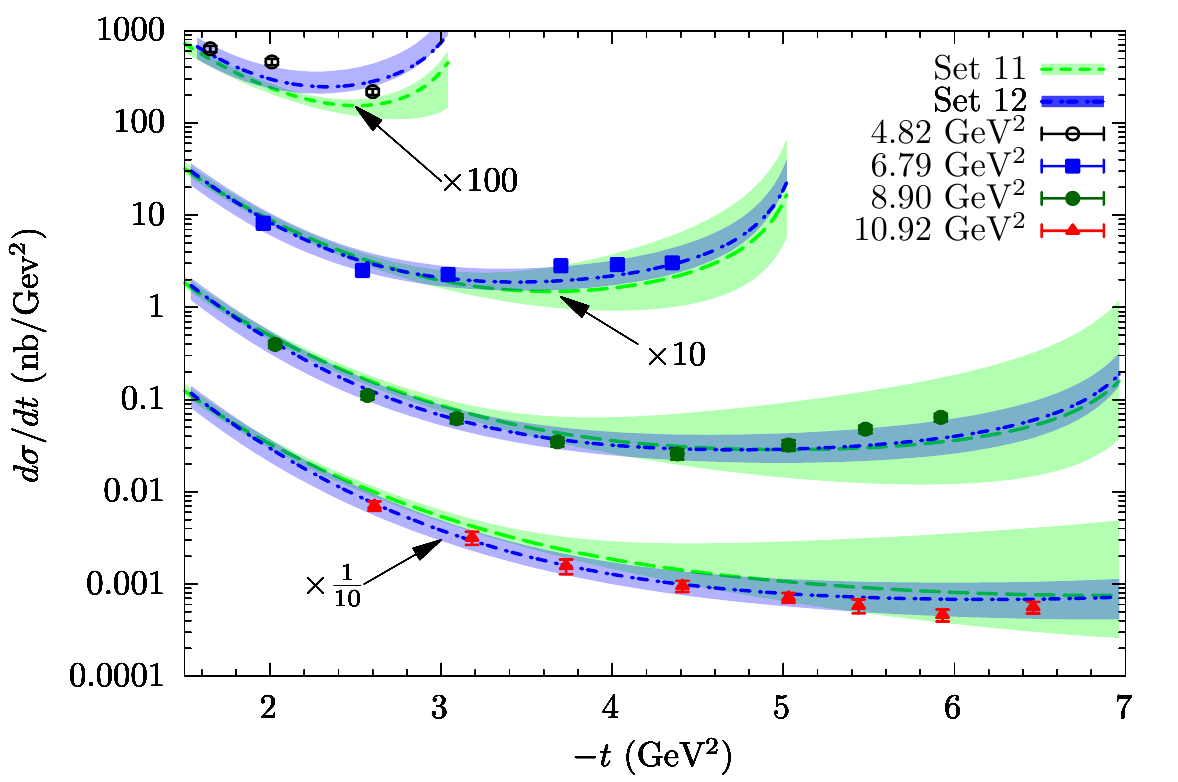}   
    \caption{Same as Fig.~\ref{fig:WACS}, but for Sets 11 and 12.}
\label{fig:WACS2}
\end{figure}

After making sure about the phenomenological framework, we considered that maybe the high $\chi^2$ values of the WACS data are due to some tensions between them and other data presented in the analysis. So, we repeated the analysis by removing different data sets one by one to find those that are not compatible with the WACS data. After a thorough investigation, we found that there is only a tension with the $ G_M^p $ data. As the first step, we included three new normalization factors for the WACS, {AMT07} and {Mainz} data (just as for the CLAS data) to resolve the observed tension if the problem came from the normalization. However, the situation did not change considerably which shows that the data are incompatible because of their different treatments at various kinematics. As the next step, by removing the {AMT07} and {Mainz} data from the analysis, we found a good description of the WACS data (we called the extracted GPDs Set 12 in Tables~\ref{tab:chi23} and~\ref{tab:par22}). In this situation, the $\chi^2$ value of the WACS data becomes 120 for 25 data points, which is in agreement with the previous study~\cite{Hashamipour:2020kip} in which only the axial FFs and WACS data have been included in the analysis (Note that the small errors of the WACS data in the denominator of $\chi^2$ make its value mathematically large, and such a large $\chi^2$ cannot be attributed to the poor description of data, because our new comparisons in Fig.~\ref{fig:WACS2} show the good quality of the fit). Our results show that the WACS data are in fair consistency with the proton and neutron electromagnetic FF data of {YAHL18} analysis, the nucleon radii data, and the axial FF data. However, there is a hard tension with the {AMT07} and {Mainz} $ G_M^p $ data. Figure~\ref{fig:GMp2} shows a comparison between our results for $ G_M^p/\mu_p G_D $ calculated using the GPDs of Sets 11 and 12 and the related experimental data of the {AMT07} and {Mainz}. Note again that for Set 12, these data have been removed from the analysis. As can be seen, although removing $ G_M^p $ data from the analysis makes the discretion of the WACS data acceptable, it spoils the $ G_M^p $ data description at $ -t>0.4 $ GeV$ ^2 $. According to the results obtained one may consider two possibilities: (i) the theoretical description of the WACS cross section should be revised, or (ii) the measurements of the WACS or $ G_M^p $, especially at larger values of $ -t $, should be revised. In any case, we present also the results of Set 12 in all plots in the following. 

\begin{figure}[!htb]
    \centering
\includegraphics[scale=1.3]{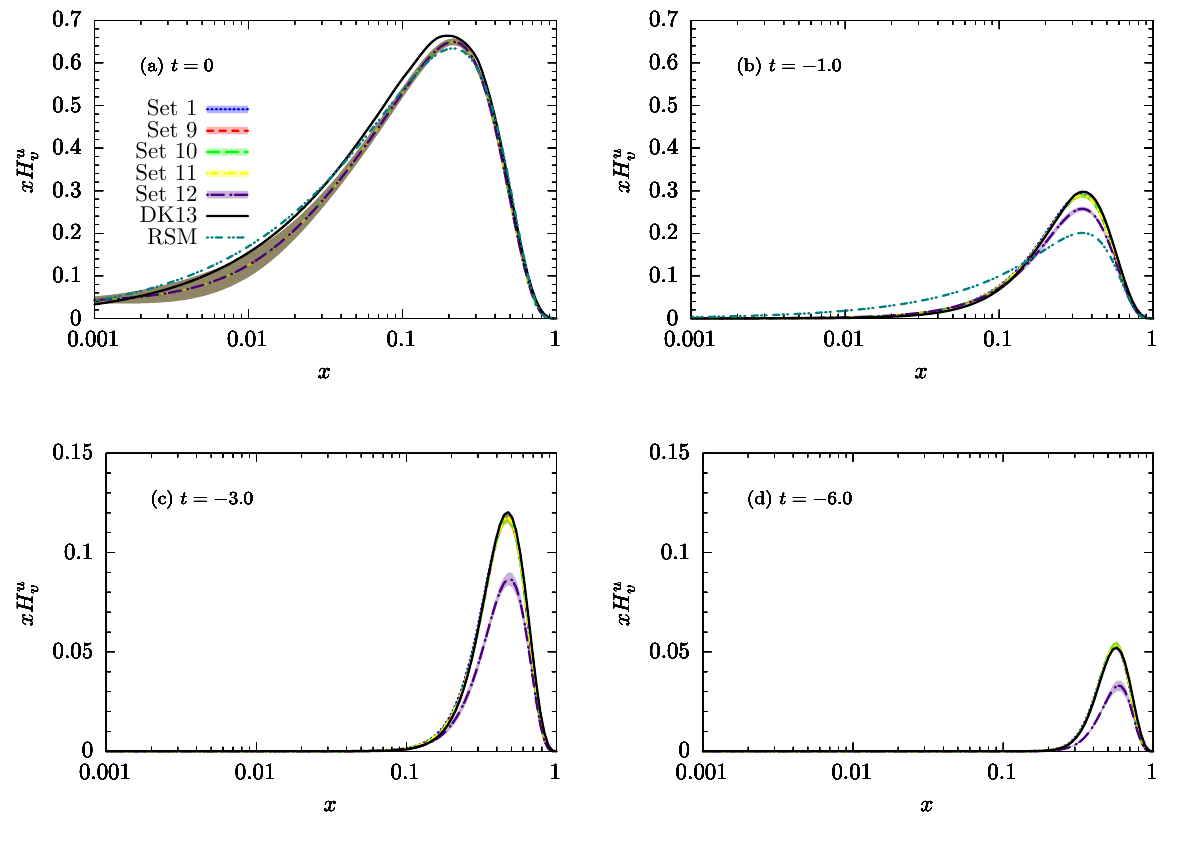}   
    \caption{A comparison between the results of Sets 9, 10, 11, and 12 for GPDs $ xH_v^u(x) $ and the corresponding results from Set 1, DK13~\cite{Diehl:2013xca}, and RSM~\cite{Kriesten:2021sqc} at four $ t $ values shown in panels (a) $ t=0$, (b) $t=-1$, (c) $t=-3$, and (d) $t=-6 $ GeV$ ^2 $.}
\label{fig:HuWACS}
\end{figure}

Now, we investigate the impact of the WACS data on the extracted GPDs in different situations described before. Figure~\ref{fig:HuWACS} shows a comparison between the results of Sets 9, 10, 11, and 12 for GPDs $ xH_v^u(x) $ and the corresponding results from Set 1, DK13~\cite{Diehl:2013xca}, and the Reggeized spectator model (RSM)~\cite{Kriesten:2021sqc} at four $ t $ values, $ t=0,-1,-3$ and $-6 $ GeV$ ^2 $. Since the RSM results are valid just for the values of $ -t $ less than unity, the corresponding ones for $ t=-3$ and $-6 $ GeV$ ^2 $  have not been plotted in this figure.
Apart from RSM and Set 12, which behave differently with $ -t $ growing, the results are very similar at all values of $ -t $ and show small uncertainties. Since $ xH_v^u(x) $ plays the most important role in the theoretical calculations of the bulk of data included in the analysis, this very good consistency confirms the universality of GPDs and thus the possibility of performing a global QCD analysis to extract them from experimental data. However, comparing Set 12 with other sets, one concludes that removing the $ G_M^p $ data from the analysis and allowing the WACS data to constrain $ H_v^u(x) $ at large $ -t $ values leads to a considerably smaller distribution.

\begin{figure}[!htb]
    \centering
\includegraphics[scale=1.3]{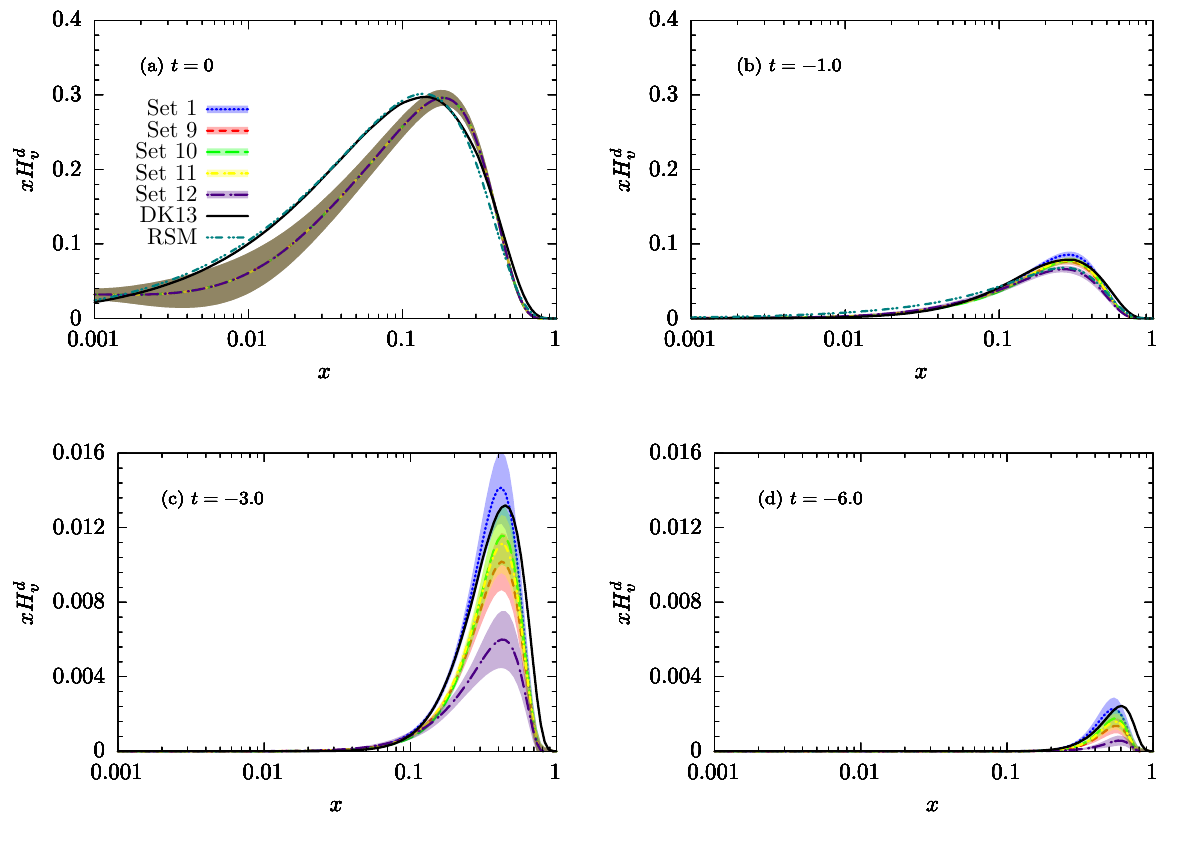}   
    \caption{Same as Fig.~\ref{fig:HuWACS}, but for GPDs $ xH_v^d(x) $.}
\label{fig:HdWACS}
\end{figure}

Figure~\ref{fig:HdWACS} shows the same results as Fig.~\ref{fig:HuWACS}, but for GPDs $ xH_v^d(x) $.
In this case, Set 9 has a smaller distribution than Sets 10 and 11, implying that the inclusion of the WACS data in the analysis leads to a significant suppression of GPDs $ xH_v^d(x) $. However, by including the CLAS data, whether considering a normalization factor (Set 11) or not (Set 10), the results become more compatible with Set 1 and DK13. Note also that there is more consistency between RSM and other sets at $t=-1 $ GeV$ ^2 $ compared with the case of GPD $ H_v^u(x) $ in Fig.~\ref{fig:HuWACS}. Just as for the up valence distribution, $ H_v^d(x) $ is suppressed significantly with $ -t $ growing by removing the $ G_M^p $ data from the analysis (Set 12). Generally, the results obtained show that the WACS data prefer smaller valence GPDs at larger values of $ -t $, as one can see the sharp falloff of the Set 12 prediction in Fig.~\ref{fig:GMp2}.

\begin{figure}[!htb]
    \centering
\includegraphics[scale=1.3]{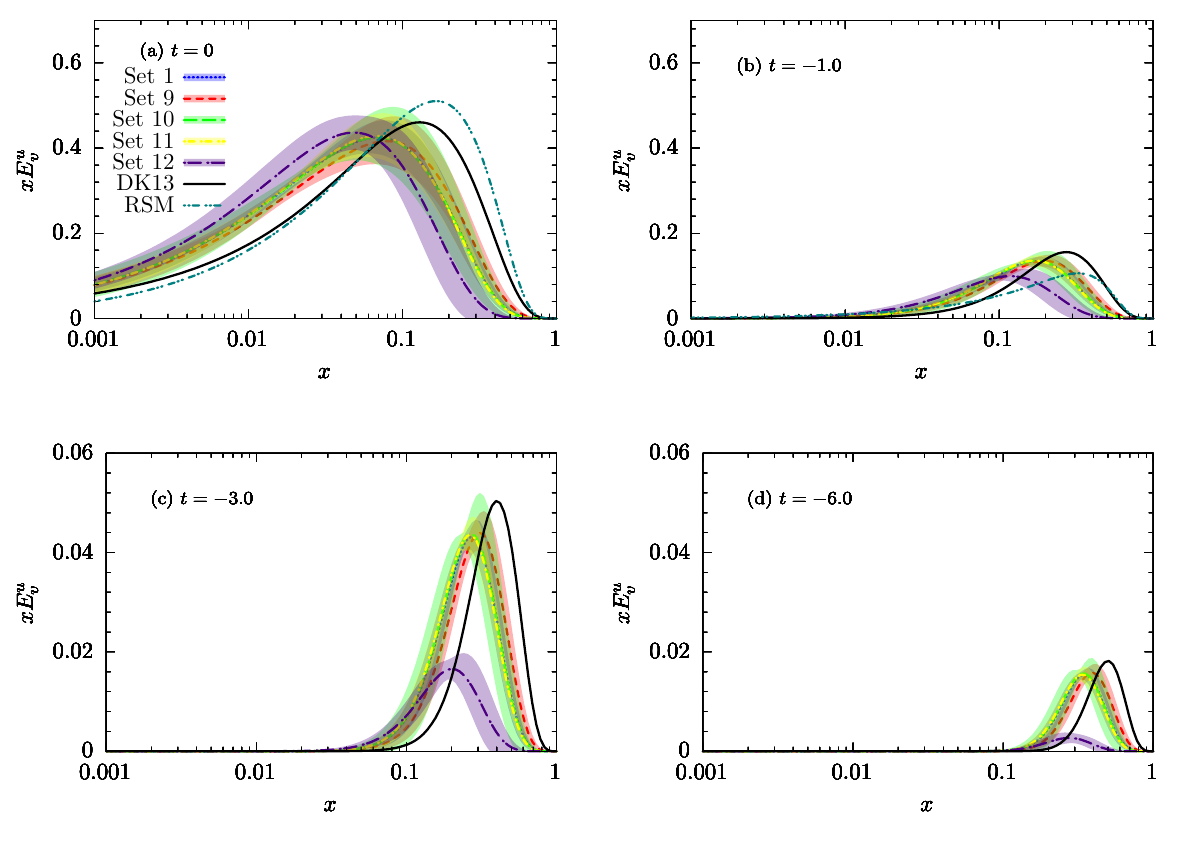}   
    \caption{Same as Fig.~\ref{fig:HuWACS}, but for GPDs $ xE_v^u(x) $.}
\label{fig:EuWACS}
\end{figure}

\begin{figure}[!htb]
    \centering
\includegraphics[scale=1.3]{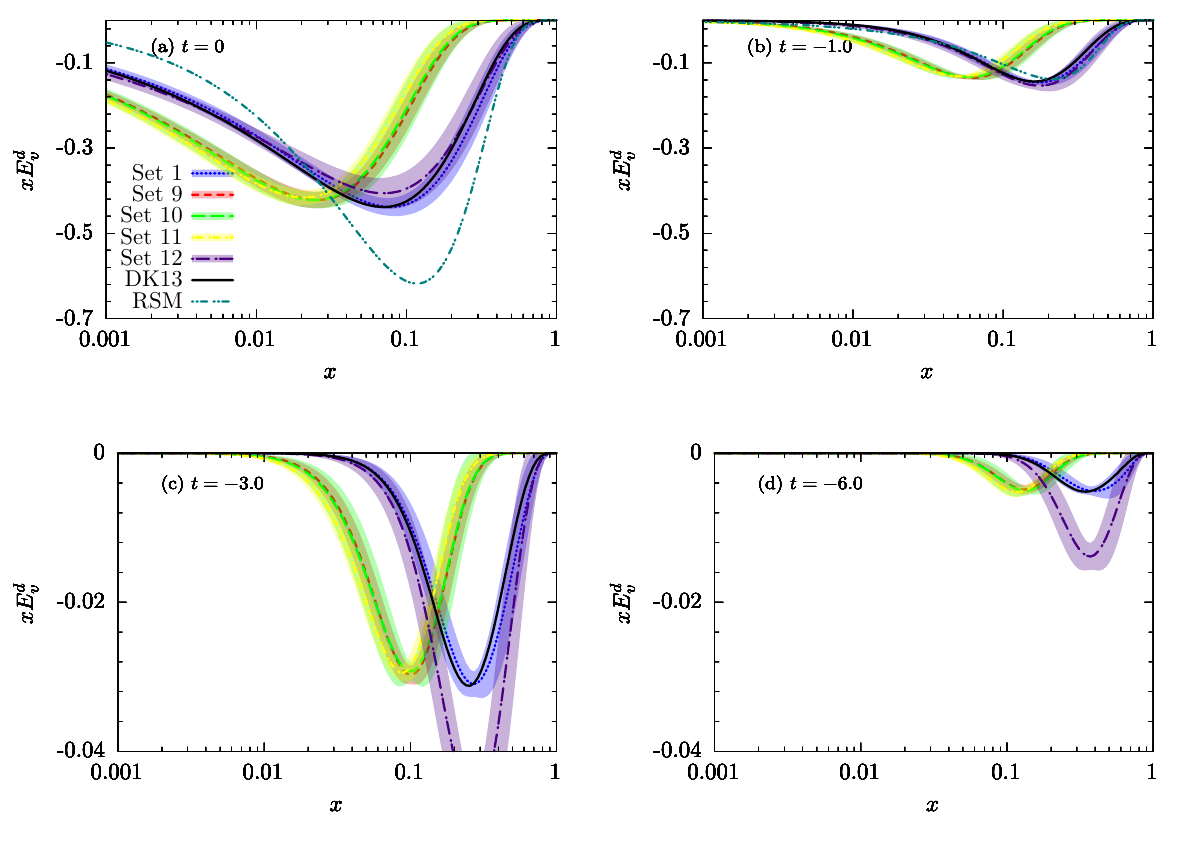}   
    \caption{Same as Fig.~\ref{fig:HuWACS}, but for GPDs $ xE_v^d(x) $.}
\label{fig:EdWACS}
\end{figure}

The corresponding results for GPDs $ xE_v^u(x) $ and $ xE_v^d(x) $ are shown in Figs.~\ref{fig:EuWACS} and~\ref{fig:EdWACS}, respectively. In the case of $ xE_v^u(x) $, our results (except for Set 12) are in good consistency with each other at all values of $ -t $, although they are different from DK13 and RSM, and located at smaller values of $ x $. Note that Sets 1, 9, and 10 contain different kinds of experimental data (Set 1 does not contain WACS or CLAS data, Set 9 contains WACS data, and Set 10 contains both WACS and CLAS data). Therefore, one can conclude that the good consistency of our results with each other and their differences from DK13 and RSM are a direct impact of the proton radii data, especially those of $\left<r_{pM}^2\right>$. However, when $ G_M^p $ data are excluded from the analysis (Set 12), the GPD $ E_v^u(x) $ is more inclined to small $ x $ at zero $ t $ and suppressed with $ -t $ growing even more than valence GPDs $ H_v^q(x) $.

\begin{figure}[!htb]
    \centering
\includegraphics[scale=1.3]{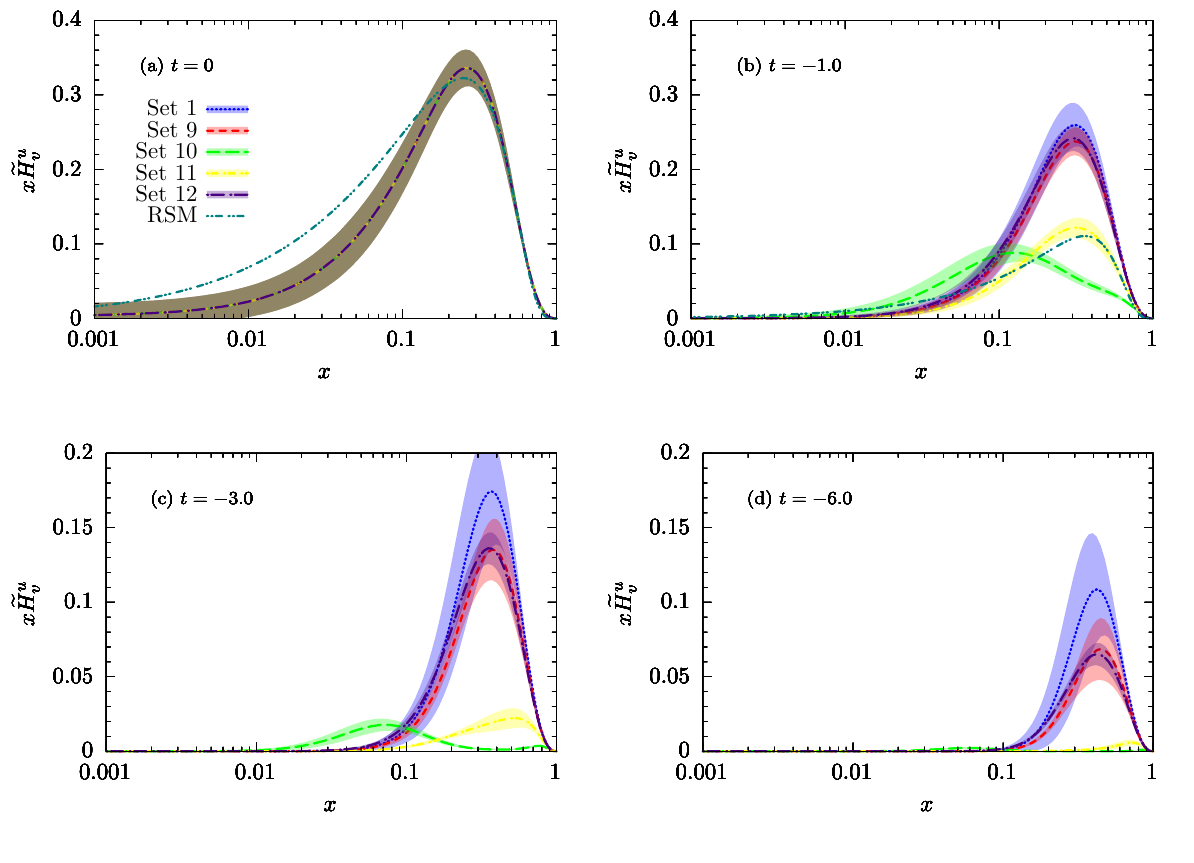}   
    \caption{A comparison between the results of Sets 9, 10, 11, and 12 for polarized GPDs $ x\widetilde{H}_v^u(x) $ and the corresponding results from Set 1 and RSM~\cite{Kriesten:2021sqc} at four $ t $ values shown in panels (a) $ t=0$, (b) $t=-1$, (c) $t=-3$, and (d) $t=-6 $ GeV$ ^2 $.}
\label{fig:HTuWACS}
\end{figure}

\begin{figure}[!htb]
    \centering
\includegraphics[scale=1.3]{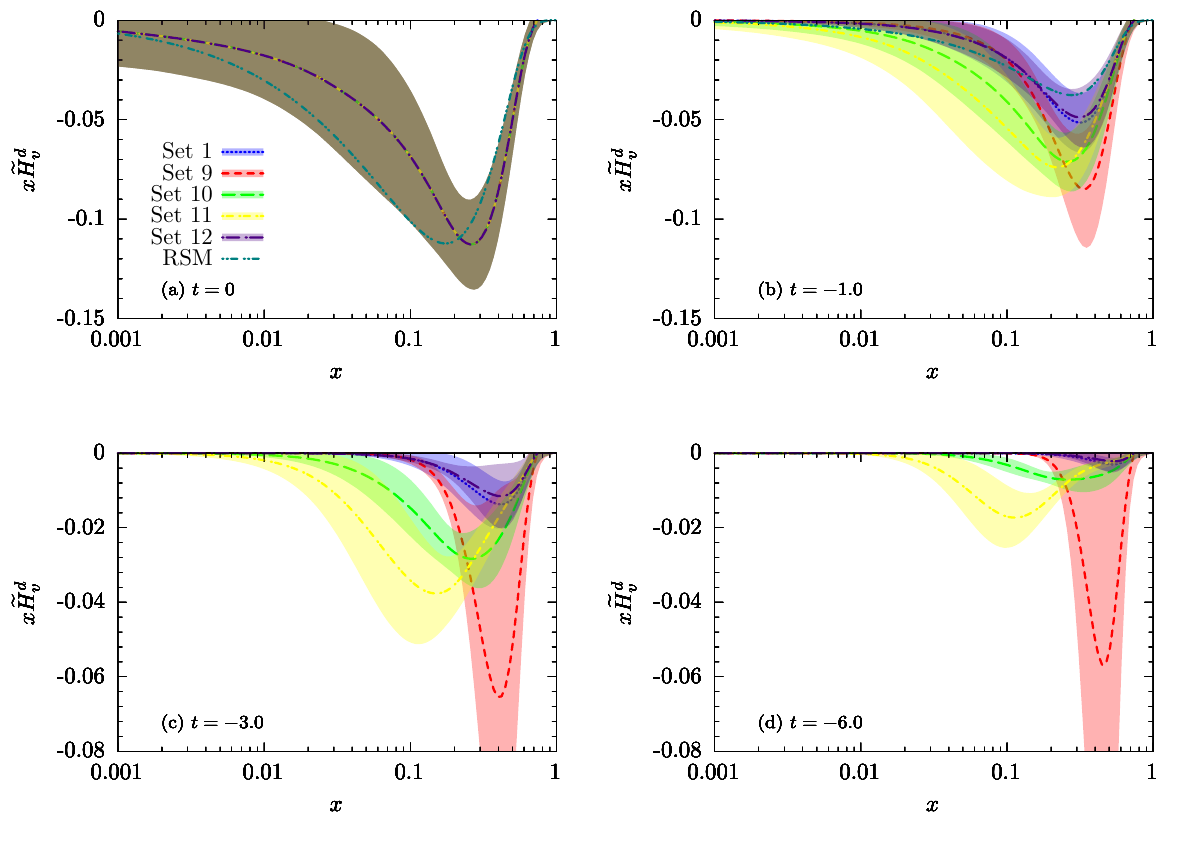}   
    \caption{Same as Fig.~\ref{fig:HTuWACS}, but for polarized GPDs  $ x\widetilde{H}_v^d(x) $.}
\label{fig:HTdWACS}
\end{figure}

The situation is somewhat different in the case of $ xE_v^d(x) $. As can be seen from Fig.~\ref{fig:EdWACS}, Set 1 and DK13 are absolutely compatible with each other, while they are considerably different from other results at all values of $ -t $, except for Set 12 at smaller values of $ -t $ and RSM at $t=-1 $ GeV$ ^2 $. Although RSM is significantly different from other results, it becomes consistent with Set 1 and DK13 with $ -t $ growing. As mentioned in Sec.~\ref{sec:four-two}, the CLAS data do not significantly affect $ xE_v^d(x) $, such that Sets 1, 4, and 5 have the same results. Considering this fact, it can be inferred that the significant difference observed between Set 1 and other sets in Fig.~\ref{fig:EdWACS} is a direct impact of the WACS data. Now, the question is why
$ xE_v^d(x) $ is impacted significantly by the WACS data but $ xE_v^u(x) $ is not changed considerably, whereas both of them contribute to the WACS calculation, especially at large $ -t $. 
As can be seen from Fig.~\ref{fig:WACS}, Set 1 cannot describe the WACS data well.
So, GPDs should be modified to fit these data as well. On the other hand, both $ H_v^q $ and $ E_v^q $ are tightly constrained by the FFs and radii data, especially for flavor $ u $. Therefore, the fit program automatically tries to change those GPDs that have more freedom in order to achieve a better fit quality for the WACS data. This is the reason that the WACS data have had a greater impact on
$ xE_v^d(x) $ (and also $ x\widetilde{H}_v^d $; see Figs.~\ref{fig:HTuWACS} and~\ref{fig:HTdWACS}, and compare Sets 1 and 9).
Another point that should be mentioned is that Sets 9, 10, and 11 lead to the same results for $ xE_v^d(x) $, which clearly implies the lack of influence of CLAS data on $ xE_v^d(x) $.  Note that excluding the $ G_M^p $ data from the analysis makes $ E_v^d(x) $ more compatible with Set 1, which does not contain the WACS or CLAS data, although it increases at larger values of $ -t $. The results obtained clearly explain the tension observed between the WACS and $ G_M^p $ data. The former prefers smaller valence GPDs $ H_v^u $, $ H_v^d $, and $ E_v^u $ as $ -t $ increases, while it enhances the contribution of $ E_v^d $.

Now we are in a position to investigate the impact of WACS data on polarized GPDs. Note that there are no corresponding results from DK13 analysis in this case. We have compared the results obtained for the polarized GPD $ x\widetilde{H}_v^u(x) $ in Fig.~\ref{fig:HTuWACS}. By comparing Sets 1 and 9, one can study the pure impact of the WACS data on the extracted distributions. As can be seen, the WACS data lead to a moderate suppression of $ x\widetilde{H}_v^u(x) $ with $ -t $ growing. This was expected from Fig.~\ref{fig:WACS}, where Set 1 overestimates most of the data points, especially at $ s=10.92 $ GeV$ ^2 $. Including the original CLAS data in the analysis suppresses $ x\widetilde{H}_v^u(x) $ even more, just as with  Fig.~\ref{fig:HTu}. However, considering a normalization factor for the CLAS data does not adjust the results as before (compare Sets 1 and 11 of Fig.~\ref{fig:HTuWACS} with Sets 1 and 5 of Fig.~\ref{fig:HTu}). It is also of interest that RSM is in good agreement with Set 11 at $t=-1 $ GeV$ ^2 $, while it is different from our results at $ t=0 $. Another interesting point is that Set 12 is in good consistency with Set 1 especially at smaller values of $ -t $, while it contains the WACS and CLAS data,  and Set 1 does not. This, and also the excellent consistency between Sets 9 and 12 indicates three facts: (i) there is not any impact from $ G_M^p $ data on polarized GPD $ x\widetilde{H}_v^u(x) $ as expected; (ii) the significant suppression of Sets 10 and 11 in Fig.~\ref{fig:HTuWACS} comes mainly from CLAS data, not WACS; and (iii) since the WACS cross section contains three kinds of GPDs simultaneously, it provides a suitable framework to change and determine their contributions depending on the other datasets included in the analysis. For example, the remarkable suppression of $ H_v^u $, $ H_v^d $, and $ E_v^u $ at larger values of $ -t $ for Set 12 (e.g., comparing with Set 11) is compensated with the significant enhancement of $ E_v^d $ and $ \widetilde{H}_v^u $.

Figure~\ref{fig:HTdWACS} shows the same results as Fig.~\ref{fig:HTuWACS}, but for polarized GPDs  $ x\widetilde{H}_v^d(x) $. This figure clearly indicates that the WACS data significantly affects the down-quark distribution. According to Fig.~\ref{fig:WACS}, in order to have a good description of the WACS data,
it is necessary to suppress the theoretical predictions at most values of $ -t $ and $ s $. This leads to significant growth of $ x\widetilde{H}_v^d(x) $ in the negative considering the fact that the polarized GPDs  $ x\widetilde{H}_v^u(x) $ are well constrained from $ G_A $ data. Comparing Figs.~\ref{fig:HTdWACS} and~\ref{fig:HTd}, it can be concluded that the inclusion of the WACS and CLAS data simultaneously has a different impact on $ x\widetilde{H}_v^d(x) $  than the case in which only the CLAS data are considered. Actually, Fig.~\ref{fig:HTdWACS} shows that if one considers the original CLAS data besides the WACS data (Set 10), the results will be more compatible with Set 1, while considering the normalized CLAS data leads to results that differ more (Set 11). However, according to Table~\ref{tab:chi23}, considering a normalization factor for the CLAS data makes the description of both WACS and CLAS data better. Note also that in this case, RSM shows different results at both $ t=0 $ and $t=-1 $ GeV$ ^2 $. Moreover, excluding the $ G_M^p $ data from the analysis, as explained above, provides a flexibility to change the distribution as the WACS data like. This leads to a significant suppression of $ \widetilde{H}_v^d(x) $ for Set 12 with $ -t $ growing.

\begin{figure}[!htb]
    \centering
\includegraphics[scale=1.3]{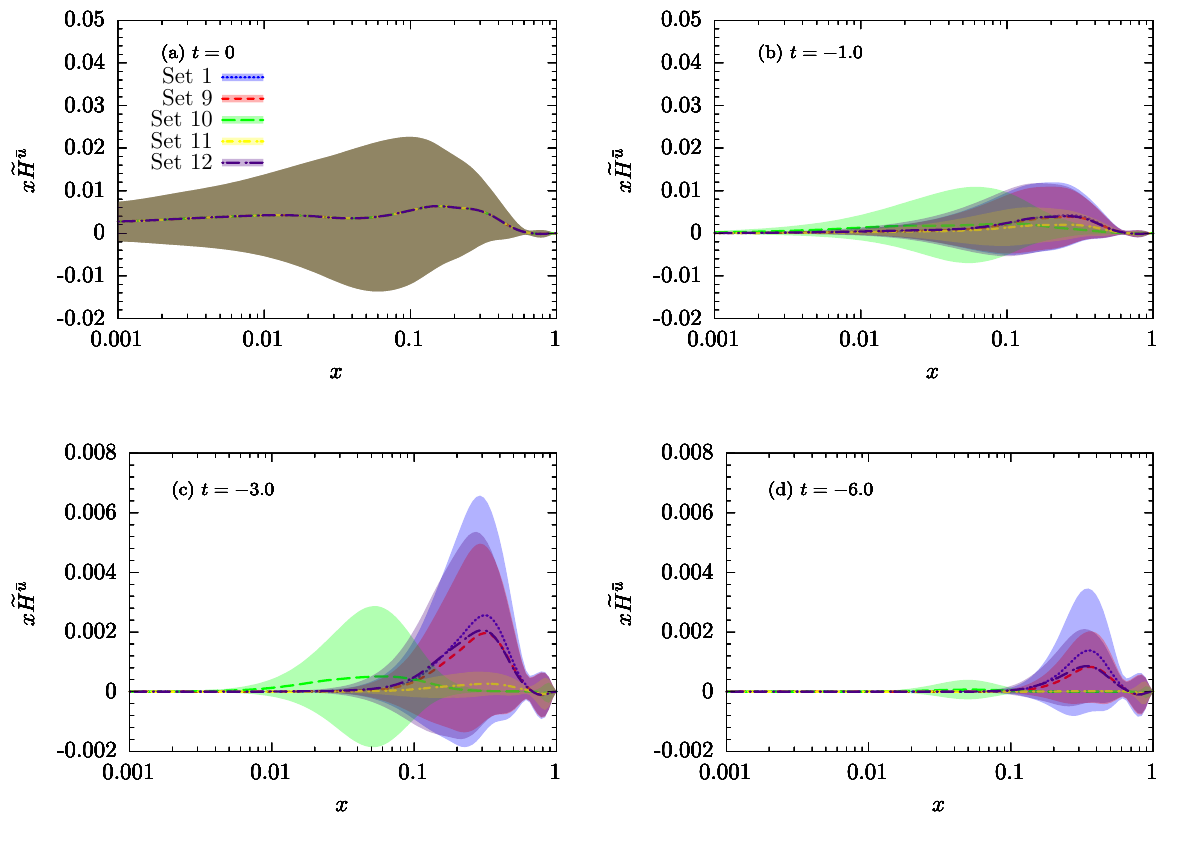}   
    \caption{Same as Fig.~\ref{fig:HTuWACS}, but for polarized GPDs  $ x\widetilde{H}^{\bar u}(x) $.}
\label{fig:HTubarWACS}
\end{figure}

\begin{figure}[!htb]
    \centering
\includegraphics[scale=1.3]{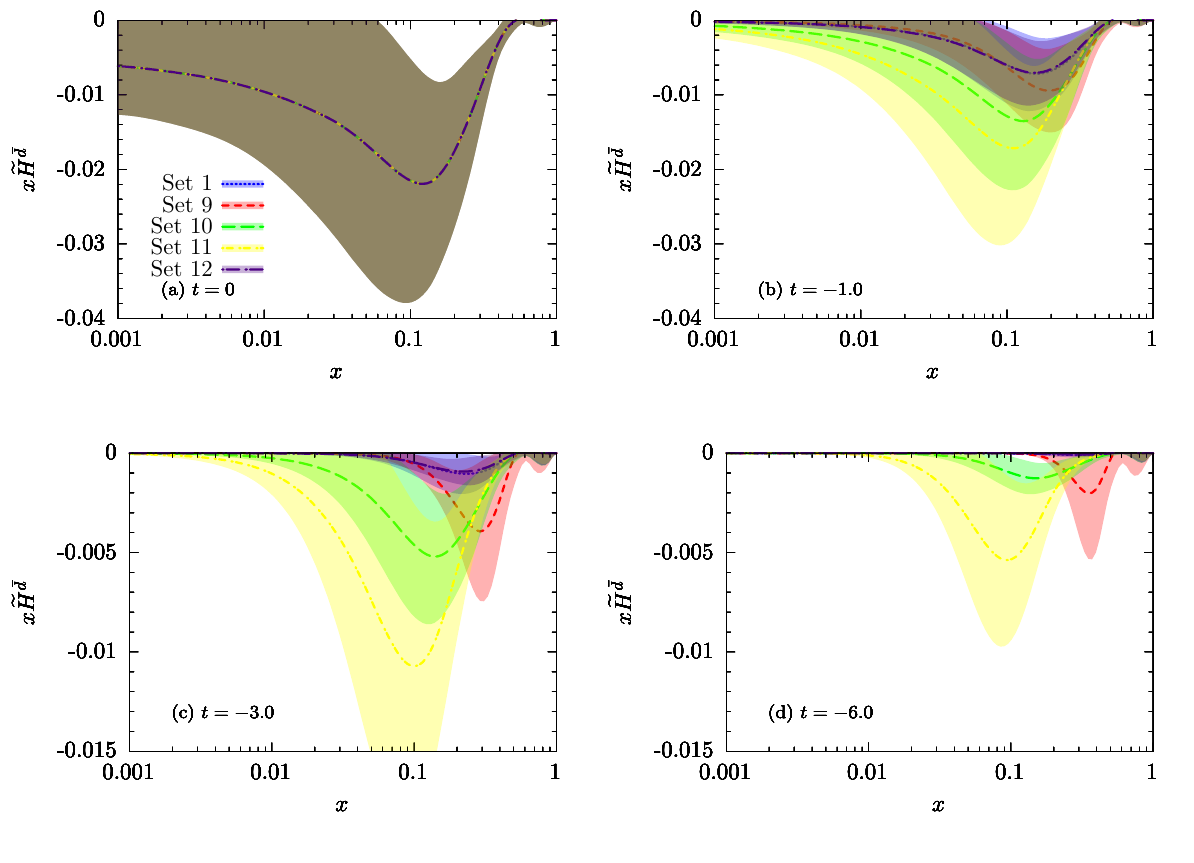}   
    \caption{Same as Fig.~\ref{fig:HTuWACS}, but for polarized GPDs  $ x\widetilde{H}^{\bar d}(x) $.}
\label{fig:HTdbarWACS}
\end{figure}

Figures~\ref{fig:HTubarWACS} and~\ref{fig:HTdbarWACS} show the same comparisons as Figs.~\ref{fig:HTuWACS} and~\ref{fig:HTdWACS}, respectively, but for the sea-quark contributions $ x\widetilde{H}^{\bar u}(x) $ and $ x\widetilde{H}^{\bar d}(x) $. Note that there are not any results from RSM~\cite{Kriesten:2021sqc} in this case. As one can see, the sea-quark polarized GPDs behave almost the same as the valence polarized GPDs by including the WACS and CLAS data in the analysis. The only considerable difference is that the down sea-quark distribution has less been affected than the corresponding one for the valence quark after the inclusion of just the WACS data (Set 9).

In order to investigate further the tension observed between the WACS and $ G_M^p $ data, it will be informative to compare the predictions of Sets 11 and 12 for the WACS cross section besides the experimental data. Figure~\ref{fig:WACS2} shows such a comparison just as for Fig.~\ref{fig:WACS}, but here we have added the results of Set 12 and only kept Set 11. As can be seen, excluding the $ G_M^p $ data from the analysis improves the description of the WACS significantly. As pointed out before, this indicates that either the theoretical calculations  or the experimental measurements of WACS and $ G_M^p $ at larger values of $ -t $ should be revised. Another point that can be inferred from Fig.~\ref{fig:WACS2} is that the exclusion of $ G_M^p $ data leads to increases and decreases in the uncertainties of the WACS cross section at smaller and larger values of $ -t $, respectively. This was expected, since the $ G_M^p $ data at small $ -t $ have considerable impact on unpolarized GPDs. On the other hand, in the absence of $ G_M^p $ at large $ -t $ (i.e., with the release of tension), the low uncertainties of the WACS data cause GPDs to be more constrained this region.

\subsection{Comparison with other quantities}\label{sec:four-four}

In this subsection, just as in our previous study~\cite{Hashamipour:2021kes}, we are going to calculate the gravitational FF $ M_2 $ as well as the angular momentum $ J^q $  carried by quarks inside the nucleon using GPDs of Sets 1, 9, 11, and 12 obtained in previous subsections and compare their results with the corresponding ones from the light-cone QCD sum rules (LCSRs)~\cite{Azizi:2019ytx} and lattice calculations~\cite{LHPC:2007blg}. 

At zero skewness, the gravitational FF $ M_2 $ is related to the GPDs $ H^q(x,t) $ through the following formula~\cite{Polyakov:2002yz}:
\begin{equation}
 M_2(t)=\int_{-1}^1 dx x \sum_q H^q(x,\xi=0,t).
\label{Eq7}
\end{equation}
In contrast, the angular momentum $ J^q $ carried by a specific quark flavor $ q $ inside the nucleon contains both GPDs $ H^q(x,t) $ and $ E^q(x,t) $ as follows:
\begin{equation}
J^q(t)=\frac{1}{2} \int_{-1}^1 dx x [ H^q(x,t) +  E^q(x,t)],
\label{Eq8}
\end{equation}
that turns into the famous Ji sum rule~\cite{Ji:1996ek,Goloskokov:2008ib} at $ t=0 $. One can obtain the total angular momentum of the proton $ J^p $ carried by quarks by making it sum over all flavors. 

In Eq.~(\ref{Eq8}), both the valence and sea-quark sectors are contributed. In our previous study~\cite{Hashamipour:2021kes}, the sea-quark contributions of GPDs had not been determined, since the analysis
just contained the FFs and radii data. Thus, in order to calculate $ M_2 $ and $ J^q $, we had used the profile functions of the valence quarks for the sea-quarks too as a reasonable assumption. However, in the present study, we have also determined the sea quark contributions of GPDs from data (see Sec.~\ref{sec:four-three}). Therefore, the results which are presented here will be more precise compared with our previous study. However, we present the results of Set 1 (for both $ M_2 $ and $ J^{q(p)} $) without considering the sea-quark contributions to show their impacts on the results as well.

\begin{figure}[!htb]
    \centering
\includegraphics[scale=0.9]{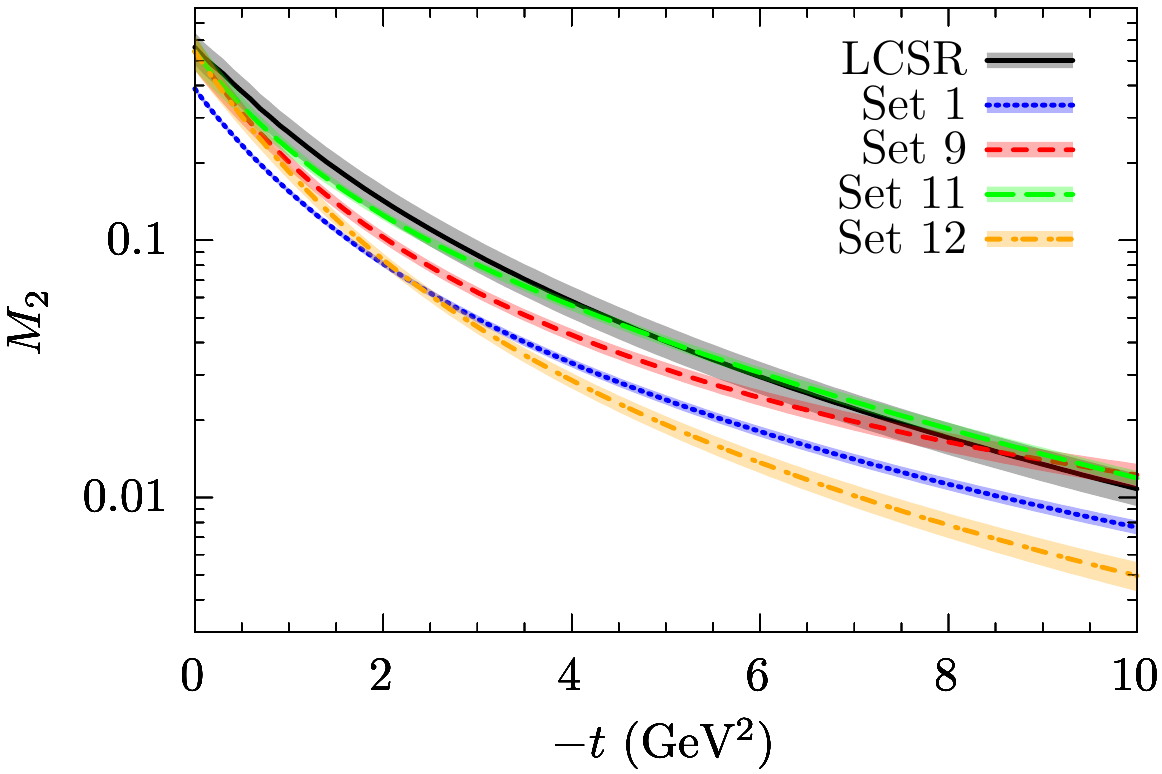}    
    \caption{A comparison between our results obtained for $ M_2(t) $, using Sets 1, 9, 11, and 12 and the corresponding results from LCSR~\cite{Azizi:2019ytx}.}
\label{fig:M2t10}
\end{figure}
In Fig.~\ref{fig:M2t10}, we have compared our results obtained for $ M_2 $ using the GPDs of Sets 1, 9, 11, and 12 and the corresponding results calculated using LCSR~\cite{Azizi:2019ytx} as a function of $ -t $. Note that here (and when we present the results of $ J^p $) we have evolved the LCSR result
to $ \mu=2 $ GeV using the renormalization group equations, including the mass renormalization in order to make it comparable with our results. As can be seen, Set 1, which does not include the contribution of the sea quarks, has a considerable difference from Sets 9 and 11. This clearly indicates the importance of the sea-quark contributions which are accessible through the WACS and AFF data. One can also conclude that the precise determination of GPDs at zero skewness is not possible except by performing a simultaneous analysis of all related experimental data. {Note that although there are some tensions between them, the results of the analysis including all data are in relatively good agreement with the pure theoretical calculations.}. This can be further confirmed when the results of Sets 9, 11, and 12 are compared with the results of LCSR.  
One can see that Set 11, which has been obtained by the inclusion of all data, has a good consistency with LCSR at all values of $ -t $, especially considering the uncertainties {(in contrast to Sets 9 and 12 that do not contain the CLAS $ G_A $ and the $ G_M^p $ data, respectively)}. It is also amazing that the result extracted from the experimental data is in such significant agreement with the pure theoretical calculations. Another interesting point is that Set 12 (which has been obtained by excluding the $ G_M^p $ data from the analysis) diverges significantly from other sets and also LCSR as $ -t $ increases. This indicates the importance of the $ G_M^p $ data at larger values of $ -t $. Actually, if one relies only on the WACS data, an egregious difference in the theoretical predictions of gravitational FF $ M_2 $ results. This convinces one to revise the theoretical description of the WACS process (or, more unlikely, the WACS and $ G_M^p $ measurements at larger $ -t $).

\begin{figure}[!htb]
    \centering
\includegraphics[scale=0.9]{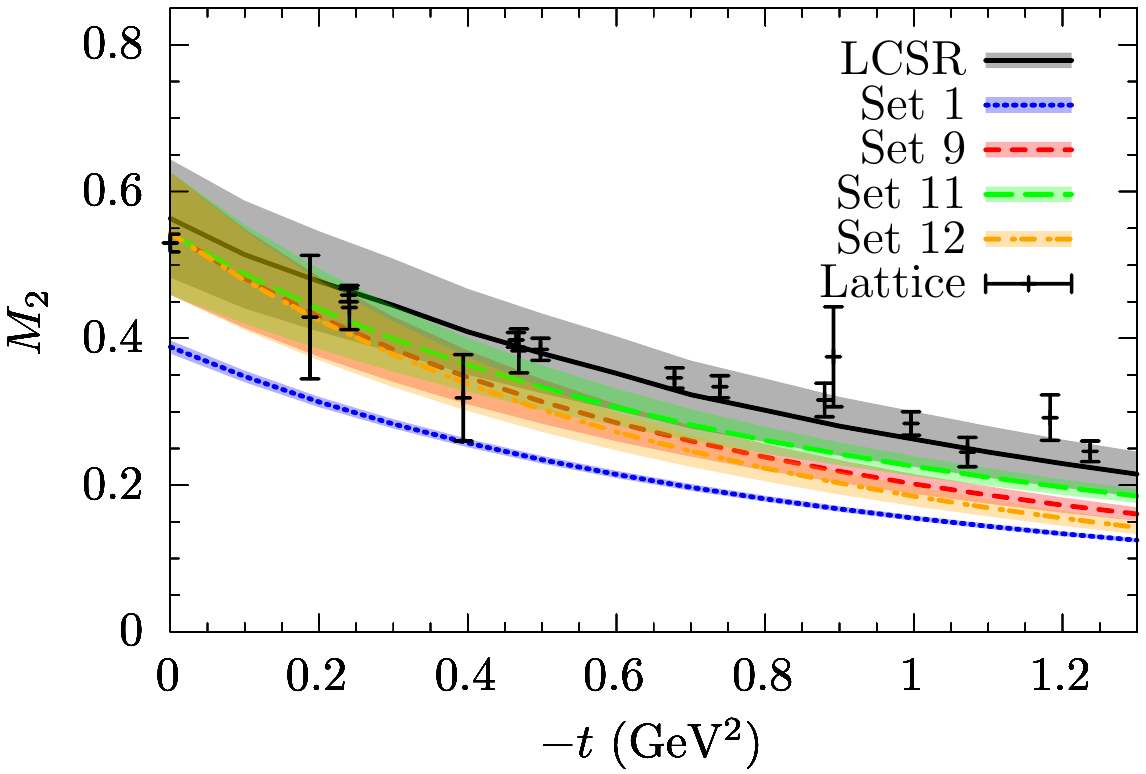}    
    \caption{Same as Fig.~\ref{fig:M2t10}, but in the interval $ 0 < -t < 1.3 $ GeV$ ^2 $. The lattice results~\cite{LHPC:2007blg} have also been presented.}
\label{fig:M2t1}
\end{figure}
In order to make the comparison between different results more clear at small $ -t $, we have replotted the results of Fig.~\ref{fig:M2t10} in Fig.~\ref{fig:M2t1} but in the interval $ 0 < -t < 1.3 $ GeV$ ^2 $, and have included also the lattice results taken from Table 15 of Ref.~\cite{LHPC:2007blg}. This figure shows the good consistency between our results (especially those obtained from the analyses containing the WACS and AFF data) and the corresponding ones from LCSR and lattice at small values of $ -t $. 

\begin{figure}[!htb]
    \centering
\includegraphics[scale=0.9]{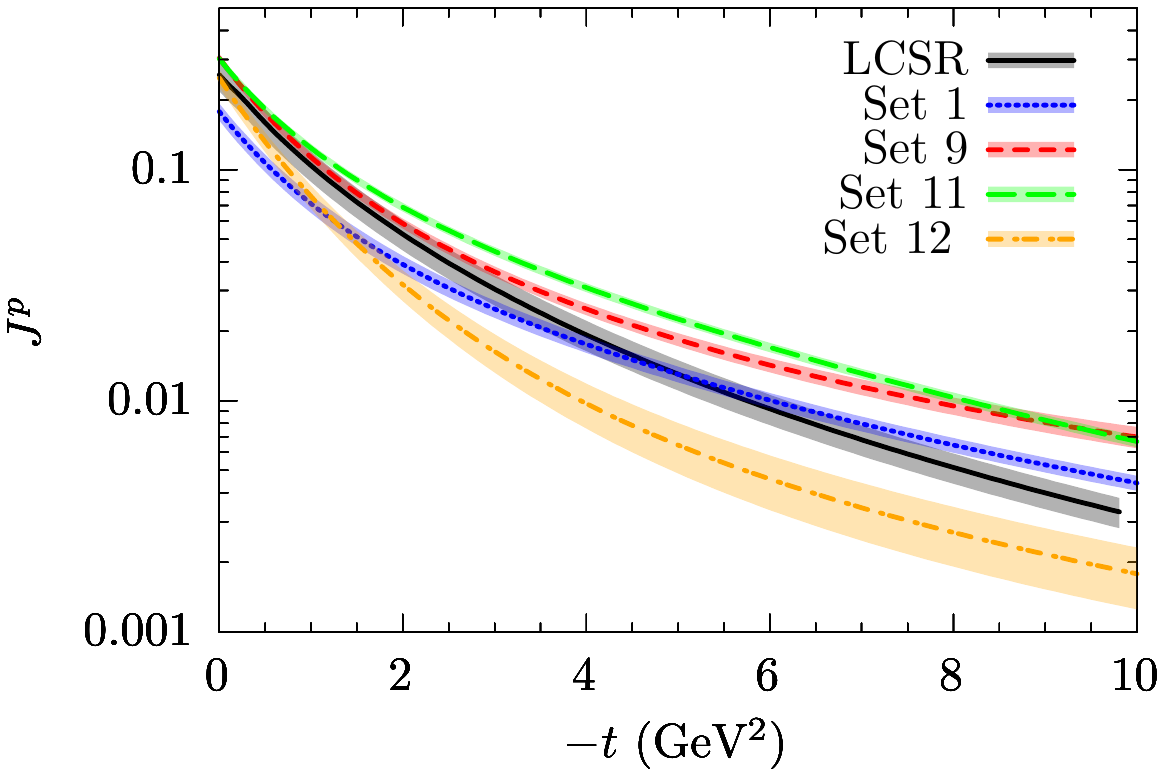}
    \caption{Same as Fig.~\ref{fig:M2t10}, but for the proton total angular momentum $ J^p $.}
\label{fig:JFF}
\end{figure}
\begin{figure}[!htb]
    \centering
\includegraphics[scale=0.9]{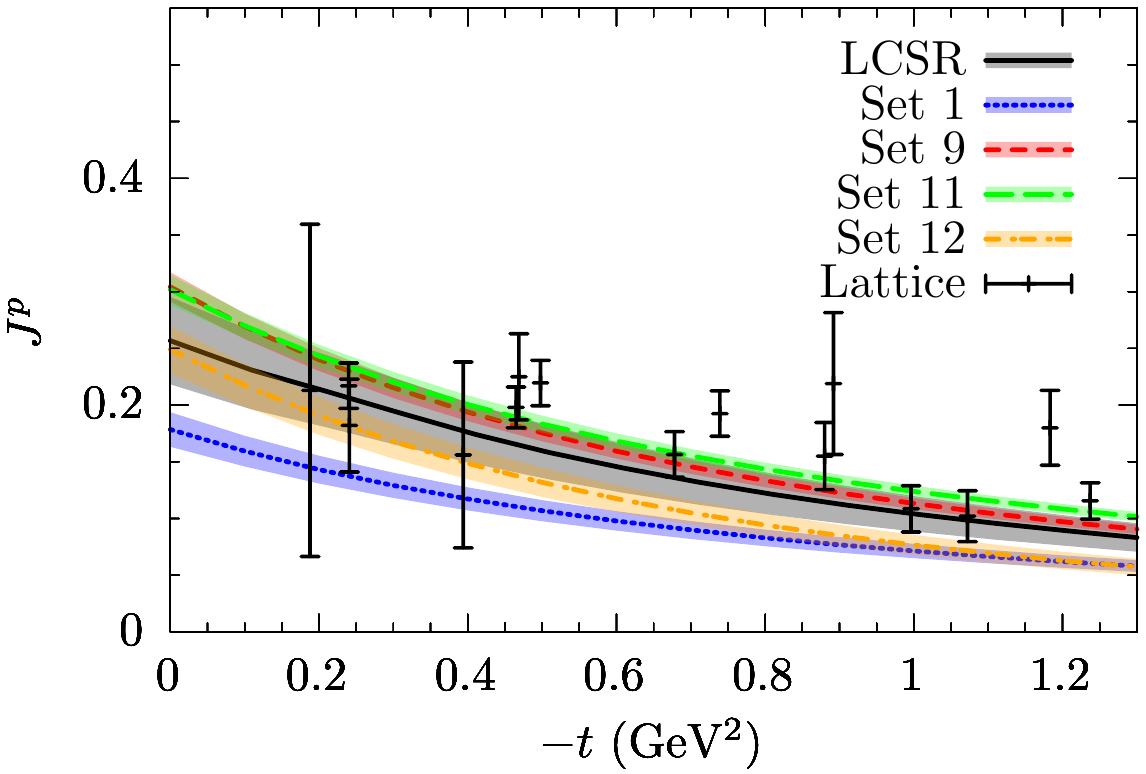}    
    \caption{Same as Fig.~\ref{fig:JFF}, but in the interval $ 0 < -t < 1.3 $ GeV$ ^2 $. The lattice results~\cite{LHPC:2007blg} have also been presented.}
\label{fig:JqFF}
\end{figure}
Now, we are in a position to calculate the proton total angular momentum $ J^p $ as a function of $ -t $, which is obtained by making a sum over all flavor contributions calculated using Eq.~(\ref{Eq8}). As mentioned before, we consider both the valence and sea-quark contributions. Figure~\ref{fig:JFF} shows the results obtained using Sets 1, 9, 11, and 12 and compares them with the corresponding results obtained from LCSR. In this case, the differences between different results are more significant compared with the case of $ M_2 $. This can be attributed to the presence of GPDs $ E^q $ in the calculation of $ J^q $, while $ M_2 $ is calculated just using GPDs $ H^q $. Note also that GPDs $ H^q $ have been better constrained than GPDs $ E^q $. The large difference of Set 1 from Sets 9 and 11 indicates again the important role of the sea-quark contributions. As can be seen, the consistency of Sets 9 and 11 with LCSR is not as good as in the case of $ M_2 $, especially at medium and large values of $ -t $. However, the consistency is very good at small values of $ -t $. This can also be seen from Fig.~\ref{fig:JqFF}, where we have again compared our results with the corresponding ones of LCSR and lattice in the interval $ 0 < -t < 1.3 $ GeV$ ^2 $. Note also that Set 12 has a considerably suppressed prediction, just as for $ M_2 $, that indicates again the importance of the $ G_M^p $ data in constraining the unpolarized valence GPDs.

\begin{figure}[!htb]
    \centering
\includegraphics[scale=0.9]{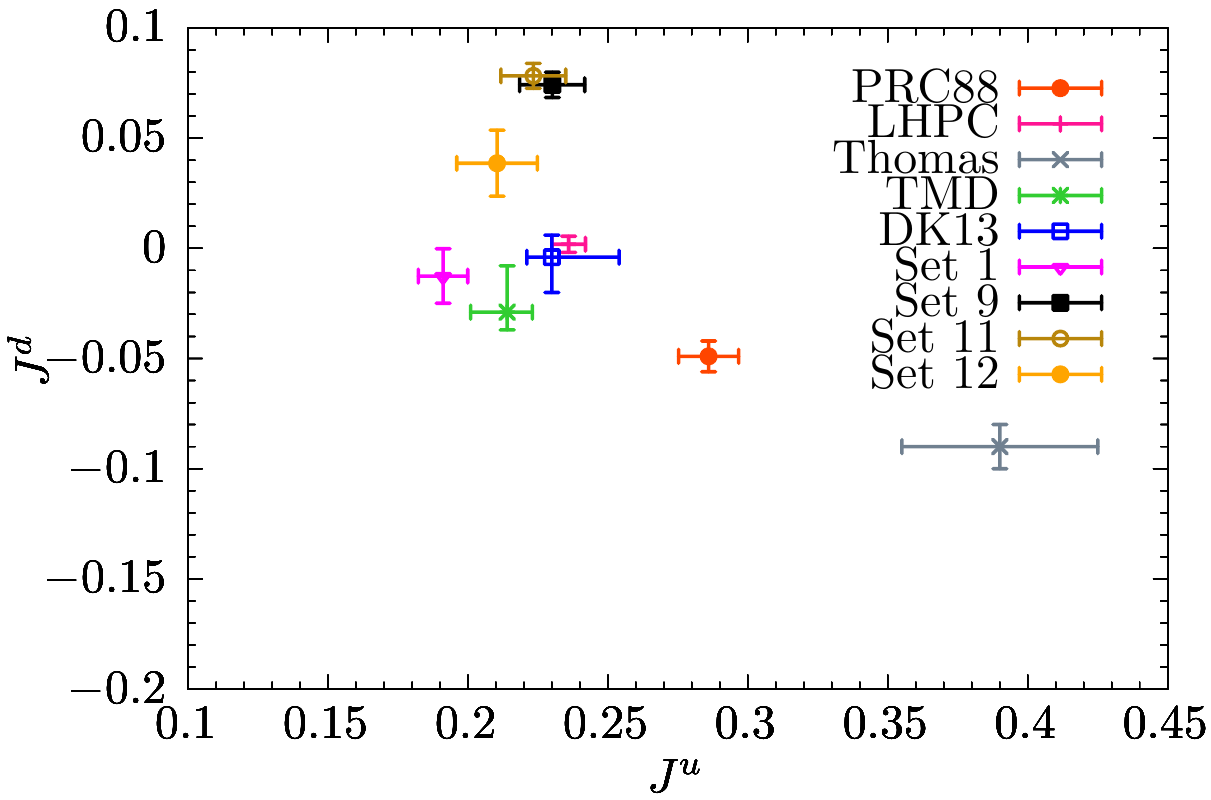}    
    \caption{A comparison between our results for $ J_v^u $ and $ J_v^d $ at the limit $ t=0 $ and $ \mu=2 $ GeV calculated using Sets 1, 9, 11, and 12 and the corresponding results from {PRC88}~\cite{Gonzalez-Hernandez:2012xap}, {LHPC}~\cite{LHPC:2010jcs}, {Thomas}~\cite{Thomas:2008ga},  {TMD}~\cite{Bacchetta:2011gx}, and DK13~\cite{Diehl:2013xca}. }
\label{fig:JuJd}
\end{figure}
It is also of interest to compare our results for the Ji sum rule with the corresponding ones obtained from other studies. In Fig.~\ref{fig:JuJd}, we have compared our predictions for $ J_v^u $ and $ J_v^d $ at limit $ t=0 $ and $ \mu=2 $ GeV calculated using Sets 1, 9, 11, and 12 with results taken from PRC88~\cite{Gonzalez-Hernandez:2012xap}, LHPC~\cite{LHPC:2010jcs}, Thomas~\cite{Thomas:2008ga},  TMD~\cite{Bacchetta:2011gx}, and DK13~\cite{Diehl:2013xca}.
Unlike our previous study~\cite{Hashamipour:2021kes}, the results obtained here (except for Set 1) are not in good agreement with other groups even with DK13, which has the most similar theoretical and phenomenological framework to our work. Let us explore this issue further. In the case of Set 1, our result for $ J^d $ is the same as DK13 to a large extent, but there is a little difference for $ J^u $. 
This can be explained by considering panels (a) of Figs.~\ref{fig:Hu}-\ref{fig:Ed}, where the GPDs $ H^q $ and $ E^q $ have been plotted at $ t=0 $ (note again that both $ H $ and $ E $ are contributed to $ J^q $). As can be seen, there are some differences between Set 1 and DK13 for $ E^u $, while both of them predict the same results for $ E^d $. Considering also the small differences between our results and DK13 for $ H^u $ and $ H^d $ (especially at medium and large $ x $) one can conclude that the difference observed between Set 1 and DK13 for $ J^u $ in Fig.~\ref{fig:JuJd} is mainly due to the difference between them for $ E^u $ in Fig.~\ref{fig:Eu}. However, one should also consider the lake of the sea-quark contributions in our calculations of $ J^q $ using Set 1.

Another point can be inferred from Fig.~\ref{fig:JuJd} is that the inclusion of the WACS data in the analysis leads to a considerable increase in $ J^d $ ($ J^u $ is also increased but not as severely as $ J^d $). Although the value of $ J^u $ is closer to DK13 than before (Set 1), $ J^d $ begins to differ after considering new data and departs from zero range. Note that since Sets 1, 9, and 11 have almost the same predictions for $ xH_v^u $ and $ xE_v^u $ at $ t=0 $ (see Figs.~\ref{fig:HuWACS} and~\ref{fig:EuWACS}), the increase in the value of $ J^u $ from Set 1 to Sets 9 and 11 can be attributed to considering the sea-quark contributions in the last two cases. In the case of $ J^d $, the large difference of the prediction of Set 1 from the corresponding ones of Sets 9 and 11 becomes understandable by considering Fig.~\ref{fig:EdWACS}, where $ xE_v^d(x) $ has been changed dramatically after the inclusion of the WACS data in the analysis. Actually, in this case, the integrate of $ xH_v^d $ overcomes the corresponding one of $ xE_v^d $, which leads to a positive value for $ J^d $. Note also that although Set 12 has a very different prediction at large values of $ -t $ according to Fig.~\ref{fig:JFF}, its result for $ J_v^d $ at $ t=0 $ is more compatible with DK13 than Sets 9 and 11. This again shows the problem with the WACS data at large $ -t $.

%

\section{Summary and conclusion}\label{sec:five} 
 
In this work, following the previous studies performed to determine the polarized GPDs $ \widetilde{H}^q $~\cite{Hashamipour:2019pgy,Hashamipour:2020kip} and unpolarized GPDs $ H^q $ and $ E^q $~\cite{Hashamipour:2021kes}, we determined these three kinds of GPDs with their uncertainties at zero skewness ($ \xi=0 $) by performing a simultaneous analysis of all available experimental data of the nucleon FFs, nucleon charge and magnetic radii, proton AFF, and WACS cross section for the first time {and investigated whether there is any tension between these data}. To this aim, we first performed a base fit considering the world electron scattering data presented in Ref.~\cite{Ye:2017gyb} ({YAHL18}), the world $ G_M^p $ measurements from the {AMT07} analysis~\cite{Arrington:2007ux} and {Mainz} data~\cite{A1:2013fsc}, as well as the nucleon radii data taken from the Review of Particle Physics~\cite{ParticleDataGroup:2018ovx}. We also investigated the impact of excluding the {AMT07} and {Mainz} data in turn on final GPDs. As a result, we found that the {Mainz} data have the most impact on GPD $ E_v^u $, while the {AMT07} data have an important role in constraining the whole $ -t $ behavior of the up-quark unpolarized GPDs, especially $ H_v^u $. Overall, we concluded that the inclusion of both {AMT07} and {Mainz} data in the analysis is preferred, though it leads to an almost large $ \chi^2 $ for them. 

We investigated also the impact of CLAS Collaboration measurements of $ G_A $~\cite{CLAS:2012ich} on GPDs as a separate study. These data are more accurate and cover higher values of $ -t $ compared with the older world data of $ G_A $. We showed that the original CLAS data have a great deal of impact on polarized GPDs $ \widetilde{H}^q $, as expected. To be more precise, they need a significant suppression of GPDs $ \widetilde{H}^q $ at larger values of $ -t $ to be well fitted. Moreover, we found that there is a crucial tension between the CLAS data and the other world data of $ G_A $. We indicated that the CLAS data have some problems with normalization, such that it is not possible to obtain a simultaneously good description of all $ G_A $ data without normalization. By normalizing the CLAS data, we showed that the results become more compatible with the corresponding ones that have been obtained by excluding the CLAS data.
For the case of unpolarized GPDs $ H_v^q $ and $ E_v^q $, the significant changes are happening just for the down-quark distributions; the CLAS data (whether original or normalized) do not considerably affect the up-quark unpolarized GPDs.

As a further investigation, we studied the impact of the JLab Hall A
Collaboration data~\cite{Danagoulian:2007gs} of WACS on GPDs. We indicated that including the WACS data in the analysis can provide new and important information about the three kinds of GPDs, especially at large values of $ -t $. They can also provide some information about the sea-quark contributions of GPDs. Moreover, we found that there are some tensions between the CLAS and WACS data. So, we performed three independent analyses: (i) by excluding the CLAS data, (ii) by including the original CLAS data, and (iii) by including the normalized CLAS data.  In this way, we investigated both the pure impact of the WACS data on GPDs and the possibility of the simultaneous analysis of the WACS and CLAS data as well as its impact on GPDs. As a result, we found that the best $ \chi^2 $ belongs to the analysis in which the CLAS data have been excluded, although the results do not show a very desirable description of the WACS data even after
excluding the CLAS data. However, we showed that utilizing a normalization factor for the CLAS data makes the situation better. Overall, the inclusion of the WACS data in the analysis makes the description of the neutron data worse. They have the greatest impact on the $ d $ flavor of GPDs namely  $ H^d$,  $ E^d$, and $ \widetilde{H}^d $ although they lead also to a moderate suppression of $ \widetilde{H}^u $ with $ -t $ growing. On the other hand, including the original CLAS data in the analysis dramatically changes the polarized profile functions $ \widetilde{f}^q $ and thus GPDs $ \widetilde{H}^q $, while their impacts become less drastic in most cases after considering a normalization factor. Overall, the sea-quark GPDs behave almost as the valence polarized GPDs by including the WACS and CLAS data in the analysis.

By performing a through investigation on the poor description of the WACS data, we found that there is a hard tension between them and the {AMT07} and {Mainz} $ G_M^p $ data, while they are in a fair consistency with the proton and neutron electromagnetic FF	 data of {YAHL18} analysis, the nucleon radii data, and the axial FF data. By excluding the $ G_M^p $ data from the analysis, we indicated that the description of the WACS data becomes acceptable, though it spoils the $ G_M^p $ data description at $ -t>0.4 $ GeV$ ^2 $. According to the results obtained, we concluded that there are two possibilities: (i) the theoretical description of the WACS cross section should be revised, or (ii) the measurements of the WACS or $ G_M^p $, especially at larger values of $ -t $, should be revised.

As the last step, we calculated the gravitational FF $ M_2 $ and the the proton total angular momentum $ J^p $ using the extracted GPDs and compared their results with the corresponding ones from the LCSR~\cite{Azizi:2019ytx} and lattice calculations~\cite{LHPC:2007blg}. We showed that our results are in relatively good agreement with LCSR and lattice QCD when all experimental data are included in the analyses and the sea-quark contributions are considered too. {Although there are some remarkable tensions between the CLAS and WACS data as well as the WACS and $ G_M^p $ data, removing any of these data sets from the analyses leads to more deviations of the results from the pure theoretical predictions}. However, the results obtained for th Ji sum rule~\cite{Ji:1996ek,Goloskokov:2008ib} that presents the angular momentum $ J^q $ carried by individual quarks inside the nucleon are not in good agreement with other studies for the case of the down contribution $ J^d $. This is due to the significant impact of the WACS data on the down GPDs, as mentioned before. On the other hand, by excluding the $ G_M^p $ data from the analysis, the results obtained for $ M_2 $ and $ J^p $ diverge significantly from the LCSR as $ -t $ increases. Actually, if one relies only on the WACS data, an egregious difference in the theoretical predictions of $ M_2 $ and $ J^p $ results. This convinces one to revise the theoretical description of the WACS process (or, more unlikely, the WACS and $ G_M^p $ measurements at larger $ -t $ values).
 These facts clearly show the importance of considering the WACS and AFF data as well as the sea-quark contributions. We emphasize also that the precise determination of GPDs at zero skewness is not possible except by performing a simultaneous analysis of all related experimental data {despite the tension observed between some of them}. Such an analysis also makes the phenomenological results obtained 
from the experimental data more consistent with the pure theoretical calculations, {although it is not possible to find at this stage a set of universal GPDs that provides a desirable description of these data simultaneously.} 

According to the results of the present study, the precise determination of GPDs at zero skewness through the global analysis of all available experimental data remains an open and interesting subject that requires crucial development in the theoretical calculations, phenomenological frameworks, and experimental information.  Future programs like those planned at JLab~\cite{Accardi:2020swt,Hague:2021xcc} or the future Electron-Ion Collider (EIC)~\cite{Schmookler:2022gxw} can shed new light on this issue.

%
\section*{ACKNOWLEDGMENTS}
M.~Goharipour and K.~Azizi are thankful to Iran Science Elites Federation (Saramadan) for the partial financial support provided under Grant No. ISEF/M/400150. 
M.~Goharipour and H.~Hashamipour thank the School of Particles and Accelerators, Institute for Research in Fundamental Sciences (IPM), for financial support provided for this research.

%
\vspace{1cm}
\textit{Note Added.}
The GPDs extracted in this study with their uncertainties in any desired
values of $ x $ and $ t $ are available upon request.

%


\end{document}